\documentclass[a4paper,11pt]{article}
\pdfoutput=1 

\usepackage{jheppub} 
\makeatletter
\DeclareRobustCommand*{\bfseries}{%
  \not@math@alphabet\bfseries\mathbf
  \fontseries\bfdefault\selectfont
  \boldmath
}
\makeatother

\usepackage{calc}
\usepackage{rotating}
\usepackage[english]{babel}
\usepackage{graphicx}
\usepackage{subfig}
\usepackage{float}
\usepackage{amsmath}
\usepackage{amssymb}
\usepackage{amsthm}
\usepackage{latexsym}
\usepackage{dcolumn}
\newcommand{\newc}{\newcommand*}

\long\def\begincomment#1\endcomment{%
        \begingroup\sf\baselineskip12pt#1\endgroup}

\newc{\etal}{\textrm{et al.}} 
\newc{\eg}{\textrm{e.g.}} 
\newc{\ie}{\textrm{i.e.}}
\newc{\etc}{\textrm{etc}}
\newc\vs{\textrm{vs.}}
\newc{\cl}{\rm {C.L.}}

\newc{\ev}{\ensuremath{\,\mathrm{eV}}}
\newc{\kev}{\ensuremath{\,\mathrm{keV}}}
\newc{\mev}{\ensuremath{\,\mathrm{MeV}}}
\newc{\gev}{\ensuremath{\,\mathrm{GeV}}}
\newc{\tev}{\ensuremath{\,\mathrm{TeV}}}
\newc{\MeV}{\mev} 
\newc{\TeV}{\tev}
\newc{\invpb}{\ensuremath{/\text{pb}}}
\newc{\invfb}{\ensuremath{\,\text{fb}^{-1}}}
\newc\nb{\ensuremath{\,\mathrm{nb}}} \newc\pb{\ensuremath{\,\mathrm{pb}}} \newc\fb{\ensuremath{\,\mathrm{fb}}}
\newc\pc{\ensuremath{\,\mathrm{pc}}}
\newc\kpc{\ensuremath{\,\mathrm{kpc}}}
\newc\mpc{\ensuremath{\,\mathrm{Mpc}}}
\newc\ps{\ensuremath{\,\mathrm{ps}}} 
\newc\cmeter{\ensuremath{\,\mathrm{cm}}} 
\newc\meter{\ensuremath{\,\mathrm{m}}} 
\newc\kmeter{\ensuremath{\,\mathrm{km}}}
\newc\second{\ensuremath{\,\mathrm{s}}}
\newc\msecond{\ensuremath{\,\mathrm{ms}}}
\newc\nsecond{\ensuremath{\,\mathrm{ns}}}
\newc\psecond{\ensuremath{\,\mathrm{ps}}}

\newc{\chisqmin}{\ensuremath{\chi^2_{\mathrm{min}}}}
\newc{\Delchisq}{\ensuremath{\Delta\chi^2}}
\newc{\chisq}{\ensuremath{\chi^2}}
\newc{\like}{\ensuremath{\mathcal{L}}}

\newc\lsim{\ensuremath{\mathrel{\rlap{\lower4pt\hbox{\hskip1pt$\sim$}}\raise1pt\hbox{$<$}}}}
\newc\gsim{\ensuremath{\mathrel{\rlap{\lower4pt\hbox{\hskip1pt$\sim$}}\raise1pt\hbox{$>$}}}}
\newc{\VEV}[1]{\ensuremath{\langle #1 \rangle}}
\newc{\dl}{\ensuremath{\stackrel{\leftarrow}{D}}}
\newc{\dr}{\ensuremath{\stackrel{\rightarrow}{D}}}

\newc{\bcenter}{\begin{center}}    \newc{\ecenter}{\end{center}}
\newc{\bfl}{\begin{flushleft}}    \newc{\efl}{\end{flushleft}}
\newc{\bfr}{\begin{flushright}}    \newc{\efr}{\end{flushright}}

\newc{\bi}{\begin{itemize}}
\newc{\ei}{\end{itemize}}
\newc{\bed}{\begin{description}}
\newc{\eed}{\end{description}}
\newc{\ben}{\begin{enumerate}}
\newc{\een}{\end{enumerate}}

\newc{\be}{\begin{equation}}
\newc{\ee}{\end{equation}}
\newc{\bea}{\begin{eqnarray}}
\newc{\eea}{\end{eqnarray}}
\newc{\ra}{\rightarrow}

\newc{\alphas}{\ensuremath{\alpha_s}}
\newc{\alphatwo}{\ensuremath{\alpha_2}}
\newc{\alphaone}{\ensuremath{\alpha_1}}
\newc{\alphai}[1]{\ensuremath{\alpha_{#1}}}
\newc{\alphaem}{\ensuremath{\alpha_{\mathrm{em}}}}
\newc{\alphaeff}{\ensuremath{\alpha_{\mathrm{eff}}}}
\newc{\sineff}{\ensuremath{\sin \theta_{\mathrm{eff}}}}
\newc{\sinsqeff}{\ensuremath{\sin^2 \theta_{\mathrm{eff}}}}
\newc{\dalphahad}{\ensuremath{\Delta \alpha_{\mathrm{had}}}}
\newc{\yt}{\ensuremath{h_t}} \newc{\yb}{\ensuremath{h_b}} \newc{\ytau}{\ensuremath{h_{\tau}}}
\newc\mz{\ensuremath{M_Z}} 
\newc\mw{\ensuremath{m_W}}
\newc\mZ{\mz}        \newc\mW{\mw}
\newc\mhsm{\ensuremath{ m_{H_{\mathrm{SM}}}}}
\newc{\mtop}{\ensuremath{ m_t}}               \newc{\mtpole}{\ensuremath{ M_t}}
\newc{\mbottom}{\ensuremath{ m_b}} 
\newc{\mtau}{\ensuremath{ m_{\tau}}}
\newc{\mt}{\mtpole}
\newc{\mb}{\mbottom} 
\newc{\rgg}{\ensuremath{R_{h}(\gamma\gamma)}}
\newc{\rzz}{\ensuremath{R_{h}(ZZ)}}
\newc{\rtwogg}{\ensuremath{R_{h_2}(\gamma\gamma)}}
\newc{\rtwozz}{\ensuremath{R_{h_2}(ZZ)}}
\newc{\ronegg}{\ensuremath{R_{h_1}(\gamma\gamma)}}
\newc{\ronezz}{\ensuremath{R_{h_1}(ZZ)}}
\newc{\rsiggg}{\ensuremath{R_{h_\textrm{sig}}(\gamma\gamma)}}
\newc{\rsigzz}{\ensuremath{R_{h_\textrm{sig}}(ZZ)}}
\newc{\llbar}{\ensuremath{\ell\bar{\ell}}}
\newc{\tauptaum}{\ensuremath{ \tau^+\tau^-}}
\newc{\qqbar}{\ensuremath{ q\bar{q}}} \newc{\ppbar}{\ensuremath{ p\bar{p}}}
\newc{\bbbar}{\ensuremath{ b\bar{b}}} \newc{\ttbar}{\ensuremath{ t\bar{t}}}
\newc{\ffbar}{\ensuremath{ f\bar{f}}} \newc{\tautaubar}{\ensuremath{ \tau\bar{\tau}}}
\newc{\mchi}{\ensuremath{m_{\chi}}}
\newc{\squark}{\ensuremath{\tilde{q}}}
\newc{\slepton}{\ensuremath{\tilde{l}}}
\newc{\gluino}{\ensuremath{\tilde{g}}} 
\newc{\wino}{\ensuremath{\tilde{W}}}
\newc{\bino}{\ensuremath{\tilde{B}}}
\newc{\mgluino}{\ensuremath{{m_{\gluino}}}}
\newc{\tone}{\ensuremath{{\tilde{t}_1}}}

\newc{\sthw}{\ensuremath{ \sin\theta_W}}              \newc{\cthw}{\ensuremath{\cos\theta_W}}
\newc{\tanthw}{\ensuremath{ \tan\theta_W}}              \newc{\cotthw}{\ensuremath{\cot\theta_W}}
\newc{\ssqthw}{\ensuremath{\sin^2 \theta_W}}
\newc{\msbar}{\ensuremath{\overline{MS}}} \newc{\drbar}{\ensuremath{\overline{DR}}}
\newc{\mtmtsmmsbar}{\ensuremath{ m_t(m_t)^{\msbar}_{{\mathrm{SM}}}}}
\newc{\mtmtsmdrbar}{\ensuremath{ m_t(m_t)^{\drbar}_{{\mathrm{SM}}}}}
\newc{\mtmtmssmdrbar}{\ensuremath{ m_t(m_t)^{\drbar}_{{\mathrm{SUSY}}}}}
\newc{\mbmbmsbar}{\ensuremath{ m_b(m_b)^{\msbar} }}
\newc{\mbmbsmmsbar}{\ensuremath{ m_b(m_b)^{\msbar}_{{\mathrm{SM}}}}}
\newc{\mbmzsmmsbar}{\ensuremath{ m_b(\mz)^{\msbar}_{{\mathrm{SM}}}}}
\newc{\mbmzsmdrbar}{\ensuremath{ m_b(\mz)^{\drbar}_{{\mathrm{SM}}}}}
\newc{\mbmzmssmdrbar}{\ensuremath{ m_b(\mz)^{\drbar}_{{\mathrm{SUSY}}}}}
\newc{\mtaumzsmmsbar}{\ensuremath{ m_{\tau}(\mz)^{\msbar}_{{\mathrm{SM}}}}}
\newc{\mtaumzsmdrbar}{\ensuremath{ m_{\tau}(\mz)^{\drbar}_{{\mathrm{SM}}}}}
\newc{\mtaumzmssmdrbar}{\ensuremath{ m_{\tau}(\mz)^{\drbar}_{{\mathrm{SUSY}}}}}
\newc{\alphasmzms}{\ensuremath{\alpha_s(M_Z)^{\overline{MS}}}}
\newc{\alphaimzms}[1]{\ensuremath{\alpha_{#1}(M_Z)^{\overline{MS}}}}
\newc{\alphaemmz}{\ensuremath{\alpha_{\mathrm{em}}(M_Z)^{\overline{MS}}}}

\newc{\mzero}{\ensuremath{{m_0}}}
\newc{\mhalf}{\ensuremath{ m_{1/2}}}
\newc{\tanb}{\ensuremath{\tan\beta}}
\newc{\azero}{\ensuremath{ A_0}}
\newc{\bzero}{\ensuremath{ B_0}}
\newc{\signmu}{\ensuremath{\rm{sgn}\,\mu}}
\newc{\mueff}{\ensuremath{\mu_{\rm{eff}}}}
\newc{\lam}{\ensuremath{{\lambda}}}
\newc{\kap}{\ensuremath{{\kappa}}}
\newc{\alam}{\ensuremath{{A_{\lambda}}}}
\newc{\akap}{\ensuremath{{A_{\kappa}}}}
\newc{\hs}{\ensuremath{ H_s}}      
\newc{\mhs}{\ensuremath{ m_{H_s}}} 
\newc{\mgut}{\ensuremath{ M_{\rm GUT}}}
\newc{\mplanck}{\ensuremath{ M_{\rm P}}}      \newc{\mpl}{\ensuremath{ M_{\rm Pl}}}
\newc{\msusy}{\ensuremath{ M_{\rm SUSY}}}      \newc{\ms}{\ensuremath{ M_{\rm S}}}
 \newc{\mhl}{\ensuremath{m_\hl}} 
 \newc{\mhone}{\ensuremath{m_{h_1}}} 
 \newc{\mhtwo}{\ensuremath{m_{h_2}}} 
 \newc{\mglu}{\ensuremath{m_{\tilde g}}} 
 \newc{\mul}{\ensuremath{m_{\tilde{u}_L}}} 
 \newc{\mtone}{\ensuremath{m_{\tilde{t}_1}}} 
 \newc{\ma}{\ensuremath{m_A}} 
 \newc{\maone}{\ensuremath{m_{a_1}}} 
 \newc{\matwo}{\ensuremath{m_{a_2}}}
 \newc{\hone}{\ensuremath{h_1}}
 \newc{\htwo}{\ensuremath{h_2}}
 \newc{\aone}{\ensuremath{a_1}}
 \newc{\atwo}{\ensuremath{a_2}}
 \newc{\mhu}{\ensuremath{ m_{H_u}}}       
 \newc{\mhd}{\ensuremath{ m_{H_d}}}
 \newc{\mhusq}{\ensuremath{ m_{H_u}^2}}       
 \newc{\mhdsq}{\ensuremath{ m_{H_d}^2}}
 \newc{\mhuew}{\ensuremath{ m^{\ast}_{H_u}}}       
 \newc{\mhdew}{\ensuremath{ m^{\ast}_{H_d}}}
 \newc{\mhuewsq}{\ensuremath{ m^{\ast\, 2}_{H_u}}}       
 \newc{\mhdewsq}{\ensuremath{ m^{\ast\, 2}_{H_d}}}
 \newc{\hu}{\ensuremath{ H_u}}       
 \newc{\hd}{\ensuremath{ H_d}}
 \newc{\barmhu}{\ensuremath{ \bar{m}_{H_u}}}
 \newc{\barmhd}{\ensuremath{ \bar{m}_{H_d}}}

 \newc{\mqthree}{\ensuremath{m_{\widetilde{Q}_3}^2}}
 \newc{\muthree}{\ensuremath{m_{\tilde{u}_3}^2}}
 \newc{\mdthree}{\ensuremath{m_{\tilde{d}_3}^2}}
 \newc{\mlthree}{\ensuremath{m_{\widetilde{L}_3}^2}}
 \newc{\methree}{\ensuremath{m_{\tilde{e}_3}^2}}
 \newc{\mqtwo}{\ensuremath{m_{\widetilde{Q}_2}^2}}
 \newc{\mutwo}{\ensuremath{m_{\tilde{u}_2}^2}}
 \newc{\mdtwo}{\ensuremath{m_{\tilde{d}_2}^2}}
 \newc{\mltwo}{\ensuremath{m_{\widetilde{L}_2}^2}}
 \newc{\metwo}{\ensuremath{m_{\tilde{e}_2}^2}}
 \newc{\mqone}{\ensuremath{m_{\widetilde{Q}_1}^2}}
 \newc{\muone}{\ensuremath{m_{\tilde{u}_1}^2}}
 \newc{\mdone}{\ensuremath{m_{\tilde{d}_1}^2}}
 \newc{\mlone}{\ensuremath{m_{\widetilde{L}_1}^2}}
 \newc{\meone}{\ensuremath{m_{\tilde{e}_1}^2}}
 \newc{\mone}{\ensuremath{M_1}}
 \newc{\monesq}{\ensuremath{M_1^2}}
 \newc{\mtwo}{\ensuremath{M_2}}
 \newc{\mtwosq}{\ensuremath{M_2^2}}
 \newc{\mthree}{\ensuremath{M_3}}
 \newc{\mthreesq}{\ensuremath{M_3^2}}
 \newc{\atau}{\ensuremath{{A_{\tau}}}}
 \newc{\at}{\ensuremath{{A_{t}}}}
 \newc{\ab}{\ensuremath{{A_{b}}}}
 \newc{\atausq}{\ensuremath{{A_{\tau}^2}}}
 \newc{\atsq}{\ensuremath{{A_{t}^2}}}
 \newc{\absq}{\ensuremath{{A_{b}^2}}}

 \newc{\dmzero}{\ensuremath{\Delta{_{m_0}}}}
 \newc{\dmhalf}{\ensuremath{\Delta{_{m_{1/2}}}}}
 \newc{\dmu}{\ensuremath{\Delta{_{\mu}}}}

 \newc{\pten}{\ensuremath{\psi_{10}}}
 \newc{\ffive}{\ensuremath{\phi_{5}}}
 \newc{\hfive}{\ensuremath{h_{5}}}
 \newc{\hbfive}{\ensuremath{h_{\bar{5}}}}
 \newc{\thet}{\ensuremath{\theta_{50}}}
 \newc{\thetb}{\ensuremath{\theta_{\,\overline{50}}}}
 \newc{\ptenhat}{\ensuremath{\hat{\psi}_{10}}}
 \newc{\ffivehat}{\ensuremath{\hat{\phi}_{5}}}
 \newc{\hfivehat}{\ensuremath{\hat{h}_{5}}}
 \newc{\hbfivehat}{\ensuremath{\hat{h}_{\bar{5}}}}
 \newc{\thethat}{\ensuremath{\hat{\theta}_{50}}}
 \newc{\thetbhat}{\ensuremath{\hat{\theta}_{\,\overline{50}}}}
 \newc{\si}{\ensuremath{\Sigma}}
 \newc{\mfive}{\ensuremath{m_5^2}}
 \newc{\mten}{\ensuremath{m_{10}^2}}
 \newc{\dfive}{\ensuremath{\Delta^2_5}}
 \newc{\dbfive}{\ensuremath{\Delta^2_{\bar{5}}}}
 \newc{\dfifty}{\ensuremath{\Delta^2_{50}}}
 \newc{\dfiftyb}{\ensuremath{\Delta^2_{\,\overline{50}}}}
 \newc{\msi}{\ensuremath{m_{\Sigma}^2}}
 \newc{\lamh}{\ensuremath{\lambda_{H}}}
 \newc{\lamhb}{\ensuremath{\lambda_{\bar{H}}}}
 \newc{\ah}{\ensuremath{A_{H}}}
 \newc{\ahb}{\ensuremath{A_{\bar{H}}}}
 \newc{\lams}{\ensuremath{\lambda_{S}}}
 \newc{\as}{\ensuremath{A_{S}}}
 \newc{\lamsig}{\ensuremath{\lambda_{\si}}}
 \newc{\asig}{\ensuremath{A_{\si}}}

 \newc{\msten}{\ensuremath{m_{16}^2}}
 \newc{\mhun}{\ensuremath{m_{126}^2}}
 \newc{\mhunb}{\ensuremath{m_{\bar{126}}^2}}
 \newc{\mthun}{\ensuremath{m_{210}^2}}
 \newc{\ahun}{\ensuremath{A_{\bar{126}}}}
 \newc{\yhun}{\ensuremath{Y_{\bar{126}}}}
 \newc{\aten}{\ensuremath{A_{10}}}
 \newc{\yten}{\ensuremath{Y_{10}}}
 \newc{\alone}{\ensuremath{A_{\lambda_1}}}
 \newc{\altwo}{\ensuremath{A_{\lambda_2}}}
 \newc{\althree}{\ensuremath{A_{\lambda_3}}}
 \newc{\althreeb}{\ensuremath{A_{\bar{\lambda_3}}}}
 \newc{\lone}{\ensuremath{\lambda_1}}
 \newc{\ltwo}{\ensuremath{\lambda_2}}
 \newc{\lthree}{\ensuremath{\lambda_3}}
 \newc{\lthreeb}{\ensuremath{\bar{\lambda_3}}}

\newc{\sigsip}{\ensuremath{\sigma^{\rm SI}_{p}}}	\newc{\sigsin}{\ensuremath{\sigma^{\rm SI}_{n}}}
\newc{\sigsdp}{\ensuremath{\sigma^{\rm SD}_{p}}}	\newc{\sigsdn}{\ensuremath{\sigma^{\rm SD}_{n}}}
\newc{\sigsi}{\ensuremath{\sigma^{\rm SI}}}	\newc{\sigsd}{\ensuremath{\sigma^{\rm SD}}}
\newc{\sigv}{\ensuremath{\sigma v}}
\newc{\abund}{\ensuremath{ \Omega h^2}}
\newc{\omegadm}{\ensuremath{ \Omega_{{\rm DM}}}}     \newc{\abunddm}{\ensuremath{ \Omega_{{\rm DM}} h^2}} 
\newc{\omegam}{\ensuremath{ \Omega_{{\rm m}}}}       \newc{\abundm}{\ensuremath{ \Omega_{{\rm m}} h^2}}
\newc{\omegab}{\ensuremath{ \Omega_{{\rm b}}}}	\newc{\abundb}{\ensuremath{ \Omega_{{\rm b}} h^2}}
\newc{\omegatot}{\ensuremath{ \Omega_{{\rm TOT}}}}
\newc{\omegacdm}{\ensuremath{ \Omega_{{\rm CDM}}}}   \newc{\abundcdm}{\ensuremath{ \Omega_{{\rm CDM}} h^2}}
\newc{\omegalambda}{\ensuremath{ \Omega_{\Lambda}}} \newc{\abundlambda}{\ensuremath{ \Omega_{\Lambda} h^2}}
\newc{\omegarad}{\ensuremath{ \Omega_{{\rm rad}}}}  \newc{\abundrad}{\ensuremath{ \Omega_{{\rm rad}} h^2}}
\newc{\rhocrit}{\ensuremath{ \rho_{\rm crit}}}
\newc{\rhochi}{\ensuremath{ \rho_{\chi}}}
\newc{\abunchi}{\ensuremath{\Omega_\chi h^2}}
\newc{\abundlsp}{\ensuremath{\Omega_{\rm LSP}h^2}}
\newc{\abundchi}{\ensuremath{\Omega_\chi h^2}}
\newc{\tf}{\ensuremath{T_f}} \newc{\xf}{\ensuremath{x_f}}
\newc{\tr}{\ensuremath{T_R}}

\newc{\amu}{\ensuremath{ a_{\mu}}}        \newc{\amususy}{\ensuremath{ a_{\mu}^{\mathrm{SUSY}}}}
\newc{\amuexpt}{\ensuremath{ a_{\mu}^{\mathrm{expt}}}}        \newc{\amusm}{\ensuremath{ a_{\mu}^{\mathrm{SM}}}}
\newc\deltaamu{\ensuremath{\Delta a_{\mu}}} \newc{\deltaamususy}{\ensuremath{\delta a_{\mu}^{\mathrm{SUSY}}}}
\newc\gmtwo{\ensuremath{ (g-2)_{\mu}}} 
\newc{\deltagmtwomususy}{\ensuremath{\delta\left(g-2\right)_{\mu}^{\mathrm{SUSY}}}}
\newc{\deltagmtwomu}{\ensuremath{\delta\left(g-2\right)_{\mu}}}
\newc\BR{\ensuremath{\rm BR}}
\newc\bsgamma{\ensuremath{ b\rightarrow s \gamma }}
\newc\bxsgamma{\ensuremath{\overline{B}\rightarrow X_{s}\gamma}}
\newc\brbsgamma{\ensuremath{\BR\left(\bsgamma\right)}}
\newc\brbxsgamma{\ensuremath{\BR\left(\bxsgamma\right)}}
\newc\bsmumu{\ensuremath{B_s\to\mu^+\mu^-}}
\newc\brbsmumu{\ensuremath{\BR\left(B_s\to\mu^+\mu^-\right)}}
\newc\bdmmumu{\ensuremath{\overline{B}_d\to\mu^+\mu^-}}
\newc\bbbarmix{\ensuremath{\overline{B}_s\mbox{-}B_s}}      
\newc\delmbs{\ensuremath{\Delta M_{B_s}}}
\newc{\butaunu}{\ensuremath{B_u \rightarrow \tau \nu}}
\newc{\brbutaunu}{\ensuremath{\BR\left(B_u \rightarrow \tau \nu\right)}}

\newcommand*{\reffig}[1]{Fig.~\ref{#1}}
 
        \newcommand*{\refeq}[1]{Eq.~(\ref{#1})}
     \newcommand*{\refsec}[1]{Sec.~\ref{#1}}

\newcommand*{\mstopone}{\ensuremath{m_{\tilde{t}_1}}}
\newcommand*{\mstoptwo}{\ensuremath{m_{\tilde{t}_2}}}

\let\oldcite\cite
\renewcommand*{\cite}{~\oldcite}

\newcommand*{\hl}{\ensuremath{h}}

\restylefloat{figure}

\title{WIMP dark matter candidates and searches -- current status and
  future prospects}

\author[a,b]{Leszek Roszkowski,}
\author[c,a]{Enrico Maria Sessolo}
\author[d,a]{and Sebastian Trojanowski}

\affiliation[a]{National Centre for Nuclear Research,\\
  Ho{\. z}a 69, 00-681 Warsaw, Poland}
\affiliation[b]{Department of Physics and Astronomy, University of Sheffield,\\Sheffield S3 7RH, United Kingdom}
\affiliation[c]{Institut f\"ur Physik, Technische Universit\"at Dortmund,\\ D-44221 Dortmund, Germany}
\affiliation[d]{Department of Physics and Astronomy, University of California, Irvine,\\California 92697, USA}

\emailAdd{leszek.roszkowski@ncbj.gov.pl}
\emailAdd{enrico.sessolo@ncbj.gov.pl}
\emailAdd{sebastian.trojanowski@ncbj.gov.pl}

\abstract{ We review several current aspects of dark matter theory
  and experiment.  We overview the present experimental status, which
  includes current bounds and recent  claims and hints of a possible signal in a wide range
  of experiments: direct detection in underground laboratories,
  gamma-ray, cosmic ray, X-ray, neutrino telescopes, and the LHC.  We
  briefly review several possible particle candidates for a Weakly
  Interactive Massive Particle (WIMP) and dark matter that have
  recently been considered in the literature. We pay
  particular attention to the lightest neutralino of supersymmetry as it remains
  the best motivated candidate for dark matter and also shows
  excellent detection prospects. Finally we briefly review some
  alternative scenarios that can considerably alter properties and
  prospects for the detection of dark matter obtained within the
  standard thermal WIMP paradigm.
}

\dedicated{UCI-HEP-TR-2017-09\\ DO-TH 17/15}

\begin{document}
\maketitle
\flushbottom
\section{\label{sec:intro}Introduction}

One of the most important quests of contemporary physics is to
understand the nature of dark matter (DM) in the Universe.  The
long-held paradigm is that most DM is cold (CDM) and is made up of
some weakly interacting massive particles (WIMPs). The WIMP solution
to the DM problem remains attractive for a number of reasons.
Firstly, WIMPs arise naturally in a large number of theoretically
well-motivated models. Secondly,
for reasonable ranges of WIMP mass and annihilation cross section the
relic abundance of DM can be obtained through the robust mechanism of
thermal freeze-out, possibly augmented with some other production
mechanisms. Lastly, thermal
  WIMPs represent a promising target for DM experiments because a
  large fraction of their typical detection rates are within reach of
  current or planned detectors, making them testable by experiment.

We note that the concept of a WIMP as used in the literature is
somewhat ambiguous.  In general it encompasses a broad category of
hypothetical candidates coming from specific theoretical scenarios, or
their classes. In general it includes any non-baryonic massive
particle (even with a very tiny mass) that interacts with any
interaction that is either weak or sub-weak (\eg, axionic,
gravitational). In a more commonly
used sense,  WIMP refers to a particle with mass in the
range from about 2\gev\ (the so-called Lee-Weinberg
bound)\footnote{The bound was actually derived by more authors. For
  more references, see\cite{Kolb:1990vq}.} up to some 100\tev\ (a
rough unitarity bound\cite{Griest:1989wd}), whose interactions are set
basically by the weak interaction coupling of the Standard Model,
although strongly suppressed, as otherwise one would run into conflict
with upper limits on its detection cross section.

In this topical review we will focus on the latter, ``proper'' WIMP
category, and will cover the present status of some of the most popular and
robust WIMP candidates and prospects for their detection. To this end
we will also briefly survey the current experimental search situation,
focusing in particular on several recent claims, or hints, of measuring a DM signal.

The field of dark matter is very broad and remains an arena of intense
research both on the theoretical and experimental sides. Its various
aspects have been covered in a number of review articles and here we
mention but some of them. Observational evidence for dark matter can
be found, \eg, in\cite{DAmico:2009tep,DelPopolo:2013qba}. Many WIMP
particle candidates and prospects for their detection have been
covered in several papers, starting from the early comprehensive
review\cite{Jungman:1995df} (which, nearly thirty
  years later, still remains a very useful classic reference) and more recently
in\cite{Bertone:2004pz,Feng:2010gw,Bednyakov:2015uoa,Arcadi:2017kky}
and\cite{Baer:2014eja} which, although mostly devoted to non-standard WIMPs, like axinos and gravitinos, in the first chapter
contains a summary of the current views on WIMPs from a particle
physics perspective. References\cite{Conrad:2015bsa,Gaskins:2016cha}
provide a recent succinct update on indirect detection aspects of WIMP
searches.

The review is organized as follows. In the remainder of this Section
we briefly summarize observational arguments for DM. Next, in
Sec.~\ref{sec:whywimps} we present the case for the WIMP solution to
the DM puzzle. We start by outlining some general properties of WIMPs
(Subsec.~\ref{sec:generalwimps}), then discuss its production
mechanisms in the early Universe (Subsec.~\ref{sec:wimpproduction})
and finally (Subsec.~\ref{sec:candidates}) comment on several specific
WIMP candidates that have recently been discussed in the literature --
the purpose of this is to provide a broader perspective on the current
speculations about particle candidates for DM. In
Sec.~\ref{sec:experimentalsummary} we turn to the experimental
searches and briefly review the current situation both in direct
detection (DD) in underground searches (Subsec.~\ref{sec:dd}) and
(Subsecs~\ref{sec:gammarays} -- \ref{sec:xray}) in several modes of
indirect detection (ID), focusing in particular on some recent claims
of DM detection. Finally, in Subsec.~\ref{sec:lhcdm}, we summarize the
searches for DM-like particles at the LHC. We devote
Sec.~\ref{sec:neutralino} to the arguably most popular WIMP candidate,
the lightest neutralino of supersymmetry (SUSY) by reviewing and
updating its properties and prospect for detection in light of recent
progress in DM searches and also of SUSY searches and Higgs boson
discovery at the LHC. While most of the Section deals with well known
SUSY frameworks and standard assumptions, we conclude it by presenting
some recent works on relaxing them and discuss ensuing
implications. In Sec.~\ref{sec:conclusions} we provide a summary and
outlook.

\subsection{Evidence for dark matter}   

Over the last decades observational evidence for the existence of
large amounts of DM in the Universe has been steadily
mounting, and is well now described in several
reviews\cite{DAmico:2009tep,DelPopolo:2013qba}. Here we merely briefly
summarize some of the better known arguments.

The first claim about the existence of DM is usually
attributed to Zwicky's original paper on the Coma
Cluster\cite{Zwicky:1933gu}.\footnote{Earlier speculations were made
  by Kapteyn\cite{Kapteyn:1922}, Oort\cite{Oort:1932} and Jeans\cite{Jeans:1922}. For a historical development,
  see\cite{Bertone:2016nfn}.}  The
cluster consists of more than a thousand galaxies.  Careful analysis of the
movement along their
gravitational orbits 
led to the
conclusion that there should be a large amount of non-luminous matter
contained in the cluster. Zwicky referred to it as ``dunkle
  Materie'' (dark matter) and apparently thought it was just ordinary
matter in a non-shining form.

One of the most widely recognized arguments for the existence of DM is
based on galaxy rotation curves, i.e., the relation between
orbital velocity and radial distance of visible stars or gas from the
center of a galaxy.  It was first noted in the late 1930s that the outer
parts of the M31 disc were moving with unexpectedly high
velocities\cite{Babcock:1939}, an observation that was then confirmed
more than thirty years later\cite{Rubin:1970zza,Roberts:1975}.
According to these observations the velocities of distant stars in M31
remain roughly constant over a wide range of distances from the center of the
galaxy, in contradiction with expectations based on the distribution
of visible matter in the galaxy.  Similar results were later
obtained\cite{Rubin:1980zd} for various other spiral galaxies.

\begin{figure}[t]
\centering
\subfloat[]{%
\label{fig:a}%
\includegraphics[width=0.396\textwidth]{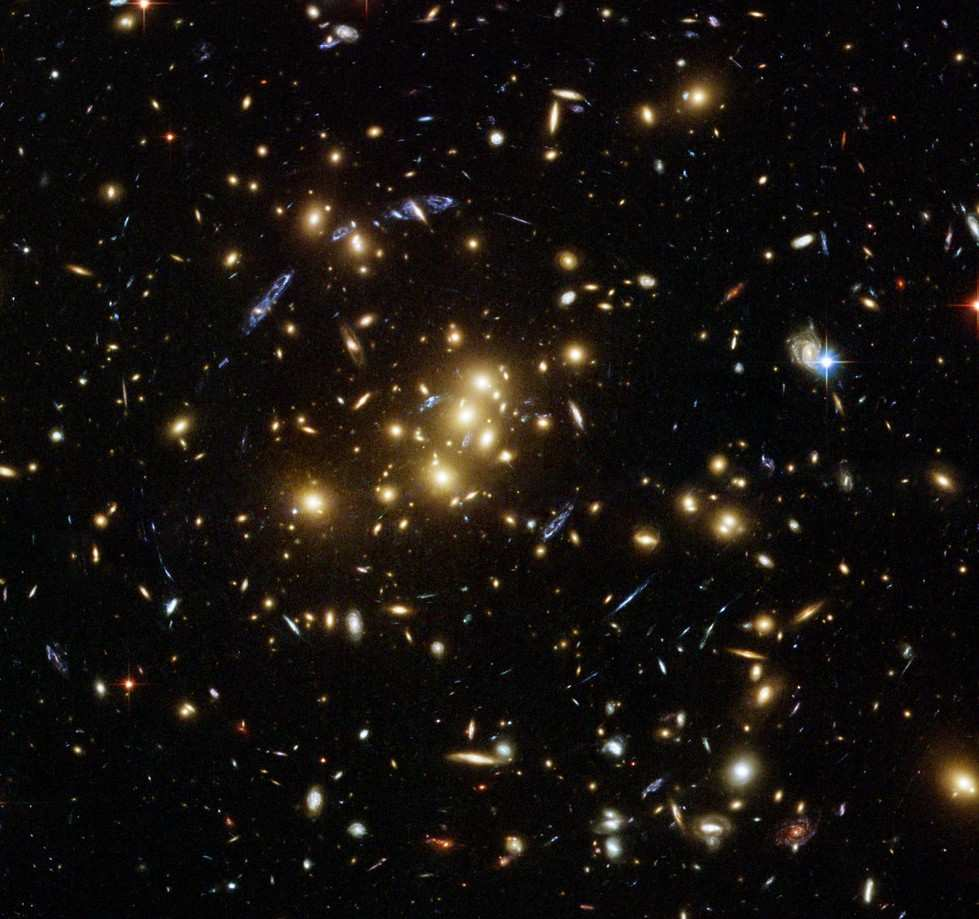}
}%
\hspace{0.02\textwidth}
\subfloat[]{%
\label{fig:b}%
\includegraphics[width=0.554\textwidth]{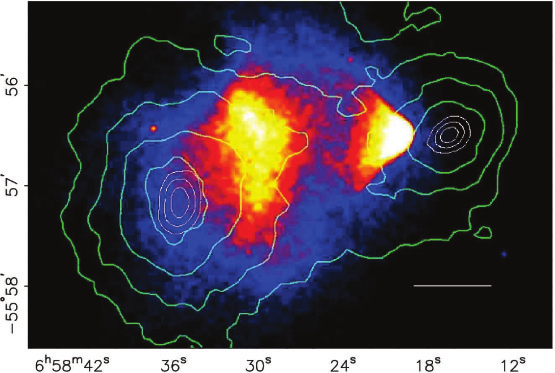}
}%
\caption{\footnotesize (a) Strong gravitational lensing around galaxy cluster CL0024+17. Taken from Ref.\cite{Massey:2010hh}. 
(b) Bullet Cluster mass density contours (green) and the 
distribution of baryonic matter. Taken from 
Ref.\cite{Clowe:2006eq}.}
\label{fig:lens}
\end{figure}

The existence of DM is also supported by data coming from
gravitational lensing.
Gravitational lensing, or the bending of light in a strong
gravitational field (for a review see,
\eg,\cite{Massey:2010hh}), is most easily observed when
light passes through a very massive and/or dense object, like a galaxy
cluster or the central region of a galaxy.  Light rays can bend around
the object, or lens, leading to a distortion of the image
of the light source, as can be seen in \reffig{fig:lens}(a). This
effect is commonly known as strong lensing.  The size and
shape of the image can be used to determine the distribution of mass
in the lens which can then be compared with the visible
mass. 

When the lens is not as massive as in the case of strong lensing, or
when light travels far from the core of the galaxy or cluster, the
effect is much weaker. However, it can still be analyzed even in the
case of individual stars.  Microlensing effects of this kind
were proposed\cite{Petrou:1981,Paczynski:1985jf} to look for DM
in the Milky Way in form of Massive Compact Halo Objects (MACHOs),
which should cause an occasional brightening of stars from nearby
galaxies.  This strategy led to an exclusion of MACHOs with masses in
the range $0.6\times 10^{-7}$ to $ 15M_{\odot}$ as the dominant form
of DM in the Galaxy\cite{Tisserand:2006zx}.

Perhaps the most spectacular argument for the existence of dark matter
in clusters can be found in the Bullet Cluster. It consists of two
clusters of galaxies which have undergone a head-on collision\cite{Clowe:2006eq}. The hot-gas clouds (observed
through their X-ray emission) that contain the majority of the
baryonic mass in both clusters have been decelerated in the collision
while the movement of the galaxies and the dark matter halos in clusters remained almost intact.
Analysis of the gravitational lensing effect shows that the
center of mass for both clusters is clearly separated from the gas
clouds, as can be seen in \reffig{fig:lens}(b).  One can thus infer
the presence of a large amount of additional mass in both
clusters. The Bullet Cluster is the first known example of a system
where the dark matter and the baryonic component have been separated
from each other.

Studies of weak gravitational lensing of large scale structures (LSS)
provide further evidence for DM. In this context the effect is usually
called  cosmic shear. It causes systematic distortions of
the positions of distant galaxies, though the impact is very subtle
($\sim0.1\% - 1\%$). Tangential shear is usually analyzed in terms of
two-point (or even three-point) correlation functions that on the
other hand can be related to the DM mass density correlation
functions. The latter quantity is a Fourier transform of the matter
power spectrum and can be used to determine the matter density (both
ordinary and dark) of the Universe; see, \eg,\cite{Fu:2014loa}.

\begin{figure}[t]
\centering
\subfloat[]{%
\label{fig:a}%
\includegraphics[width=0.55\textwidth]{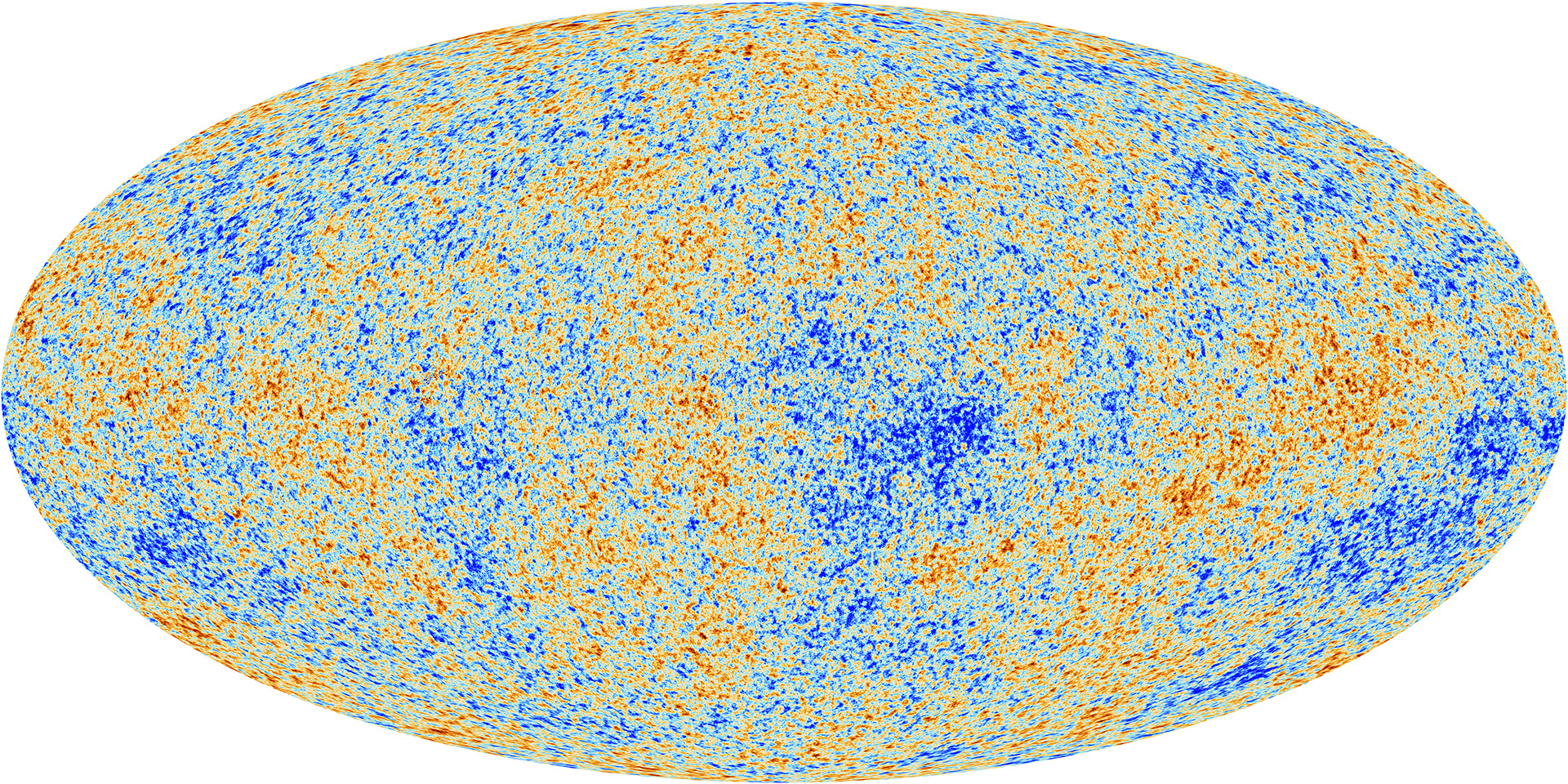}
}%
\hspace{0.02\textwidth}
\subfloat[]{%
\label{fig:b}%
\includegraphics[width=0.405\textwidth]{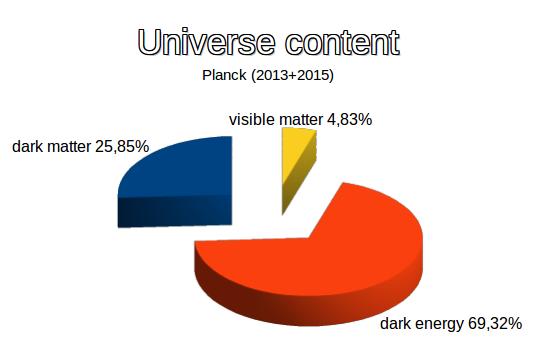}
}%
\caption{\footnotesize (a) Temperature anisotropy of the CMB after the
  first results released by the Planck
  Collaboration\cite{Ade:2013zuv}.  (b) Total mass-energy distribution
  in the Universe.}
\label{fig:massenergycontent}
\end{figure}

Last but not least, a crucial role in determining the DM abundance
in the Universe is played by studies of cosmic microwave background
(CMB) radiation. The CMB radiation seen today originates from the
decoupling and recombination epoch.  Small inhomogeneities in the
distribution of its temperature correspond to fluctuations of the
matter density in the early Universe that subsequently gave rise to
the observed large structures. The power spectrum of temperature
anisotropies (see Fig.~\ref{fig:massenergycontent}(a)) when expanded
in terms of spherical harmonics depends on cosmological parameters can
then be obtained by fitting the resulting spectrum, with some
underlying assumption of cosmological model, \eg, the
$\Lambda$CDM model.

The current  values\cite{Ade:2015xua} of the relic
abundance, that is the ratio of the density to the critical density,
of baryonic matter $\Omega_b$, and the corresponding quantity for the
non-baryonic DM component, $\Omega_{\textrm{DM}}$ that were obtained
by WMAP and more recently by PLANCK by fitting the six-parameter
$\Lambda$CDM model are: 
\begin{eqnarray}
\Omega_{b}\,h^2 &=& 0.02226(23),\\
\Omega_{\textrm{DM}}\,h^2 &=& 0.1186(20),
\label{eqOmegaDM}
\end{eqnarray}
where $h=H_0/100\kmeter \mpc\second  =0.678(9)$\cite{Ade:2015xua} is the
reduced Hubble constant, with $H_0$ denoting the Hubble constant today. 

The remaining dominant contribution, $\Omega_\Lambda \approx 0.692$, accounts
for the so-called dark energy (for a recent review see,
\eg,\cite{Li:2011sd,Weinberg:2012es}).  A schematic cartoon showing different
contributions to the mass-energy content of the Universe is shown in
\reffig{fig:massenergycontent}(b).

Further information about the amount of matter and dark energy
components of the Universe can be derived from analyses of baryon
acoustic oscillations (BAO, periodic fluctuations in the density of
baryonic matter that originated from the opposite effects of
gravitational attraction and radiation pressure), supernovae type Ia,
or from the Lyman-$\alpha$ forests (neutral hydrogen
clouds seen in absorption in quasar spectra).  In the case of
elliptical galaxies and galaxy clusters another important piece of
evidence for the existence of DM comes from the X-ray emission from
hot gas (for further discussion see,
\eg,\cite{DelPopolo:2013qba}).


\section{WIMPs as dark matter\label{sec:whywimps}}   

\subsection{General properties\label{sec:generalwimps}}

One clear conclusion about DM  that one can draw from
observational evidence is that DM is made up of some particles should be electrically
neutral.\footnote{One should not forget that in principle the DM
  puzzle could be explained in terms of something else than particles
  but such approaches suffer from problems. This includes modifying
  gravity (MOND)\cite{Milgrom:1983ca} which still needs to invoke DM
  in order to explain all data\cite{Angus:2007mn}, or cosmic
  fluid\cite{Bharadwaj:2003iw} which is also increasingly challenged
  by observations, see, \eg,\cite{Velten:2012uv}.} DM should interact
with ordinary matter preferably only weakly (or sub-weakly), where weak
can be taken to mean interacting via the weak nuclear force or just
having a (sub)weak but non negligible coupling to the Standard Model
particles.


Dark matter self-interactions cannot be too strong in order to be
compatible with constraints on structure formation and observations of
galaxy cluster systems such as the Bullet Cluster with current limits
of order $\sigma/m < 0.7
\textrm{cm}^2/\textrm{g}$\cite{Randall:2007ph}. Note, however, that
this limit can be satisfied even for strongly interacting dark matter
(SIMP) with $\alpha_\chi\sim 1$ in the dark sector and the correct
relic density obtained thanks to DM mass-dependent
$3\rightarrow 2$ processes\cite{Hochberg:2014dra}.
Moreover, to be in agreement with CMB
data, most of the DM should be non-baryonic
in nature.

It has been suggested that DM could be made up of primordial black
holes (PBHs)\cite{Carr:1974nx} 
 which, if formed
before the period of Big Bang Nucleosynthesis (BBN) would not violate
determinations of ordinary matter abundance, thus weakening the
argument for the need of non-baryonic DM today. The idea of PBHs as DM
has recently been revived with a suggestion that the first detection
of gravitational waves by LIGO\cite{Abbott:2016blz} could be
potentially explained in terms of two coalescing
PBHs\cite{Bird:2016dcv}. However, there are many limits on PBHs as DM;
see\cite{Carr:2016drx,Gaggero:2016dpq}.

One simple classification of DM particles is based on how relativistic they are around the time when they fall out of thermal equilibrium in the early Universe, i.e., when they decouple from thermal plasma. Hot dark matter (HDM) in the mass range of up to a few tens of eV,
which was still relativistic at the time of decoupling, due to large
mean free path did not cluster to form clumps as small as galaxies,
and does it reproduce the observed Universe in numerical simulations
of LSS formation (see, \eg,\cite{Frenk:2012ph}). It is incompatible
with data from the
LSS\cite{Abazajian:2005xn,dePutter:2012sh,Lukash:2012tq} and
deep-field observations\cite{Tremaine:1979we,White:1984yj}, which
constrain the allowed average velocity of the DM particles from above.
For these reasons, HDM can only contribute a small fraction,
determined by the properties of the CMB, of the total DM density. A
familiar (and known to exist) example of possible HDM is neutrinos
with a tiny mass.

In contrast, cold dark matter (CDM) behaves very
differently. Non-baryonic CDM decoupled from thermal plasma at
freeze-out and its density perturbations started growing linearly at
the onset of the epoch of matter dominance.  This provided early
potential wells (seeds), thus triggering and catalyzing the growth of
the density perturbations of baryonic matter after it decoupled from
radiation some time later. This is the basic reason why CDM generally
proves successful in reproducing observations in numerical simulations
of LSS, despite well known problems with potentially predicting too
few substructures (missing satellites
  problem)\cite{Moore:1999nt,Klypin:1999uc} or too dense regions
towards the center of the largest DM subhalos obtained in simulations
when comparing to the brightest observed dwarf satellite galaxies
(too-big-to-fail problem)\cite{BoylanKolchin:2011de,BoylanKolchin:2011dk} (for
recent review see, \eg,\cite{DelPopolo:2016emo}).  Cold DM, as opposed
to warm or hot DM is also preferred by the properties of the CMB.

Warm dark matter (WDM), in the mass
range of a few keV, is a possible form of DM which is intermediate
between HDM and CDM. It was still relativistic at the time of
decoupling but fluctuations corresponding to sufficiently large halos
would not be damped by free streaming. It has been considered as a
possible way of ameliorating some apparent problems of CDM, because it
reduces the power spectrum on small scales, thus reducing the missing
satellite problem of CDM\cite{Lovell:2011rd} although this has been
disputed. It has been claimed, however, that WDM leads to
some other problems\cite{Viel:2013apy}: the cutoff in the power
spectrum $P(k)$ at large wavelength $k$ implied by WDM will also
inhibit the formation of small dark matter halos at high redshift. But
such small halos are presumably where the first stars form, which
produce metals rather uniformly throughout the early Universe as
indicated by observations of the Lyman-alpha forests. 
An almost sterile neutrino with the mass of a few keV is a popular
candidate for WDM. One has to note, though, that such sterile
neutrinos are produced in the early Universe being not in thermal
equilibrium, hence their effect on the structure formation needs to be
studied in detail before drawing robust conclusions (for a recent
review see\cite{Abazajian:2017tcc}).

An array of these and related arguments have led to establishing a
popular (and sensible) paradigm that the dominant fraction of DM is
probably cold and that it should be not only (sub)weakly interacting
but also non-relativistic and massive, or in short, it is made up of
WIMPs. Finally, the DM particles should be either absolutely stable,
or extremely long lived (for instance, a recent analysis finds a lower
bound of 160 Gyr\cite{Audren:2014bca}). This is as much as we can be fairly confident about the
nature of DM.

Non-WIMP DM candidates (for a recent review
see\cite{Baer:2014eja}) have also been vastly explored in the
literature. Among them one can distinguish an ultralight axion that
emerges from the solution to the strong CP problem. Axion DM can
closely resemble CDM when axions, upon thermalization, form a
Bose-Einstein condensate with the energy density determined by the
mechanism of bosonic coherent motions. Another interesting scenario is
to consider extremely weakly interacting massive particles (EWIMPs) as
DM candidates. Such weak interactions can naturally appear, \eg, if
they are described by non-renormalizable operators suppressed by some
high energy scale, \eg, the Planck mass, $\mplanck\approx 10^{18}\gev$, 
as it is in the case of gravitino
DM\cite{Moroi:1993mb,Bolz:2000fu,Feng:2003xh,Feng:2003uy,Pradler:2006qh,Rychkov:2007uq}
or the Peccei-Quinn scale, $f_a\approx 10^{11}-10^{12}\gev$, for the
axino DM\cite{Covi:1999ty,Covi:2001nw}.

\subsection{Production mechanisms\label{sec:wimpproduction}}

One property of CDM that is now very precisely established is its
cosmological relic abundance -- compare Eq.~(\ref{eqOmegaDM}) -- and
the fact that it has been derived from the properties of CMB at the
time of recombination or from baryonic acoustic oscillations at the
earlier time after (dark) matter dominance started suggest that DM was indeed
produced in the early Universe. Big Bang nucleosynthesis also
place limits on the production of DM from decays during the BBN
epoch\cite{Kawasaki:2004yh,Kawasaki:2004qu,Jedamzik:2006xz}. It is
generally assumed that the bulk of dark matter in the Universe was
produced between the end of inflation (actually, reheating) and some
time before BBN. 

Several mechanisms for generating sufficient amounts of DM in the
early Universe have been identified and will be briefly reviewed below.

\subsubsection{DM production from freeze-out{\label{sec:thermalWIMPs}}}

The most robust mechanism for generating the WIMP DM relic abundance
is the so-called freeze-out mechanism. In the very early and hot
Universe SM species and DM were in thermal equilibrium, with DM
particle production from annihilations balancing each other out.  As
the Universe expanded and cooled, WIMPs eventually froze out
of equilibrium with the thermal plasma. This decoupling happened when
the WIMP annihilation rate became roughly less than the expansion rate
of the Universe $\Gamma_{\textrm{ann}}\lesssim H\sim
T_f^2/\overline{M}_P$, where $T_f$ stands for the freeze-out
temperature (the index $f$ indicates that quantities are evaluated at the freeze-out time) and $\overline{M}_P$ is the reduced Planck mass.  After
freeze-out the WIMP yield, $Y_\chi=n_\chi / s$, where $n_\chi$
(denoting here generic WIMPs with the symbol $\chi$) is the number
density of DM particles and $s\sim T^3$ is the entropy density,
remained mostly constant.
 
Given the annihilation rate, $\Gamma_{\textrm{ann}} =
n_\chi\langle\sigma_{\textrm{ann}} v\rangle$, one can rewrite the
formula for today's value of the DM relic abundance in terms of the
thermally averaged product of annihilation cross section
$\sigma_{\textrm{ann}}$ and the M\o ller velocity, $v_{\textrm{M\o l}} = \sqrt{(\vec{v}_1-\vec{v}_2)^2-(\vec{v}_1\times \vec{v}_2)^2}$, at freeze-out\cite{gondolo:1990dk},
\begin{equation}
  \Omega_{\chi}h^2 \simeq \frac{m_\chi\,n_{\chi}(T_0)}{\rho_c}\,h^2 =
  \frac{T_0^3}{\rho_c}\,\frac{x_f}{\overline{M}_P}\,\frac{1}{\langle\sigma_{\textrm{ann}}v_{\textrm{M\o
        l}}\rangle_f}\,h^2,\label{eq:omega} 
\end{equation}
where $T_0\approx 2.35\times 10^{-13}\,$GeV\cite{Olive:2016xmw} is the
temperature of the Universe at present, $\rho_c\approx 8\times
10^{-47}\,h^2\,\textrm{GeV}^4$\cite{Olive:2016xmw} is the critical
energy density, $x=m_\chi /T$ and $\vec{v}_{1,2}$ are the velocities of both annihilating DM particles. 
Note that M\o ller velocity is equal to the relative velocity of two DM particles $|\vec{v}_1-\vec{v}_2|$ in the center-of-mass frame.

The value of $x_f$ can be roughly estimated by assuming that around
freeze-out the number density of WIMP DM is equal to the the
non-relativistic equilibrium number density $n_\chi \approx
n_{\chi,\textrm{eq}}\simeq g_\chi\,\left(m_\chi
  T/2\pi\right)^{3/2}\,\exp{\left(-m_\chi/T\right)}$, where $g_\chi$
is the number of degrees of freedom for the DM particles. Using
$\Omega_\chi\,h^2\approx 0.12$ one then obtains
\begin{equation}
x_f^{3/2} e^{-x_f} \approx \frac{10^{-8}\,\textrm{GeV}}{m_\chi}.
\end{equation}
This leads to $x_f\approx 30$ for $m_{\chi} \approx 100\,\textrm{GeV} - 10\,\textrm{TeV}$. 
More careful analysis shows that the appropriate value is closer to
$x_f\approx 25$\cite{Nihei:2001qs}.

Finally we put the estimated value of $x_f$ back into Eq.~(\ref{eq:omega}) and find
\begin{equation}
\langle\sigma_{\textrm{ann}}v\rangle_f \approx 3\times 10^{-26}\,\textrm{cm}^3/\textrm{s},
\label{eq:sigmavvalue}
\end{equation}
for which the correct value of the WIMP DM relic density is obtained (see\cite{Steigman:2012nb} for a more detailed study).
For typical velocities $v\approx 0.1\,c$ one obtains a cross section of
weak strength for WIMP with mass around the electroweak scale.  
In the early days, this coincidence was found so remarkable that it
was coined as the ``WIMP miracle''. 

However, subsequent developments showed that the situation may well be
much more complex. A critique of the ``WIMP miracle'' can be found in
Ref.\cite{Baer:2014eja}. Here we merely mention that in specific well
motivated cases the relic abundance can often be different from 0.12
by several orders of magnitude. For instance, for the neutralino DM of
SUSY one typically finds \abundchi\ well in excess of the correct
value, as will be discussed in Sec.~\ref{sec:neutralino}. It is also
important to note that WIMP DM particle mass is not necessarily
confined to the electroweak scale. An argument based on unitarity
gives a generous upper bound on thermal relic mass of the order of
100\tev\cite{Griest:1989wd}. Furthermore,
on dimensional grounds, and for simplicity assuming that WIMP mass
\mchi\ is the only relevant scale, one expects
\begin{equation}
\sigma_{\textrm{ann}}\ \propto\ \frac{g^4}{m_\chi^2},
\end{equation}
where $g$ is a coupling constant responsible for the WIMP annihilation
process. One can see that Eq.~(\ref{eq:sigmavvalue}) can be then
satisfied for a wide range of masses (from 10\mev\ to 10\tev) and a
wide range of coupling constant values (from gravitational to strong)
as long as their ratio is kept fixed\cite{Feng:2008ya,Baer:2014eja}.
Even more freedom can be achieved in a more realistic scenario in
which DM-SM mediator mass and its coupling constants to the SM appear
in the annihilation cross section.

In a precise treatment, which takes into account the dynamics of
freeze-out, the DM yield after freeze-out is found by solving the
respective set of Boltzmann equations:\footnote{One assumes here
  Majorana DM particles. For Dirac particles there should appear an
  additional factor of two in the second term on the r.h.s. of
  Eq.~(\ref{eq:Boltzmannradiation}). However this plays a negligible
  role in determining the DM relic density; see, \eg,\cite{Erickcek:2015jza}. In practice one usually neglects the second
  term on the r.h.s. of Eq.~(\ref{eq:Boltzmannradiation}), which leads
  to $\rho_R\,a^4\sim\textrm{const}$, where $a$ is the scale
  factor. This condition, along with the Friedmann equation in the
  radiation dominated epoch, sets the temperature dependence on
  time when solving Eq.~(\ref{eq:BoltzmannDM}).}
\begin{eqnarray}
\frac{d\rho_R}{dt} & = &  -4H\rho_R + \langle\sigma_{\textrm{ann}} v\rangle \,\langle E\rangle\big(n_\chi^2-n_{\chi,\textrm{eq}}^2\big),\label{eq:Boltzmannradiation}\\
\frac{dn_\chi}{dt} & = &  -3Hn - \langle\sigma_{\textrm{ann}} v\rangle\big(n_\chi^2 - n_{\chi,\textrm{eq}}^2\big),\label{eq:BoltzmannDM}
\end{eqnarray}
where $\rho_R$ is the radiation energy density and $\langle E\rangle$
is the average energy of annihilating DM particles. As can be deduced
from Eq.~(\ref{eq:BoltzmannDM}) (and even more evidently seen from a
simplified solution~(\ref{eq:omega})), the larger is
$\langle\sigma_{\textrm{ann}} v_{\textrm{M\o l}}\rangle$ at freeze-out
the longer WIMPs $\chi$ stay in thermal equilibrium and therefore the
lower relic abundance $\Omega_\chi h^2$ one obtains.

One should mention here the related thermal mechanism of
coannihilation\cite{Griest:1990kh}. If there is some other particle
$\chi^\prime$ in thermal equilibrium, nearly degenerate in mass with
the DM WIMP $\chi$\,, and such that it annihilates with $\chi$ more
efficiently than $\chi$ with itself, then it is the mechanism of
coannihilation that primarily determines the relic density of dark
matter. A more detailed discussion of coannihilation is postponed to
Subsec.~\ref{sec:cannihilations}, where we apply it to the case of the
neutralino of supersymmetry. It is worth mentioning, though, that the
mass degeneracy between $\chi$ and $\chi^\prime$ can not only lead to
the decrease but also to the increase of the final DM relic
density. This can happen when $\chi^\prime$ can freeze-out before
decaying completely into the DM particles, with a larger yield than
that of $\chi$, and subsequently decay to $\chi$.

\subsubsection{Other WIMP production mechanisms\label{sec:departuresDM}}

Even though the  freeze-out mechanism always contributes to
the WIMP DM abundance, in order to be the dominant process a fairly
specific range of $\langle\sigma_{\textrm{ann}} v_{\textrm{M\o
    l}}\rangle$ at freeze-out is required. However, several other
modes of WIMP production exist which can still lead to the correct
$\Omega_\chi h^2$ even when this condition is not satisfied. Here we
will merely mention some such mechanisms and their specific
implementations. A more general and exhaustive discussion can be found
in Sections~IV and~V or Ref.\cite{Baer:2014eja}.

\begin{figure}[!t]
\centering
\subfloat[]{
\label{fig:a}%
\includegraphics*[width=0.47\textwidth]{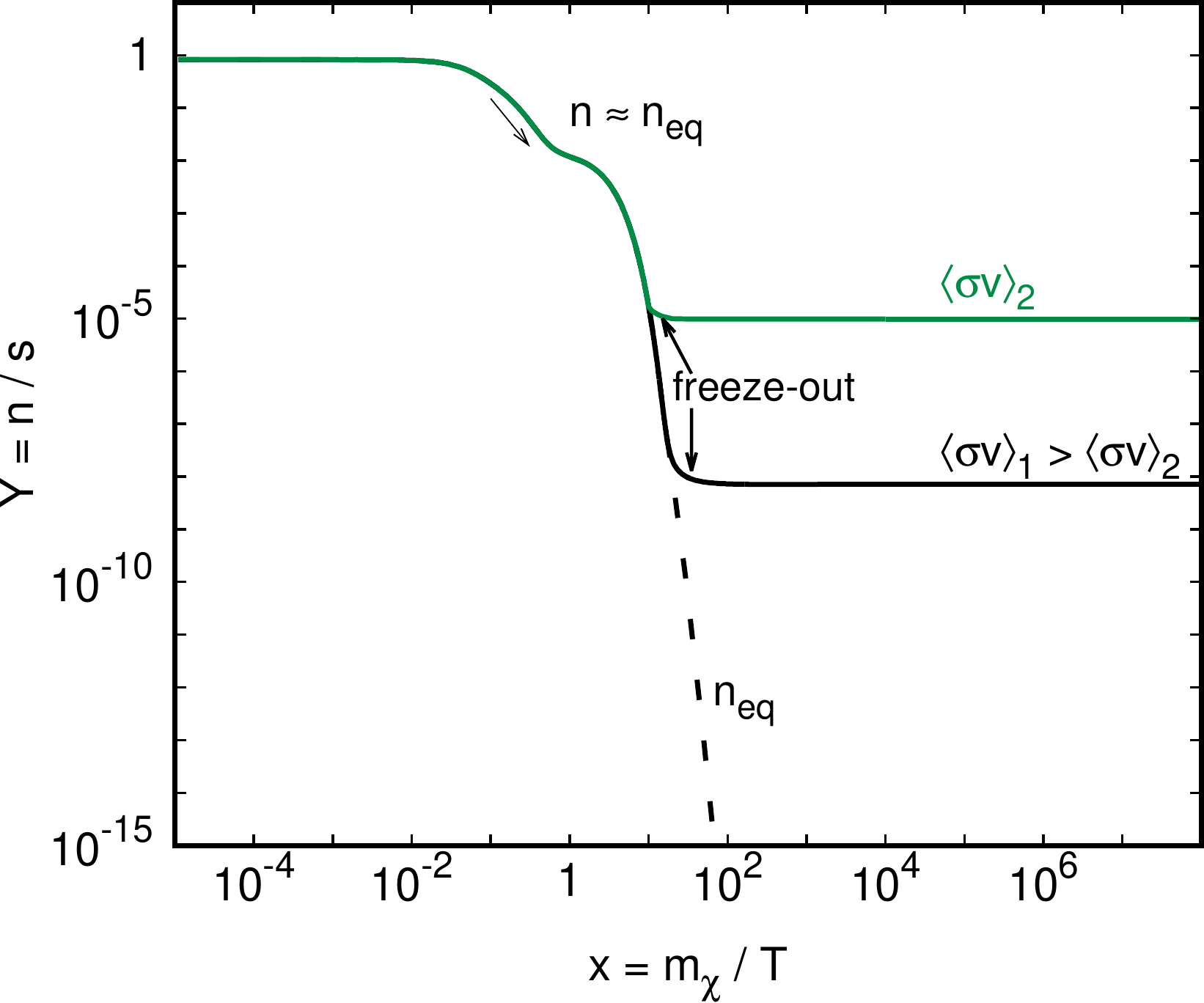}
}
\hspace{0.02\textwidth}
\subfloat[]{%
\label{fig:b}%
\includegraphics*[width=0.47\textwidth]{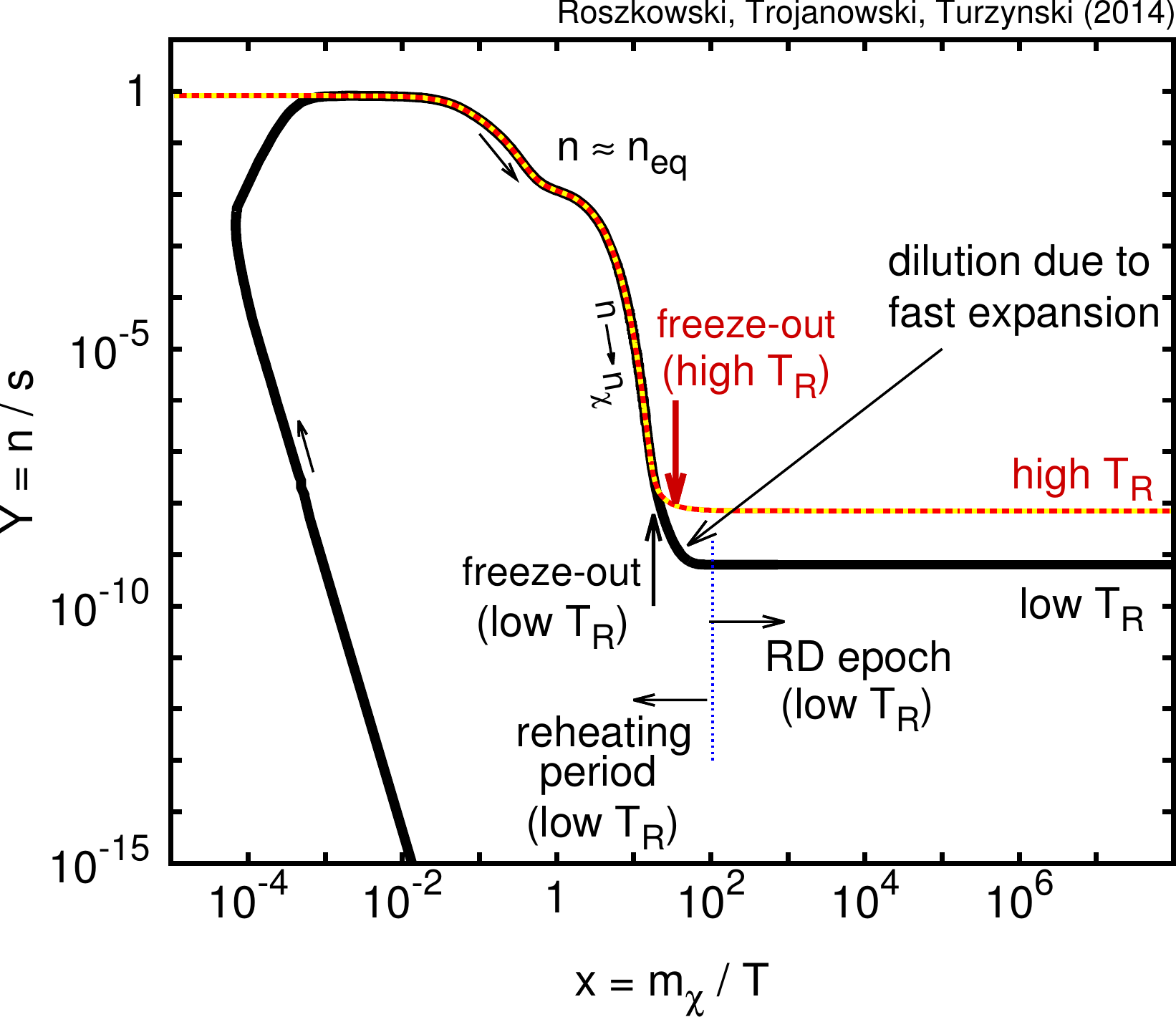}
}
\caption{\footnotesize The DM yield, $Y$, as a function of $x = \mchi/T$ for a standard freeze-out scenario. (b) Similar plot for a scenario with low reheating temperature (black solid line) compared with a standard case (high reheating temperature; indicated by dotted red-yellow line). The beginning of the RD epoch for the low $T_R$ scenario is denoted by vertical dotted blue line. Taken from Ref.\cite{Roszkowski:2014lga}. }
\label{Fig:yields}
\end{figure}

First, however, let us mention the situation when
$\langle\sigma_{\textrm{ann}} v_{\textrm{M\o l}}\rangle_f$
  is too low in which case freeze-out occurs too early and the final
relic density of DM may become too large.  In such a case the DM relic
density must be reduced.  This can be achieved by an additional 
  entropy production from out-of-equilibrium decays of some heavy
species that occur between the time of the DM freeze-out and  BBN
which marks the epoch of standard cosmological expansion. In
particular such heavy particles could dominate the energy density of
the Universe during the period of reheating, \ie, before the radiation dominated (RD) epoch
(for a discussion see\cite{Kolb:1990vq,Giudice:2000ex}). If the
reheating temperature $T_R$, i.e., the temperature at which the RD
epoch begins, is lower than the DM freeze-out temperature, $\tf$, the
additional entropy production due to, \eg, decays of an inflaton or
moduli fields, can effectively dilute away thermally overproduced DM
particles. See \reffig{Fig:yields}\subref{fig:b} and Section~\ref{sec:lowrehchi} for a discussion about neutralino DM.

If $\langle\sigma_{\textrm{ann}} v_{\textrm{M\o l}}\rangle$ is so low
that $\chi$ particles never reach thermal equilibrium after reheating
then they actually never freeze out. In this category of EWIMPs,\footnote{They are also called super-WIMPs [279] or
  FIMPs\cite{Hall:2009bx} (feebly interacting massive particles,
  although the name of frozen-in massive particles would perhaps be
  more appropriate\cite{Baer:2014eja}) in this case. See also a recent review\cite{Bernal:2017kxu}.} DM relics can
be produced through at least one of two, distinct but not mutually
exclusive, mechanisms. Firstly, some heavier particle can first 
  freeze out and then decay into EWIMPs. Note that in this case the
resulting number density of DM is still determined at
freeze-out. Alternatively, EWIMPs can be produced in scatterings
  or decays of some heavier particles in the thermal plasma.  The
production of EWIMPs from decays is most efficient at lower
temperatures, $T\sim m$, just before the Boltzmann suppression kicks
in. The production through scatterings is not accompanied by a reverse
process which is too inefficient due to low cross sections. In the
case of non-renormalizable interactions -- typical for gravitinos and
axinos -- the process is typically more efficient at high temperatures,
near the reheating temperature \tr. When interactions
are renormalizable, scatterings continuously contribute to the DM
relic density until the temperature drops down below a certain value. This
has recently been advocated under the name of so-called
freeze-in mechanism\cite{Hall:2009bx}, as the final yield
increases with $\langle\sigma_{\textrm{ann}} v_{\textrm{M\o
    l}}\rangle$. Note, however, that, strictly speaking,
freeze-in is not a new  mechanism of DM production but simply refers
to a certain type of particle physics interactions that is responsible
for generating DM.

Both mechanisms are different from the standard picture based on
freeze-out, and both can be efficient, depending on some other
quantities (\eg, the DM particle mass or the reheating temperature
\tr). This greatly relaxes the standard thermal WIMP paradigm, as has
been shown in the case of
axinos\cite{Covi:1999ty,Covi:2001nw,Choi:2011yf} (for a recent review
see\cite{Choi:2013lwa}) and
gravitinos\cite{Bolz:2000fu,Pradler:2006qh,Rychkov:2007uq} both of
which belong to the class of EWIMPs.

In contrast, if $\langle\sigma_{\textrm{ann}}
  v_{\textrm{M\o l}}\rangle_f$ is larger than the canonical value
from Eq.~(\ref{eq:sigmavvalue}), then the thermal yield of WIMPs is
too low to produce the DM relic abundance. The DM particles can be
additionally produced in late-time decays (after DM freeze-out) of
some heavier species. Examples include moduli field
(see\cite{Moroi:1999zb,Acharya:2008bk} and references therein),
Q-balls (see, \eg,\cite{Fujii:2002kr}), the inflaton field (see,
\eg,\cite{Lyth:1995ka}) or cosmic strings\cite{Jeannerot:1999yn}. Such
a non-thermal production of DM is often associated with the
aforementioned additional entropy production that also partially
reduces the increase of $\Omega_\chi h^2$. As mentioned above,
late-time decays can also occur for another species $\chi^\prime$, almost
mass degenerate with the DM particles. In principle, coannihilations
reduce both $Y_\chi$ and $Y_{\chi^\prime}$. However, it is possible that the
yield of $\chi^\prime$ after freeze-out is larger even than the yield of
$\chi$ calculated in absence of any mass degeneracy. In this case the
final DM relic density can be increased with respect to the
non-degenerate scenario.

In the (partially) asymmetric DM (ADM)
scenario\cite{Graesser:2011wi,Iminniyaz:2011yp} one can accommodate
an otherwise too low $\Omega_\chi h^2$ by assuming that one component of the
DM relic density from freeze-out is accompanied by one which is set by
an initial asymmetry between DM and their antiparticles, in a way
analogous to the mechanism of baryogenesis. 
It is worth noting that, since in the ADM scenario the DM is not its own antiparticle and the abundance of $\chi$ and $\bar{\chi}$ 
particles can be highly asymmetric nowadays, the expected indirect detection rates from $\chi\bar{\chi}$ annihilations are typically 
suppressed with respect to, e.g., Majorana DM. 
A more detailed discussion can be found in, \eg, Ref.\cite{Petraki:2013wwa}.

Among various other possible ways to deal with too large values of
$\langle\sigma_{\textrm{ann}} v_{\textrm{M\o l}}\rangle_f$ one should also
mention an increased expansion rate of the Universe prior
to, and around, the DM freeze-out due to a dynamics of the dark energy
content of the Universe (see, \eg,\cite{Salati:2002md} for  a
discussion for quintessence). This leads to an earlier decoupling of
the $\chi$ particles and therefore larger $\Omega_\chi h^2$.

It is clear that the mechanism of freeze-out, while remaining
attractive and robust, provides only one of several possible ways of
generating WIMP relics in the early Universe. As we shall see later,
this will have implications for prospects for DM searches.


\subsection{Examples of  WIMP candidates\label{sec:candidates}}

To give a taste of the particle physics context of DM candidates, in
this subsection we give examples of some specific WIMPs that either have
withstood the test of time, or where there has been some fairly recent
activity. 

\begin{figure}[!t]
\centering
\includegraphics*[width=10cm]{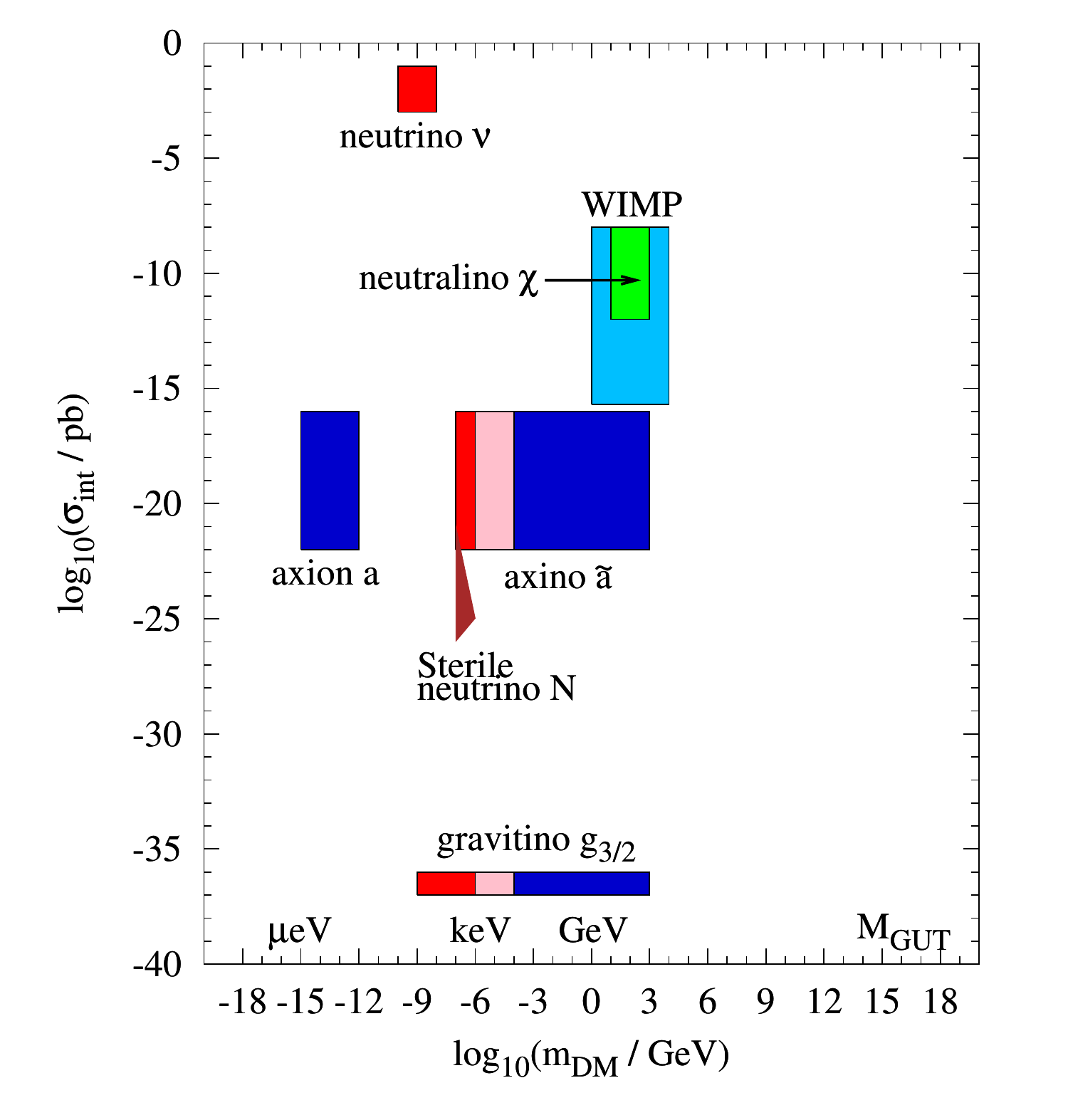}

\caption[Dark matter candidates]{\footnotesize Typical ranges of the cross section
  of DM interactions with ordinary matter as a function of DM mass is
  shown for some of DM candidates that are strongly motivated by
  particle physics.  The red, pink and blue colors represent HDM, WDM
  and CDM, respectively. Adapted from Ref.\cite{Baer:2014eja,Roszkowski:2004jc}.}
\label{Fig:DMcandidates}
\end{figure}

First, to give a wider context, we start with a broad brush picture of
different classes of DM candidates that have emerged from particle
physics. In \reffig{Fig:DMcandidates}, adapted
from\cite{Baer:2014eja} (and originally from\cite{Roszkowski:2004jc}), 
we show where different DM candidates lie in
terms of their masses and typical detection cross
section.\footnote{For another broad-range of DM candidates see Fig.~1 
or Ref.\cite{Gelmini:2016emn}.} The red, pink and blue colors
represent HDM, WDM and CDM, respectively.  Both axes span several orders
of magnitude making clear that a large range of interaction strengths and
masses become allowed by going beyond the thermal WIMP
paradigm. Otherwise one would remain confined basically only to the
light blue rectangle labeled ``WIMP''. Such ``proper'', or thermal,
WIMPs have a larger interaction cross section than axions, axinos,
gravitinos and sterile neutrinos making them a promising target for
experiments.  In contrast, neutrinos have a much larger cross
section but would only constitute HDM which is disfavored, as
discussed earlier.

Within this class of  WIMPs from thermal freeze-out a large number of particle
candidates for DM have been proposed in the literature, and new ones
(sometimes in a reincarnated form) appear on a frequent basis. For one,
this means that it is actually fairly easy to invent DM-like WIMPs. On the other hand, it is fair to say that, from the
perspective of today's (or foreseeable) experimental sensitivity, it
would be very difficult to distinguish many (perhaps most?) of them from
each other or from well established and popular candidates, like the
lightest neutralino of SUSY. Furthermore, one should not forget that
the underlying frameworks that predict many, if not most, of them, while interesting, very
often lack a deeper or more complete theoretical basis and instead
invoke some sort of ``dark portal''. For instance, many such approaches lack a UV
completion, do not address other serious questions in particle physics or
cosmology, \etc. 

{\em The lightest neutralino} of low energy supersymmetric models has long
been recognized as an attractive WIMP candidate for
DM\cite{Goldberg:1983nd,Ellis:1983ew}. 
It is the lightest mass state of a combination of the
superpartners of the neutral gauge bosons and Higgs particles. The neutralino is
characterized by a mass range from about 2\gev\ to $10^4$\gev\ and can span
a large range of the interaction cross section depending on its
composition.  
The relic density and detection prospects of the lightest neutralino have been
extensively studied in a large number of papers 
giving rise to several distinct scenarios.
Since the neutralino remains by far the most popular WIMP and detection
cross section of many other candidates often fall into the ballpark of
the case of the neutralino, we will devote to this case a detailed
discussion, which we postpone to Section~\ref{sec:neutralino}.

{\em Two Higgs Doublet Models} (2HDM) extend the Higgs sector by the
addition of another doublet giving rise to additional charged and
neutral Higgs bosons\cite{Deshpande:1977rw}, see\cite{Branco:2011iw}
for a review of 2HDM phenomenology.  In the simplest extension, 
the inert doublet model
(IDM)\cite{Barbieri:2006dq,Ma:2006km,Goudelis:2013uca}, a
$\textbf{Z}_2$ symmetry is imposed under which all the SM fields are
even while the additional Higgs doublet is odd.  The neutral
additional Higgs boson states, $H^0$ or $A^0$ can then play the role
of DM if one of them is the lightest odd particle.  The DM couples to
the SM via the Higgs boson leading to possible signals from
spin-independent scattering in direct detection experiments.
Alternatively the DM can be provided by an additional field coupled to
the extended Higgs sector and stabilized by the $\textbf{Z}_2$ see for
example\cite{Goh:2009wg,Aoki:2008av,Boucenna:2011hy}.

{\em Little Higgs} models represent a class of possible solutions to the naturalness
problem\cite{ArkaniHamed:2001nc,ArkaniHamed:2002pa,ArkaniHamed:2002qx,ArkaniHamed:2002qy,Kaplan:2003uc,Schmaltz:2002wx,Chang:2003un}
(see\cite{Schmaltz:2005ky} for a comprehensive review).  In little
Higgs models the Higgs fields are the Goldstone bosons of global
symmetry broken at the cut off scale.  The Higgs becomes massive due
to symmetry breaking at the electroweak scale, however, the mass is
protected by the approximate global symmetry and is free from 1-loop
corrections from the cutoff scale.  This allows little Higgs models to
remain natural with a cutoff scale of up to 10\tev. In little Higgs
models  new heavy states are introduced to act as partners of the 
top quark and gauge bosons, such that the divergences are cancelled at the 1-loop
level.  To have a DM candidate in little Higgs models there must be a
remaining $\textbf{Z}_2$ symmetry often called
T-parity\cite{Low:2004xc,Hubisz:2004ft,Hubisz:2005tx}.  The new heavy
partner fields are odd under the symmetry while the SM particles are
even, this results in lightest partner field being stable.  The DM
candidate can come in the form of additional scalars which can be charged
under $SU(2)$ or a singlet\cite{BirkedalHansen:2003mpa,Martin:2006ss}.
Alternatively, little Higgs models can have extended gauge sectors with
a heavy partner to the photon which can act as a WIMP giving an
example of vector DM\cite{Hubisz:2004ft,Freitas:2009jq,Wang:2013yba}.

{\em The twin Higgs}
mechanism\cite{Chacko:2005pe,Chacko:2005un,Barbieri:2005ri,Craig:2013fga}
was introduced to solve the naturalness problem by positing a twin
sector, which is a copy (or partial copy) of the SM related via a
$\textbf{Z}_2$ symmetry.  The combined Higgs sector posses an
approximate $U(4)$ symmetry, and the lightest Higgs then comprises the
pseudo-Goldstone boson of the broken $U(4)$.  A WIMP DM candidate can arise in
twin Higgs models as a member of the twin sector\cite{Garcia:2015loa}
see also\cite{Hedri:2013ina,Poland:2008ev}.  Alternatively the DM can
form part of the scalar sector\cite{Dolle:2007ce} with similar
properties to the IDM.  The twin DM couples to the Standard Model via
the Higgs leading to a spin-independent scattering cross section that
can be searched for with direct detection
experiments\cite{Garcia:2015loa,Hedri:2013ina,Poland:2008ev}.
Annihilation of twin DM can also produce indirect detection signals if
it has a sizable annihilation fraction to SM states.

{\em Sneutrinos} are an example of a non-thermal WIMP. Sneutrinos are
the scalar partners of the neutrinos in SUSY models. For the
sneutrinos that were originally proposed as a DM
candidate\cite{Hagelin:1984wv,Ibanez:1983kw} as partners of the SM
left handed neutrinos, the spin-independent scattering cross section is
generically of weak interaction strength and already firmly ruled out by
direct detection experiments unless the sneutrino makes up only a
subdominant component of the
DM\cite{Falk:1994es,Hall:1997ah,Arina:2007tm}.

For sneutrino DM to be viable there must be either mixing between the
sneutrinos partners of left handed neutrinos and those of the right
handed
neutrinos\cite{ArkaniHamed:2000bq,Arina:2007tm,Belanger:2010cd,Dumont:2012ee,Kakizaki:2015nua},
or lepton violating mass terms that split the sneutrino eigenstates
such that elastic scattering via the $Z$-boson is not
possible\cite{Hall:1997ah,KlapdorKleingrothaus:1999bd,Kolb:1999wx}.
Typically purely right-handed sneutrinos have too small coupling to
the SM to be a WIMP candidate but may be a viable non-WIMP candidate;
see for example\cite{Asaka:2005cn}. The mixed left-right handed
sneutrino DM has recently been reanalyzed in the context of
constrained SUSY with updated LHC results in\cite{Banerjee:2016uyt}.

Models with compactified {\em universal extra dimensions} (UED) possess a DM candidate from the tower of Kaluza-Klein (KK) states.
In these models SM particles can propagate in the new compactified dimension\cite{Appelquist:2000nn} and particles with quantized momenta in the new spatial dimension appear as heavier copies of the standard model particles and the lightest KK state is stable.
The lightest KK state is often a heavy copy of the hypercharge gauge boson and can have the properties of a WIMP\cite{Cheng:2002ej,Servant:2002aq,Kakizaki:2006dz,Hooper:2007qk}.
Run 1 of the LHC ruled out many of the minimal UED models; see\cite{Cornell:2014jza,Servant:2014lqa} for the status of minimal UED theories post LHC Run 1.

A plethora of other DM candidates exist in the literature and it is
clear that it will be up to experiment to identify which of them (if
any) is the choice made by Nature. It is also possible to study WIMP
DM properties within a framework of generic portals between the dark
sector and the SM without specifying other contents of models that are
not directly related to DM interactions (see\cite{Arcadi:2017kky} and
references therein). Many such possibilities has been explored in the
literature including scalar, fermion or vector DM (see, e.g.,~\cite{Karam:2015jta,Karam:2016rsz}) with various kinds
of mediators that can either belong to the SM, \ie, the Higgs or $Z$
bosons, or can themselves belong to the hidden sector. Such an
approach might have less predicting power than well-defined models
described above. However, the advantage is that within this framework
one can study DM-specific features of many definite models at a time
and therefore derive more general conclusions. In particular, it can
be checked that $Z$ boson-portal models are already excluded by the
current limits with the exception of the case of Majorana DM particles,
while for the Higgs boson-portal the only allowed regions for
$\mchi\lesssim 1\tev$ can be obtained for the resonance scenario in
which $\mchi\approx m_h/2$ (see a discussion in\cite{Arcadi:2017kky}).

\section{Experimental  situation \label{sec:experimentalsummary}}

For nearly three decades now the experimental search for DM has continued
its intense activity and generally impressive progress that has led to
several strongly improved bounds on WIMP interactions. At the same
time, with improving sensitivity in some cases new effects have been
identified which could be interpreted as caused by DM. We will pay
attention to a possible WIMP (or WIMP-like) interpretation of these
effects when in this section we make an updated survey of the current
situation in several modes of DM searches.

\subsection{Direct detection: limits and anomalies\label{sec:dd}}
One of the most important strategies to search for WIMP DM is its
possible direct detection (DD) through elastic scatterings of DM
particles off nuclei\cite{Goodman:1984dc} (for reviews see,
\eg,\cite{Lewin:1995rx,Gaitskell:2004gd,Peter:2013aha,Undagoitia:2015gya}). For
WIMPs that interact efficiently enough with baryons this process can lead
to a clear signature in low-background underground detectors. In order
to distinguish a DM recoil signal from background, today's detectors
typically employ some methods of discrimination (see below). In some
cases one looks for the annual modulation of the signal due to the
Earth's movement with respect to the DM
halo\cite{Drukier:1986tm} (for review
see\cite{Freese:2012xd}).  In addition, in the current and the next
generation of the DD experiments effort is made to see further
DM-specific features in the nuclear recoil energy distribution due to
a possible directional detection\cite{Spergel:1987kx} (for review
see\cite{Mayet:2016zxu}).

\paragraph{Formalism} 
An evaluation of a  DM event rate in DD experiments necessarily involves
factors from particle physics and nuclear physics, as well as from
astrophysics. This can be  seen from the formula for the
differential recoil event rate\footnote{Note that for directional
  detection one needs to consider a double differential rate that takes
  into account the dependence on the recoil angle
  (see\cite{Mayet:2016zxu}). Higher order corrections to the
  differential event rate can be found in\cite{Cirigliano:2012pq}.} as
a function of the recoil energy $E_r$
\begin{equation}
\frac{dR}{dE_r}(E_r) =
\left(\frac{\sigma_0}{2\mu^2\,\mchi}\right)\times F^2(E_r)\times
\left(\rhochi\int_{v\geq v_{\textrm{min}}}^{v\leq
    v_{\textrm{esc}}}{d^3v\,\frac{f(\mathbf{v},t)}{v}}\right), 
\label{dRdE}
\end{equation}
where $\sigma_0$ is the DM-nucleus scattering cross section in the zero momentum
transfer limit, \mchi\ is the DM mass, $\mu \equiv \mchi M/(\mchi +
M)$ is the reduced mass of the WIMP-nucleus system for nucleus of mass
$M$, $F(E_r)$ is the nuclear form factor of the target nucleus,
\rhochi\ is the local DM density and $v$ is the relative velocity of
DM particle with respect to the nucleus, while $f(\mathbf{v},t)$
denotes the distribution of the WIMP velocity with cut-off at the
galaxy escape velocity $v_{\textrm{esc}}$. The minimum velocity that
can result in an event with the recoil energy $E_r$ is given by
$v_{\textrm{min}} = \left(\delta +ME_r/\mu\right)/\sqrt{2M\,E_r}$,
where $\delta =0$ for elastic scatterings.

Since DM WIMPs are characterized by non-relativistic velocities, one
typically applies the limit $v\rightarrow 0$ when calculating the cross
section. In this case the corresponding cross section can be
decomposed into two contributions: the spin-independent (SI) and the
spin-dependent (SD), $\sigma^0F^2(E_r)\simeq
\sigma^{\textrm{SI}}F^2_{\textrm{SI}}(E_r) +
\sigma^{\textrm{SD}}F^2_{\textrm{SD}}(E_r)$, where
$\sigma^{\textrm{SI}}$ and $\sigma^{\textrm{SD}}$ are given at zero
momentum transfer, $q$, while the dependence on $q$ is encoded in the
form factors.\footnote{Recently, increased attention has been paid to
  studies of more general scenarios for scatterings of the DM
  particles on nuclei in which this interaction is described in terms
  of extended set of effective operators that go beyond pure SI and SD
  cases (see, \eg,\cite{Fan:2010gt,Fitzpatrick:2012ix}). See also~\cite{Hoferichter:2015ipa,Klos:2013rwa} for related discussion.}
In the absence of isospin violating interactions between DM and
nucleons one obtains $\sigma^{\textrm{SI}} =
\sigsip(\mu^2/\mu_p^2)\,A^2$ where $\sigsip = (4\mu_p^2/\pi)\,f_p^2$
and $\mu_p$ is the reduced mass of the WIMP-proton system; for a discussion in presence of isospin violation see, e.g.,~\cite{Crivellin:2013ipa}. The limits
for the SI cross section are typically presented in the
(\mchi, \sigsip) plane. Note the characteristic $\sim A^2$ dependence
(coherent enhancement) that results in an increased differential
recoil event rate for heavier target nuclei.  The lack of coherent
enhancement in SD cross section results in typically lower
differential recoil event rates than in the SI case and therefore
weaker exclusion limits for \sigsdp\ than for \sigsip. In addition, it
is important to note that spin-zero isotopes do not give any signal in
DM searches based on SD.

\paragraph{Experiments: limits and anomalies} Scatterings of DM
particles off nuclei can be detected via subsequently produced light
(scintillation photons from excitation and later de-excitation of
nuclei), charge (ionization of atoms in a target material) or heat
(phonons in crystal detectors). Using one or a combination of two such
discrimination techniques is now often employed to disentangle
a potential WIMP signal from nuclear recoils and background electron
recoils. This is possible due to different quenching factors that describe the difference between the recorded signal and the actually 
measured recoil energy. The electron recoils constitute the
background of the experiment and can come from, \eg,
$\gamma$-radiation from natural radioactivity or $\beta$-decays that
take place in the detector surrounding materials, on its surface or
even inside the detector. Other sources of background, \eg,
neutrons or $\alpha$-decays, can be associated with nuclear recoils
that can mimic the WIMP signal. Therefore they need to be either
screened out or rejected at the level of  signal
analysis.
A particularly challenging type of such a background that will be very
important for future detectors, especially for DM mass below $10$ GeV,
comes from coherent elastic neutrino-nucleus
scatterings\cite{Freedman:1973yd} and cause the existence of the
so-called coherent solar neutrino background\cite{Cabrera:1984rr,Billard:2013qya}.

Depending on the choice of signal detection technique a variety of
target materials can be employed in DD searches. Light signal from
DM-nucleus scattering can be collected, \eg, by using
scintillating crystals.\footnote{Signal in single-phase liquid noble
  gas detectors also comes entirely from scintillation light emitted
  by ionized or excited dimers (for a review
  see\cite{2013JInst8R4001C}).}  A well-known example of such a
detector is that of the DAMA/LIBRA experiment\cite{Bernabei:2008yh}
operating at the LNGS laboratory in Italy, which for two decades have
been reporting to see an annually modulated DM-like signal, currently
with the significance at the level of
$9.3\sigma$\cite{Bernabei:2013xsa}. The estimated mass of the DM
particles from this measurement would range between $10$ to $15$ GeV
or between $60$ to $100$ GeV depending on the actual nucleus involved
in the scattering process (sodium or iodine, respectively). However,
the DM interpretation of these results is in strong tension with
null results published by some other collaborations: the first XENON1T
limits\cite{Aprile:2017iyp,Aprile:2018dbl}, the final LUX\cite{Akerib:2016vxi} and the
PandaX-II\cite{Tan:2016zwf} limits, as well as, in the low mass
region, with the limits from CDMSlite\cite{Agnese:2015nto} and
XMASS\cite{Abe:2015eos}, which excluded the annual modulation of DM
interpretation of the effect claimed by DAMA/LIBRA. Non-DM
explanations were also considered, including an unknown source of
background, as well as possible errors in data collection and
processing (for a review see\cite{Davis:2015vla}). In addition, other
experiments employing similar detection strategy have been proposed to
verify the DAMA/LIBRA results. In particular, the results of the
KIMS-CsI experiment\cite{Kim:2012rza} disfavor interpretation of
DAMA/LIBRA signal in which the DM particles scatter off iodine
nuclei. This could be circumvented in specific scenarios,
\eg, for Magnetic Inelastic DM (see, however, recent XENON1T limits\cite{Aprile:2017kek}), in models with dominant
WIMP inelastic spin-dependent coupling to protons if different
quenching factors are assumed in both
experiments\cite{Barello:2014uda,DelNobile:2015lxa} (for an extensive
discussion see also\cite{Gelmini:2016emn}; for recent limits see\cite{Aprile:2017ngb}) or leptonically interacting DM particles that induce electron recoils\cite{Kopp:2009et} (see, however, Ref.\cite{Aprile:2017yea} and references therein).

Charge (ionization) signal from DM-nuclei scatterings can be
efficiently measured by low-temperature ultra-low background germanium
detectors\cite{Ahlen:1987mn}.\footnote{Another important example of
  detection technique that focuses on ionization signal from DM-nuclei
  scatterings is used in gaseous detectors employed in directional
  dark matter searches (for review see\cite{Mayet:2016zxu}).} This
technique has been employed by the CoGeNT Collaboration leading to the
claim of an observation of an annually modulated signal in their
data\cite{Aalseth:2012if}. The signal could be consistent with
WIMP DM-hypothesis with mass about $7-10$ GeV, although it only had
$2.8\sigma$ significance and it was not confirmed in later searches in
the similar mass range. On the other hand, the observed excess of
events may be fully explained when an improved background treatment is
applied, as pointed out in\cite{Aalseth:2014jpa,Davis:2014bla}. The
DM interpretation of the CoGeNT data has also been disfavored by other
germanium detectors, \eg, CDEX\cite{Yue:2014qdu} and
MALBEK\cite{Giovanetti:2014fhx}.  A halo-independent analysis
performed in\cite{DelNobile:2013cta} for light ($\lsim10\gev$) WIMPs
showed a strong tension between the DM interpretation of the annual
modulation of DAMA/LIBRA and CoGENT events when compared with the
CDMS-II silicon data.

Phonon signal coming from DM-nuclei scatterings in crystals can
provide another important experimental signature in DM DD
searches.
This
technique is particularly useful when looking for low mass DM due to a
very good energy threshold. Moreover, one typically further improves
the treatment of the background in such experiments by using cryogenic
bolometers with additional charge or scintillation light readouts. In
2013 CDMS-Si detector results were published\cite{Agnese:2013rvf}
reporting observation of $3$ WIMP-candidate events with only $0.19\%$
probability for the background-only hypothesis. The highest likelihood
occurred for WIMP-DM with $8.6$ GeV mass. However, these results were
not confirmed by the germanium CDMS-II\cite{Agnese:2014xye} and
SuperCDMS\cite{Agnese:2013jaa} detectors and there is no plausible DM
halo function for which this tension could be alleviated unless one
assumes, \eg, exothermic DM with Ge-phobic interactions as
discussed in\cite{Gelmini:2016pei}.

DM-nuclei scatterings also can be detected via heat signal in
experiments based on superheated fluids used as a target material. DM
particle passing through a detector can then be visualized thanks to
initiated process of bubble creation.

Another DM-like signal was found in the data obtained by the CRESST-II
Collaboration\cite{Angloher:2011uu} in 2011. An excess in the expected
number of events was observed in two mass ranges around $10\gev$ and
$25\gev$ with the significance at the level of $4.2\sigma$ and
$4.7\sigma$, respectively. However, as it was pointed out
in\cite{Kuzniak:2012zm} and confirmed in a later study by the
collaboration\cite{Angloher:2014myn}, the excess was mainly due to a
missing contribution to the background (see also\cite{Gelmini:2016emn}
and\cite{Davis:2015vla} for an updated discussion).

\begin{figure}[t]
\centering
\includegraphics[width=\textwidth]{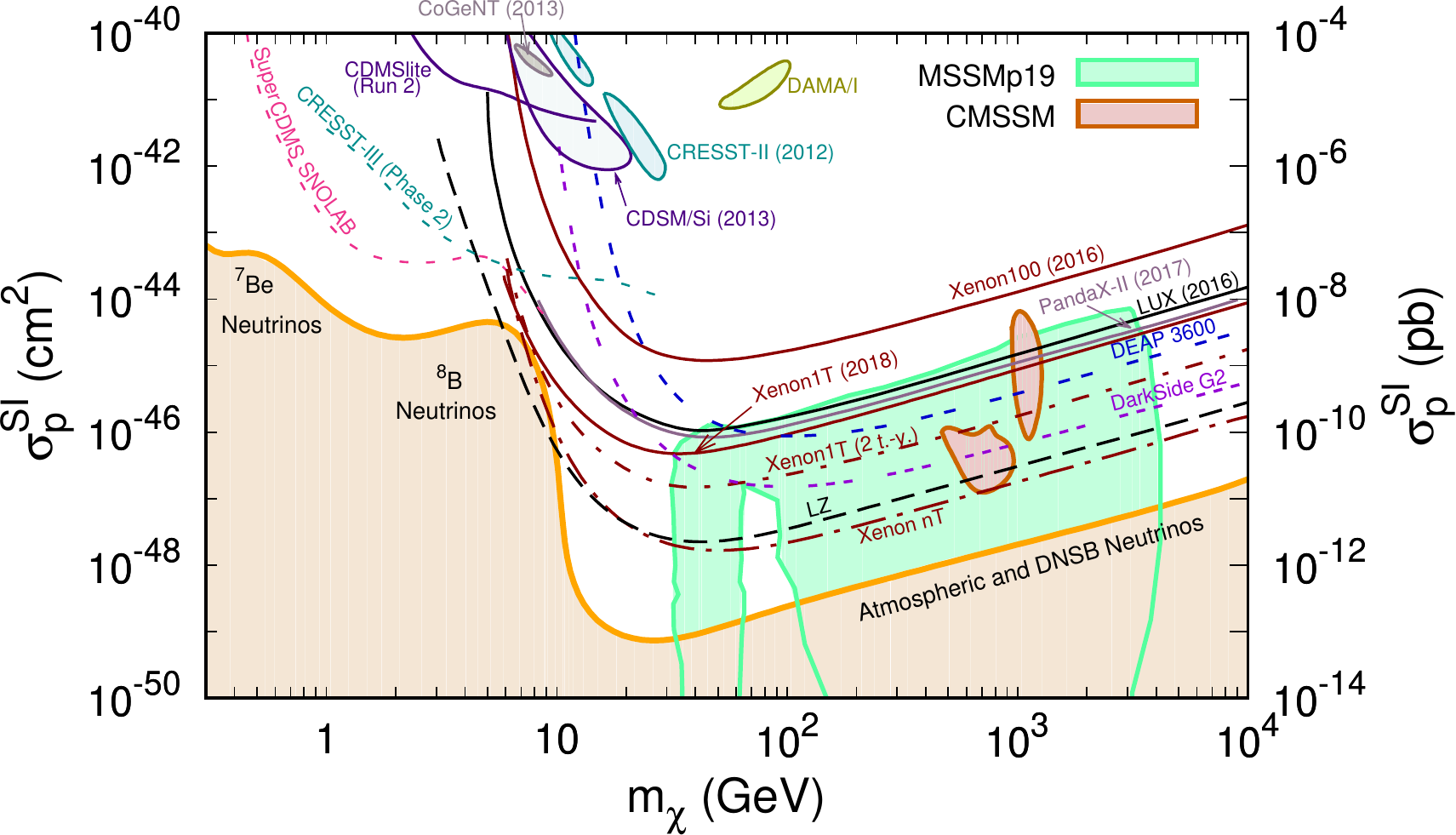}
\caption{\footnotesize Current and future limits on DM direct
  detection spin-independent cross section, \sigsip, as a function of
  DM mass, \mchi. The current limits are shown with solid black
  (LUX\cite{Akerib:2016vxi}), gray (PandaX-II\cite{Tan:2016zwf}),
  brown (XENON100\cite{Aprile:2016swn} and
  XENON1T\cite{Aprile:2018dbl}) and violet
  (CDMSlite-II\cite{Agnese:2015nto}) lines. Future projections
  correspond to CRESST-III (Phase 2)\cite{Strauss:2016sxp} (light
  blue), DarkSide G2\cite{Aalseth:2015mba} (violet triple-dashed
  line), DEAP3600\cite{Amaudruz:2014nsa} (blue double-dashed line),
  LZ\cite{Szydagis:2016few} (black long-dashed line), SuperCDMS at
  SNOLAB\cite{Agnese:2016cpb} (pink short-dashed line) as well as
  XENON1T/nT\cite{Aprile:2015uzo} (brown dash-dotted lines). We also
  show the 95\%~C.L. region for the 19-parameter version of the MSSM
  (green shaded area)\cite{Roszkowski:2014iqa} and posterior plot for the allowed parameter
  space of the CMSSM (brown area enclosed with the solid brown
  line)\cite{Roszkowski:2014wqa}. The shaded areas on top of the plot correspond to the favored
  regions for DM interpretations of anomalies reported in the
  literature by the CDMS-Si\cite{Agnese:2013rvf} (blue),
  CoGeNT\cite{Aalseth:2012if} (gray) CRESST-II\cite{Angloher:2011uu}
  (light blue) and DAMA/LIBRA\cite{Bernabei:2008yh} (light green)
  collaborations. The shaded area below the solid orange line on the
  bottom of the plot corresponds to the irreducible neutrino
  background\cite{Billard:2013qya}. 
\label{fig:DDlimits}
}
\end{figure}

The most stringent current limit on \sigsip\ for large DM mass comes
from null results of DM searches in dual phase (liquid-gas) xenon
detectors: XENON1T\cite{Aprile:2017iyp,Aprile:2018dbl} -- the most recent (and currently the strongest) --
the final LUX result\cite{Akerib:2016vxi} and that of
PandaX-II\cite{Tan:2016zwf}, both of which improved the previous limits of XENON100
collaboration\cite{Aprile:2016swn}; see Fig.\ref{fig:DDlimits}.  The strongest up-to-date
exclusion lines for spin-dependent cross section, \sigsdp, from DD
experiments were published by the PICO
collaboration\cite{Amole:2015pla,Amole:2016pye} (see also
LUX\cite{Akerib:2016lao} and XENON100\cite{Aprile:2016swn}
results). However, \sigsdp\ can also be effectively constrained by
neutrino telescopes as will be discussed in Section~\ref{sec:neutrinos}.

A further significant improvement is expected from the currently
running XENON1T, later from XENONnT\cite{Aprile:2015uzo},
LZ\cite{Szydagis:2016few}, and eventually from planned xenon detectors,
\eg, DARWIN\cite{Aalbers:2016jon}, and for argon as a target
material: DEAP3600\cite{Amaudruz:2014nsa}, ArDM\cite{Calvo:2015uln} and DarkSide
G2\cite{Aalseth:2015mba}. In the
low mass regime large part of the (\mchi,\,\sigsip) parameter space
will be probed by the future germanium and silicon detectors in the
SuperCDMS experiment operating at SNOLAB\cite{Agnese:2016cpb}.
  
A summary of the above discussion about experimental results is
presented in \reffig{fig:DDlimits} where we show current and expected
future limits on \sigsip\ as a function of the DM mass. 
Anomalies
reported in the past by some of the experimental collaborations were
not confirmed and are probably due to some non-DM effects that occur
either inside or outside the detectors. However, the upcoming years
should deliver new data covering regions in the (\mchi,\,\sigsip)
plane well below the current limits and, hopefully, eventually producing a
genuine DM signal.

It is
  important to note that the limits presented in the (\mchi,\,\sigsip)
  plane can vary depending on the underlying assumptions about
  relevant astrophysical quantities, \eg, the local DM
  density and the DM velocity distribution (see,
  \eg,\cite{Green:2011bv} and references therein). The
  dependence on the velocity distribution is typically
  weak\cite{Kamionkowski:1997xg}, but can become more important,
  \eg, if the detector is sensitive only to the tail of the
  distribution\cite{McCabe:2010zh}. Alternatively, the limits can be
  shown in a DM halo-independent way\cite{Drees:2008bv} (see
  also\cite{DelNobile:2014sja} and references therein) if a positive
  signal is measured by at least two different targets.

\paragraph{Implications for WIMP models}
Direct detection searches play a vital role in constraining various
models of WIMP as DM. For instance, early negative results from the
Heidelberg-Moscow experiment\cite{Beck:1993sb} led to an exclusion of
the scenario in which the majority of DM was composed of left-handed sneutrinos in the MSSM\cite{Falk:1994es}, as discussed in
Section~\ref{sec:candidates}. Since then many other theoretical
candidates have been constrained by null results of searches for the
DM particles in DD experiments.

Limits from DD have also been derived on effective contact operators
describing possible interactions between DM and the SM particles (for
studies related to DD see, \eg,
\cite{Fan:2010gt,Fitzpatrick:2012ix}). One can then translate the
usual DD limits shown in the (\mchi,\,\sigsip) plane into the actual
limits on the coefficients of the operators that contribute to
\sigsip, while the other coefficients remain free and can,
\eg, help to achieve the proper value of the DM relic
density. Stronger constraints can be obtained when both DD and ID
searches are taken into account (see, \eg,\cite{Liem:2016xpm}).

Another phenomenological approach consist in ``expanding'' the contact
operator approach by introducing specific mediators (``portals'')
between the DM sector and the SM particles in a framework of so-called
simplified models. For studies related to DM DD see,
\eg,\cite{DiFranzo:2013vra,Alves:2013tqa,Buckley:2014fba,Choudhury:2015lha,Matsumoto:2016hbs}. It
has been pointed out that gauge invariance and perturbative
unitarity need to be carefully taken into account when constructing
simplified models of DM
interactions\cite{Bell:2015sza,Kahlhoefer:2015bea}. For further
discussion about the effective theory approach (EFT) and simplified
models see, \eg,\cite{DeSimone:2016fbz} and references
therein.

\subsection{Gamma rays: limits and Galactic Center excess\label{sec:gammarays}}

Gamma-rays represent a promising channel in which to search for dark
matter (for reviews see,
\eg,\cite{Cirelli:2010xx,Conrad:2015bsa,Gaskins:2016cha}).
WIMPs are expected to annihilate at present leading to the possibility
of detecting annihilation products, in particular a spectrum of
gamma-rays.  Since gamma-rays are not deflected in their journey from
the emission point to detection on Earth, the direction of the source
can be determined, thus allowing specific targets or regions of DM
annihilation to be identified.  This can be compared to the situation
with charged cosmic rays which are deflected by the magnetic field of
the Galaxy.  It is therefore possible to use the expected morphology
of the DM signal to discriminate it from background\cite{Ando:2005xg}.

\paragraph{Gamma rays from DM} 
The spectrum of gamma rays expected from a DM annihilation depends on
particles produced in the final state.  Typically one assumes that the
DM annihilates to Standard Model particles, which must account for at
least some fraction of the annihilations for a WIMP produced through
thermal freeze-out.  Gamma-ray emission from DM annihilations can be
of two types: a continuous spectrum generated by the decay,
hadronization and final state radiation of the SM particles produced,
and spectral features in the form of gamma ray lines and internal
bremsstrahlung.

The continuous spectrum of gamma rays for a specific SM final state
can be estimated via Monte Carlo simulation using standard event
generators\cite{Bahr:2008pv,Gleisberg:2008ta,Sjostrand:2007gs} that
include parton showering and hadronization.  There is therefore some
uncertainty associated with the choice of event generator. Additionally
it can also be important to include polarization and electroweak
corrections, see\cite{Ciafaloni:2010ti} for a full discussion.
Gamma-ray lines appear from the processes $\chi\chi \rightarrow
\gamma\gamma$ and $\chi\chi \rightarrow Z \gamma$, since the DM must
be electrically neutral these must arise at the loop level but are
nearly impossible to mimic by astrophysical
background\cite{Bergstrom:1997fj}.  Internal bremsstrahlung can also
lead to sharp spectral feature as well as lifting chiral suppression
in some models\cite{Bringmann:2007nk}.  In addition, $\gamma$-rays can
also be produced as secondary products of DM annihilations into other
particles once the latter interact in interstellar medium,
\eg, thanks to inverse Compton scattering (ICS),
bremsstrahlung off of galactic gas, neutral pion decays originating
from interactions of hadronic cosmic rays with interstellar gas or
synchrotron emission induced by propagation in magnetic fields.

\paragraph{Targets for gamma-ray searches} 
The morphology of the DM signal can be used to discriminate DM
annihilations from the background by focusing searches in regions with
a high density of DM.  The Galactic center (GC) is expected to be the
brightest source of gamma rays from annihilating DM and has attracted
considerable attention as a target for indirect detection
experiments\cite{Gondolo:1999ef}.  The main argument for a high
density of DM in the inner regions of galaxies comes from N-body
simulations that determine the expected halo distribution (see,
\eg,\cite{Navarro:2008kc} and references therein). Dynamical
and microlensing observations act to constrain the halo
profile\cite{Iocco:2011jz} but there remains significant uncertainty
particularly in the GC, see\cite{Catena:2009mf,Nesti:2013uwa} for some
recent determinations.  These uncertainties are further compounded by
the complex background of conventional astrophysical sources of
gamma rays present in the GC\cite{Ackermann:2012rg}.

Dwarf spheroidal galaxies (dSphs) in the Local Group do not suffer
from the same problems as the GC of the Milky Way. The dSphs are
expected to be DM dominated\cite{Mateo:1998wg,McConnachie:2012vd} and
are free from the astrophysical backgrounds plaguing the GC.  In
addition the distribution of DM can be estimated from the dynamics of
stars inside the
dSphs\cite{Martinez:2009jh,Walker:2009zp,Wolf:2009tu}.  This makes
limits obtained from dSphs more robust, however the expected signal is
much lower for a single dSphs than for the GC therefore a stacking
analysis is required to obtain competitive limits.

Galaxy clusters are another promising target for gamma-ray
searches\cite{Jeltema:2008vu,Pinzke:2010st}.  The main drawback
compared to dSphs is that they suffer from large and poorly understood
astrophysical backgrounds, and the expected sensitivity depends
strongly on the DM substructure which is unknown.

Finally the full sky can be observed to place limits on annihilations
summed over all halos\cite{Abdo:2010dk}.  Since the cosmological DM
halos are in general unresolved they contribute to an approximately
isotropic background.  The isotropic gamma-ray background can be
searched for spectral features that would indicate DM
annihilations. However, large uncertainties in the backgrounds and
expected rate make setting robust limits difficult.  A related idea is
to measure the angular correlations or cross-correlations to search
for a DM signal above the expected isotropic background.  Angular
correlations can be used to search for extragalactic halos or subhalos
as well as other sources such as annihilation of DM around
intermediate mass black holes\cite{Taoso:2008qz}.  Cross correlations
can also be searched for between the diffuse extragalactic gamma-ray
signal and the cosmological DM distribution inferred in other channels
such as the CMB and structure surveys (see\cite{Xia:2015wka} and
references therein).

\begin{figure}[t]
\centering
   \includegraphics[width=3.5in]{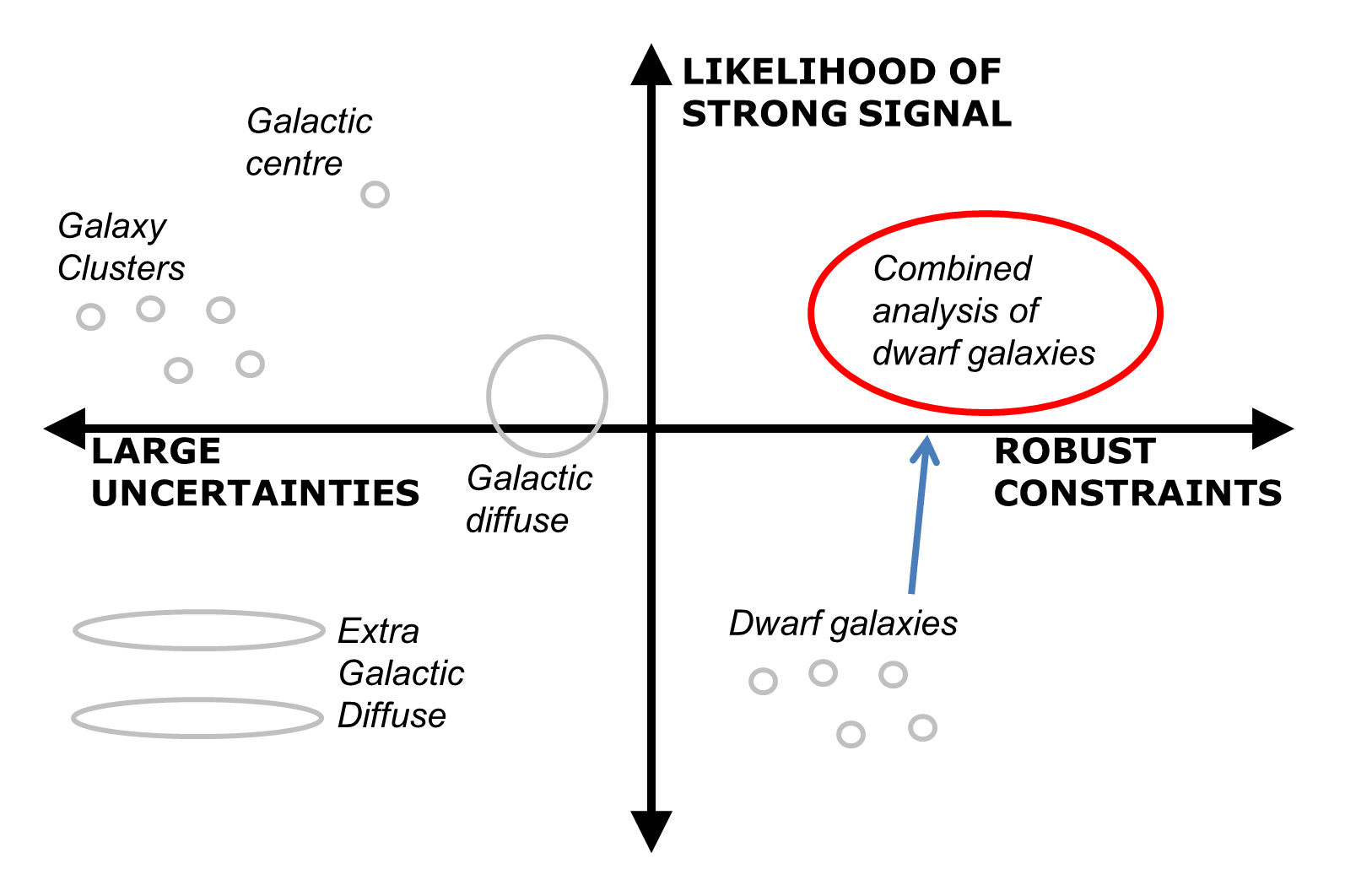} 
\caption{\footnotesize A cartoon, taken from\cite{Conrad:2014tla}, of the relative merits of different gamma-ray observation regions, projected into the expected strength of a signal and relative uncertainties that are associated with the backgrounds and signal strength.
\label{fig:targets}}
\end{figure}

Figure~\ref{fig:targets}, taken from\cite{Conrad:2014tla}, summarizes the
discussion of different targets of gamma-ray observations in terms of
the expected signal strength and uncertainty associated with an
observed signal or limit.  While the strongest case for the
observation of DM would be made by complementary observations in
several of those channels, two particular targets stand out as
currently being most relevant.  These are the GC, which has the
largest expected signal strength and is therefore the most likely
target for a signal to show up first, and combined data from dwarf
galaxies of the Local Group, which currently produce the strongest and
most robust limits.

\paragraph{Gamma-ray telescopes} 
There are two main strategies to observe gamma rays.  Since gamma rays
interact with the atmosphere direct observation can only be made by
space telescopes.  On the other hand the ground based telescopes are
able to detect gamma rays indirectly by observing the Cherenkov light
produced by the showers of charged particles produced by the gamma ray
as it hits the atmosphere.  Direct observation of gamma rays using
space telescopes has been performed by EGRET\cite{Thompson:1993zz} and
the currently running Fermi-LAT\cite{Atwood:2009ez}.  Fermi-LAT uses
pair conversion inside a tracking detector, and an electromagnetic
calorimeter to determine the energy and incident direction of the
gamma-ray.  Charged cosmic rays are rejected using an anti-coincidence
shield. Fermi-LAT is able to observe gamma rays in the range 20\mev\
to 300\gev\ and has an effective area of $\sim1\textrm{m}^2$.  It
typically scans the full sky continuously taking advantage of its
large field of view.  This means that Fermi-LAT is able to make
observations of all of the promising targets for DM annihilation.

The most promising ground based telescopes are the Imaging Air
Cherenkov Telescopes (IACTs).  When a gamma ray enters the atmosphere
it interacts creating a shower of secondary particles.  The Cherenkov
light from the shower is then captured by one or more telescopes on
the ground.

One major source of background comes in the form of air showers caused by
charged cosmic rays.  The cosmic ray background is isotropic and
greatly exceeds the gamma-ray signal.  This background must be reduced
by rejecting showers produced by cosmic rays. This can be done by
distinguishing the patterns of Cherenkov light produced by hadronic
and gamma-ray showers.  Unfortunately, showers produced by electrons
cannot be differentiated from gamma rays this way.  Monte Carlo
simulations\cite{Heck:1998vt} are used to model the Cherenkov light
from hadronic and gamma-ray showers and to estimate the remaining
background for a particular telescope configuration.  IACTs have a
higher energy threshold than space telescopes since lower energy
gamma rays produce less Cherenkov light, however they have the
advantage that the volume of atmosphere observed can be large leading
to huge (energy dependent) effective area compared to space
telescopes.  On the other hand, the field of view of IACTs is much
smaller and dedicated observations of particular target regions must
be made.  This means that potential time allowed for DM studies must
compete with the other science goals of the IACT.  Currently running
IACTs include MAGIC\cite{Aleksic:2011bx}, VERITAS\cite{Holder:2008ux}
and HESS\cite{Aharonian:2006pe}.  While the next generation telescope
will be the currently planned CTA\cite{Consortium:2010bc}.

\begin{figure}[t]
\centering
\includegraphics[width=\textwidth]{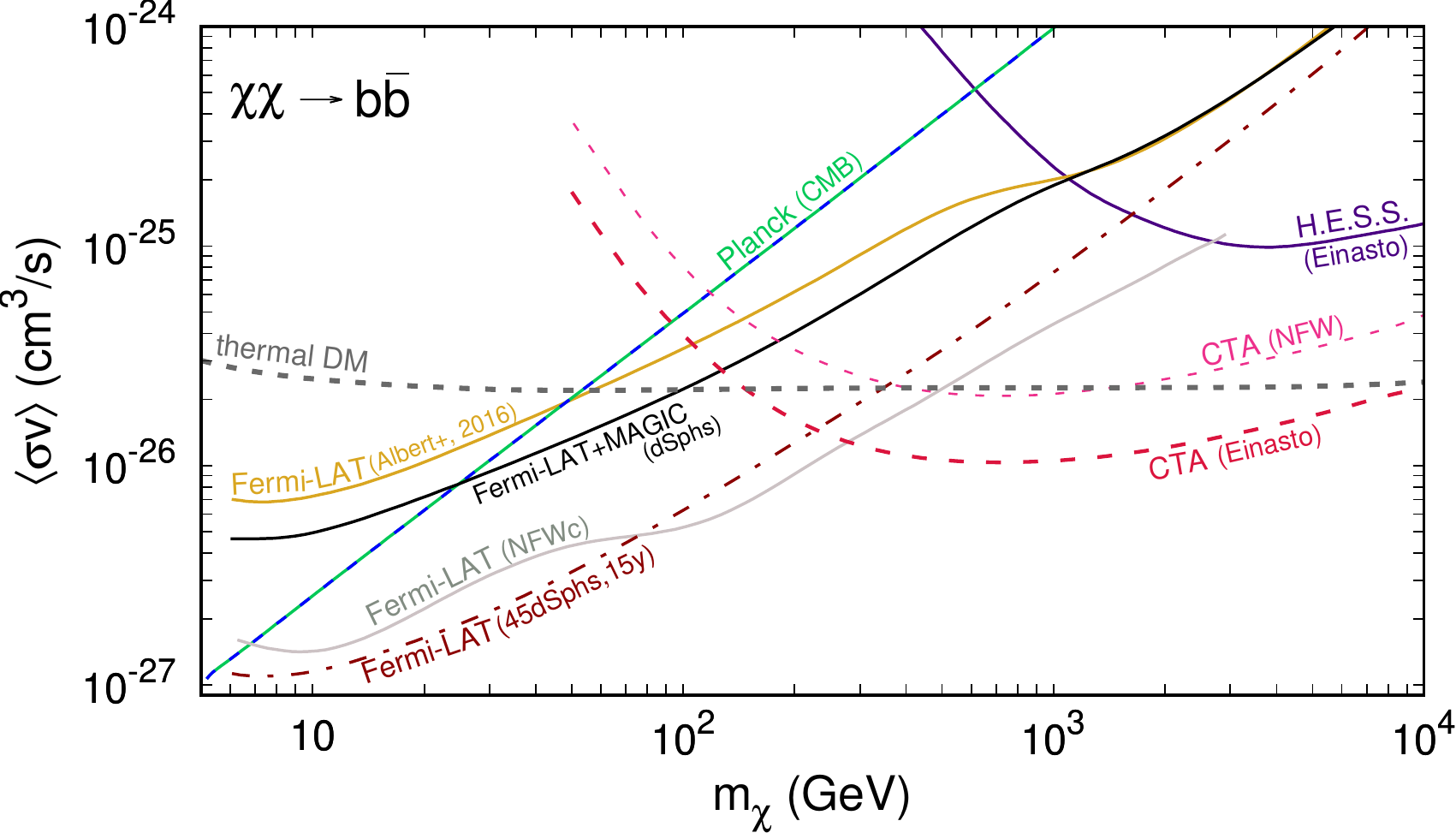}
\caption{\footnotesize Limits on the annihilation cross section for the DM particles annihilating into a $b\bar{b}$ pair. Currently the strongest limits correspond to the stacked analysis of dwarf galaxies from Fermi-LAT\cite{Fermi-LAT:2016uux} (solid golden line), combined analysis of both Fermi-LAT and MAGIC\cite{Ahnen:2016qkx} (solid black line) collaborations, H.E.S.S. observations of the GC\cite{Abdallah:2016ygi} (solid violet line) that we show for the assumed Einasto profile of the DM in the Galactic halo and the Fermi-LAT observations of the Inner Galaxy\cite{Gomez-Vargas:2013bea} (solid gray line) if the DM distribution is assumed to be consistent with the compressed NFW profile (NFWc). In the low DM mass regime the important limits come from the CMB analysis released by the Planck collaboration\cite{Ade:2015xua} which constraints \sigv\ around the time of recombination (dashed blue-green line). The future projections are shown for the stacked analysis of 45 dwarf galaxies and 15 years of data taking\cite{Charles:2016pgz} (dash-dotted brown line), as well as for the CTA collaboration\cite{Carr:2015hta} based on the assumed NFW (dashed pink line) and Einasto profiles (dashed red line). The value of the annihilation cross section that corresponds to the thermal production of WIMP DM\cite{Steigman:2012nb} is denoted with dotted gray line.
\label{fig:CurrentGammaLimits}}
\end{figure}

\paragraph{Current limits and future prospects}
We summarize some of the recent limits from Fermi-LAT and the IACTs,
as well as future projections in \reffig{fig:CurrentGammaLimits}.  On
the direct observation side the strongest limits come the stacked
analysis of dwarf galaxies\cite{Ahnen:2016qkx,Fermi-LAT:2016uux}.  The
canonical thermal annihilation rate\cite{Steigman:2012nb} is shown as
a gray dotted line and the current limits from
Fermi-LAT exclude WIMPs annihilating to $b\bar{b}$ with this cross
section below $\sim100\gev$. The limits from the
  GC\cite{Hooper:2012sr,Gomez-Vargas:2013bea} are competitive but are
  weaker due to the uncertainty in the halo profile.  Observations of
  the Galactic halo\cite{Ackermann:2012rg}, galaxy
  clusters\cite{S.ZimmerfortheFermi-LAT:2015oka} and the Fermi
  analysis of the intensity\cite{Ackermann:2015tah} and angular power
  spectrum\cite{Fornasa:2016ohl} of IGRB are weaker by about an order
  of magnitude.

Turning to the current generation of IACTs,
\reffig{fig:CurrentGammaLimits}, we first note that the Cherenkov
telescopes are sensitive to larger DM masses, whereas Fermi-LAT loses
sensitivity as the DM mass increases.  The strongest limits come from
H.E.S.S. observations of the Galactic center\cite{Abdallah:2016ygi}
which at $\sim1\tev$ become stronger than the Fermi-LAT limit.  The
limits from MAGIC's observation of the dwarf galaxy Segue
1\cite{Aleksic:2011jx} and the H.E.S.S. combination analysis of dwarf
galaxies\cite{Abramowski:2014tra} are comparable with
H.E.S.S. extending the analysis to higher masses.  The VERITAS
analysis of the Fornax galaxy cluster\cite{Zitzer:2015uta} is the
least sensitive and has large uncertainties from the modeling of the
substructure and backgrounds (see also recent VERITAS limits from joint analysis of four dSphs\cite{Archambault:2017wyh}). In the search for spectral features the
most relevant limits come from Fermi-LAT and H.E.S.S. searches for
gamma-ray lines in the GC\cite{Ackermann:2015lka,Abdalla:2016olq} and
galaxy clusters\cite{Adams:2016alz}.

In the short term, Fermi-LAT, H.E.S.S. and VERITAS are currently
taking data, which will continue to place limits on DM annihilation to
gamma rays.  In particular the best prospects for Fermi-LAT consist of
observations of dSphs with more dSphs expected to be discovered in the
future (see, \eg,\cite{Charles:2016pgz}).

Other stringent limits come from possible deviations from a standard
recombination history caused by DM-induced cascades of highly
energetic particles around this period in the evolution of the
Universe\cite{Padmanabhan:2005es}.  Depending on the actual dominant
annihilation final state and on the associated efficiency parameters
that describe the fraction of the rest mass energy contributing to CMB
distortions, the current limits sometimes reach up to the values of the
thermal annihilation cross section for low DM mass\cite{Ade:2015xua}.
It is important to remember that limits derived this way constrain the
value of the annihilation cross section around the time of
recombination and therefore they are not necessarily directly
comparable with the DM ID lines discussed above unless DM annihilation
is predominantly of $s$-wave type.

Looking a few years ahead, the CTA experiment\cite{Consortium:2010bc}
is expected to improve on the current limits from H.E.S.S.  CTA will
be the next  very large IACT consisting of arrays in the northern and southern
hemisphere, with small, medium and large telescopes giving CTA
improved sensitivity over a large energy range from $\sim100\gev$ to
tens of \tev\cite{Bernlohr:2012we}.  This results in strong expected
sensitivity of the CTA to DM annihilations based on observations of
the GC\cite{Carr:2015hta}.

Future gamma-ray space telescopes are planned, \eg,
CALET\cite{Adriani:2015cda}, GAMMA-400\cite{Topchiev:2015wva} and
HERD\cite{Huang:2015fca}.  CALET, recently launched in 2015, has
excellent energy resolution above 100\gev\ and is likely to place
strong limits on gamma-ray lines\cite{Adriani:2015cda} although will
not be competitive with Fermi-LAT for limits on the continuous
spectrum of gamma-rays. The Russian-led GAMMA-400 is in preparation and
will also mostly contribute to searches for spectral features due to
its improved energy resolution compared to Fermi-LAT.  The HERD
telescope, to be launched in 2020, on the other hand will represent an
increase in the effective area and thus should improve on the limits
set by Fermi-LAT\cite{Huang:2015fca}, although detailed studies have
not yet been performed.

\paragraph{Galactic Center Excess}
One of the most intriguing hints of DM detection that has emerged in
the last few years is a consistent excess above the expected
background observed in the diffuse gamma-rays coming from the GC in
the Fermi-LAT data\cite{Goodenough:2009gk,Hooper:2010mq}.  This
signal, known as the Galactic Center Excess (GCE), appears to be
peaked around photon energy of a few \gev\ and was consistently
confirmed by many independent analyses (see,
\eg,\cite{Hooper:2011ti,Abazajian:2012pn,Gordon:2013vta,Abazajian:2014fta,Daylan:2014rsa,Calore:2014xka,Karwin:2016tsw}
and references therein). The Fermi-LAT Collaboration released their
own analysis\cite{TheFermi-LAT:2015kwa} in which GCE is studied
employing multiple specialized interstellar emission models (IEMs)
that enabled to separate the signal coming from the region surrounding
the GC (within $\sim$1 kpc) from the rest of the Galaxy (see also~\cite{Zhou:2014lva,Huang:2015rlu} for related discussion). The excess of gamma rays
remained, however its origin is still unclear since the spectral
properties of the signal strongly depend on the assumed IEM. In
particular, it was pointed out that the photon clustering in the
observed excess could be inconsistent with the signal expected from
smooth DM
distribution\cite{Bartels:2015aea,Lee:2015fea,Clark:2016mbb}. Such a
possibility has been recently confirmed by the Fermi-LAT
collaboration\cite{Fermi-LAT:2017yoi}. However, it is also possible
that what seems to be a point-like nature of the excess is in fact a
result of uncertainties in the background
modeling\cite{Horiuchi:2016zwu}. On the other hand, recent Fermi-LAT
analysis of the GCE\cite{TheFermi-LAT:2017vmf} confirmed the existence
of the excess and studied in detail its morphology. In particular, it
was shown that the excess is likely to have at least partial
astrophysical origin, but the DM interpretation is not excluded. In
addition, the excess in the $\gamma$-ray signal with a possible DM
origin, which is consistent with the GCE, was found in the recent
Fermi-LAT study of the center of
M31\cite{Ackermann:2017nya}.

The most popular astrophysical explanation is
a population of unresolved millisecond pulsars (MSPs) in the GC.  The
gamma-ray spectrum produced MSPs is known to be compatible with a
power law exhibiting an exponential cutoff around
$2-3\gev$\cite{Cholis:2014noa}.  It was shown that the GCE can be
partly or completely explained by a population of MSPs in the
GC\cite{Abazajian:2010zy,Abazajian:2012pn}.  However there is some
disagreement as to whether MSPs are really a viable explanation for
the GCE based on, \eg, the expected distribution of X-ray
binaries\cite{Cholis:2014lta,Haggard:2017lyq} (see
also\cite{Abazajian:2012pn} for early study in this direction),
detailed analysis of the spectra shape of the GCE\cite{Hooper:2013nhl}
or distribution of globular clusters\cite{Hooper:2016rap}.
Non-equilibrium processes were also proposed as an explanation
for the GCE. In particular, a large injection of CRs into the GC at
some point in the past could produce the observed
GCE\cite{Macias:2013vya,Carlson:2014cwa,Cholis:2015dea}. The observed
excess can also be associated with X-shaped stellar over-density in
the Galactic bulge\cite{Macias:2016nev}.

As discussed above, it is still possible that DM-induced $\gamma$-rays contribute to the GCE. If the GCE is indeed originated from DM annihilations it should then
be possible to find their trace in other observation targets
or channels.  In particular a similar excess can be searched for in
the dwarf spheroidal galaxies.  Recently the Dark Energy Survey
discovered several new dwarf galaxy candidates\cite{Bechtol:2015cbp}.
In particular, some tentative evidence was proposed for a
gamma-ray excess in the direction of nearby Reticulum
II\cite{Geringer-Sameth:2015lua,Hooper:2015ula} and for a faint excess
that can come from Tucana III\cite{Li:2015kag}, both can be compatible with
the GCE. The Fermi Collaboration also performed an analysis in which no
evidence of an excess was found\cite{Drlica-Wagner:2015xua}.  
Another hint of possible DM discovery was
  reported in\cite{Liang:2016pvm} based on the analysis of Fermi-LAT
  data obtained for nearby galaxy clusters. The observed excess, if
  interpreted in terms of DM annihilation, can point towards similar
  mass range for the DM particles as the GCE. However, this requires
  assuming relatively large boost factors.

  Alternatively, one can consider looking for a corresponding excess
  in the anti-proton spectrum.  Comparisons of the GCE and limits from
  anti-protons were made in\cite{Bringmann:2014lpa,Cirelli:2014lwa}.
  However, the situation is currently unclear with different analyses
  claiming that the GCE can be ruled out or is allowed depending on
  the modeling of the anti-proton propagation (for a recent discussion
  see, \eg,\cite{Cuoco:2017rxb}).

\begin{figure}[t]
\centering
\subfloat[]{%
\label{fig:a}%
\includegraphics[width=0.465\textwidth,trim= 100mm 70mm 10mm 10mm]{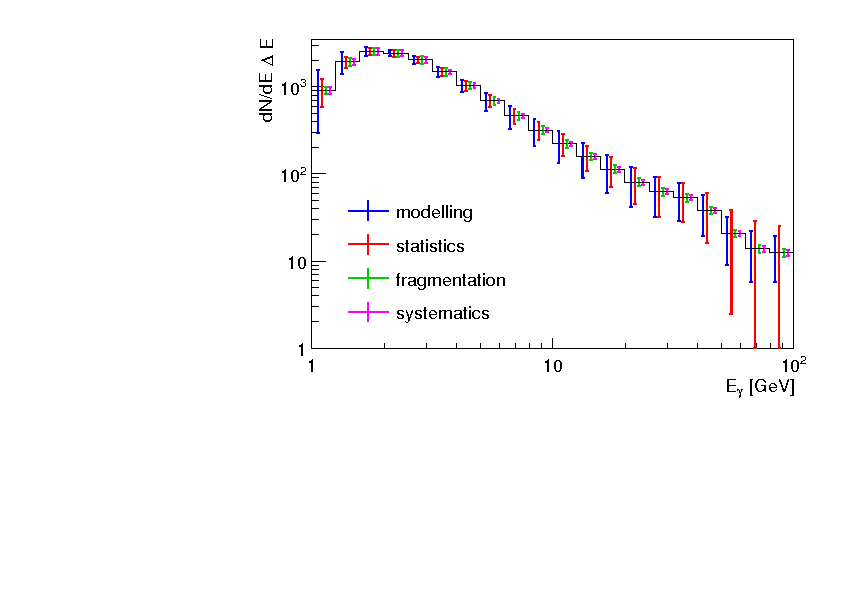}
}%
\hfill
\subfloat[]{%
\label{fig:b}%
\includegraphics[width=0.515\textwidth]{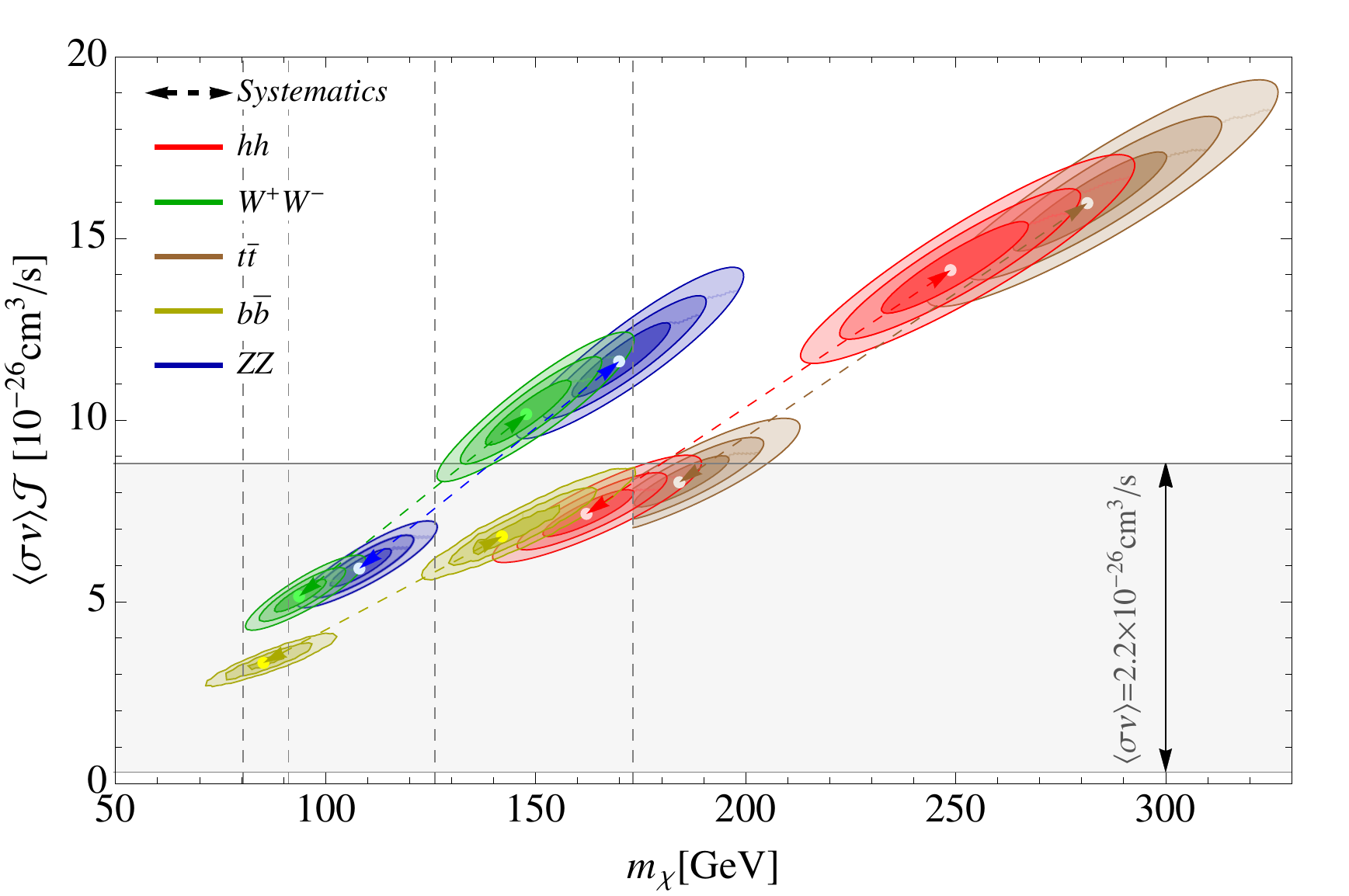}
}%
\caption{\footnotesize (a) Residual spectra of the GCE with various astrophysical and instrumental uncertainties as discussed in\cite{Butter:2016tjc}. Taken from Ref.\cite{Butter:2016tjc}.
(b) Regions of parameter space that fit two of the Fermi best fit spectra for different final states. Parameter space between the contours indicated by dashed line is likely allowed by variation of the background model\cite{Agrawal:2014oha}. Taken from Ref.\cite{Agrawal:2014oha}.}
\label{fig:fermiGCE}
\end{figure}

The GCE attracted a great deal of attention also in terms of model
building including neutralino DM (see, \eg,\cite{Cheung:2014lqa,Cahill-Rowley:2014ora,Agrawal:2014oha,Calore:2014nla,Caron:2015wda,Gherghetta:2015ysa,Bertone:2015tza,Butter:2016tjc}).
Some notable non-SUSY attempts include: multi-step cascade annihilations
where the dark matter annihilates to some intermediate states that
decay to Standard Model
particles\cite{Abdullah:2014lla,Martin:2014sxa,Boehm:2014bia},
approaches based on effective field theory and simplified models of
DM\cite{Alves:2014yha,Balazs:2014jla,Karwin:2016tsw}, Higgs portal
models\cite{Ipek:2014gua,Agrawal:2014oha,Cuoco:2016jqt}, two Higgs doublet
models\cite{Wang:2014elb,Eiteneuer:2017hoh}, models with a $Z'$\cite{Hooper:2014fda} and
vector dark matter models\cite{Ko:2014gha}.

We summarize our discussion of the GCE in \reffig{fig:fermiGCE} taken
from\cite{Agrawal:2014oha,Butter:2016tjc}. \reffig{fig:fermiGCE}(a) shows the
residual spectrum for the GCE, as well as various instrumental and
astrophysical uncertainties, as discussed in\cite{Butter:2016tjc}.
Initially, fits to the excess favored DM annihilating to $b\bar{b}$
and $\tau^+\tau^-$ with $\mchi \approx 40\gev$ and $\mchi \approx
10\gev$, respectively.  The annihilation cross section was reported as
$2-3\times 10^{-26} \textrm{cm}^3/\textrm{s}$\cite{Daylan:2014rsa} for
$b\bar{b}$ and $7-9\times 10^{-27} \textrm{cm}^3/\textrm{s}$ for
$\tau^+\tau^-$\cite{Hooper:2010mq} final states which are remarkably
similar to the annihilation cross section required for a thermal
relic.  However, these are not the only possible annihilation final
states that can fit the GCE signal. \reffig{fig:fermiGCE}(b), taken
from\cite{Agrawal:2014oha}, shows the allowed parameter space for a
WIMP annihilating to $hh$, $W^+W^-$, $ZZ$, $t\bar{t}$ and $b\bar{b}$
for two different Fermi background models.  These models favor the
lightest and heaviest DM masses.  The parameter space between the two
fits, indicated by a dashed line, is likely allowed by variation of
the background model\cite{Agrawal:2014oha}.  Annihilation to leptonic
final states also still provides a good fit to the GCE where the
effect of secondary gamma-rays is important\cite{Lacroix:2014eea}.

\paragraph{Other anomalies} Another excess in the Fermi-LAT data from
the GC was reported in 2012
by\cite{Bringmann:2012vr,Weniger:2012tx}. At that time it was found to
be consistent with the $\gamma$-ray line around $130$ GeV with
possible origin from DM annihilations or decays. However, it was soon
pointed out that the line was not associated with enough signal in the
continuum spectrum which caused some tension with popular DM
candidates trying to fit the observed excess, including the lightest
neutralino\cite{Cohen:2012me,Buchmuller:2012rc}. It has been later shown that the existence of the excess was a statistical
fluctuation as its significance has diminished in subsequent improved
analyses performed by the Fermi-LAT
Collaboration\cite{Ackermann:2015lka}.

Interesting features were identified in the spectrum of
$\gamma$-rays originating from regions around the GC in the WMAP
data\cite{Finkbeiner:2003im} (so-called WMAP haze), Fermi-LAT
data\cite{Dobler:2009xz,Su:2010qj} (Fermi bubbles) and Planck
data\cite{Ade:2012nxf} (Planck haze), with a possible common
origin. The hard edge-like structures observed for Fermi bubbles
disfavor its DM-induced origin. However, it is still possible that some
part of the WMAP/Planck haze are not due to the Fermi bubbles and
can be better fitted with DM annihilations associated with subsequent
microwave synchrotron emission (see,
\eg,\cite{Egorov:2015eta}).

A possible hint of light DM was provided by  the excess of $\gamma$-rays
observed by the INTEGRAL
collaboration\cite{Knodlseder:2003sv,Jean:2003ci}. It corresponds to
$511$ keV $\gamma$-ray line that has no widely accepted astrophysical
explanation. The signal could be compatible with MeV DM annihilating
into $e^+e^-$ pairs\cite{Boehm:2003bt}. Similar signal was found
in the direction of Reticulum II, however astrophysical explanations
of the excess observed there are favored (for further details see,
\eg,\cite{Siegert:2016ijv} and references therein).

The excess in extragalactic radio background found in the ARCADE 2
data\cite{Fixsen:2009xn,Seiffert:2009xs} could be explained by
leptophilic DM with mass typically of order $5-50$
GeV\cite{Fornengo:2011cn,Hooper:2012jc} that annihilate mostly into
electrons and/or muons inducing $\gamma$-ray production via subsequent
ICS processes. This scenario, however, can already be excluded by the
AMS positron data\cite{Fairbairn:2014nga}. On the other hand various
astrophysical explanations of the observed excess were proposed
which, however,  also often struggle to accommodate for the whole excess unless
numerous faint sources are considered\cite{Singal:2009dv} (see,
however,\cite{Fang:2015dga} for an updated discussion).


\subsection{Interlude: WIMP reconstruction from direct detection and
  gamma rays\label{sec:wimpreconstruction}}
Before we proceed with other modes of DM searches, we digress here to
consider what information about WIMP properties one could
realistically derive in case a real DM signal is actually recorded in
either direct detection or in gamma ray experiments, or -- even better
-- in both.

\paragraph{Reconstruction}
Once one day a genuine DM signal is observed, we will enter into a
new era of reconstructing  WIMP properties from experimental data. A
number of theoretical studies have already been conducted to test the quality
of a putative post-discovery reconstruction in DD experiments depending on
the DM mass, respective cross section and target material (for a review
see\cite{Peter:2013aha}). In particular, it has been pointed out that
due to diminishing differences between recoil spectra for a larger DM
mass, DD signal analysis can strongly constrain DM properties only
for $\mchi\lesssim 100$ GeV and for the values of \sigsip\ not far
below the current limits given the realistic assumptions about
achievable exposures. For larger DM mass one typically obtains a
$\sigsip/\mchi\simeq$~const degeneracy, as can be seen from Eq.~(\ref{dRdE})
for $\mchi\gg M$. 

\begin{figure}[t]
\centering
\subfloat[]{%
\label{fig:a}%
\includegraphics[width=0.44\textwidth]{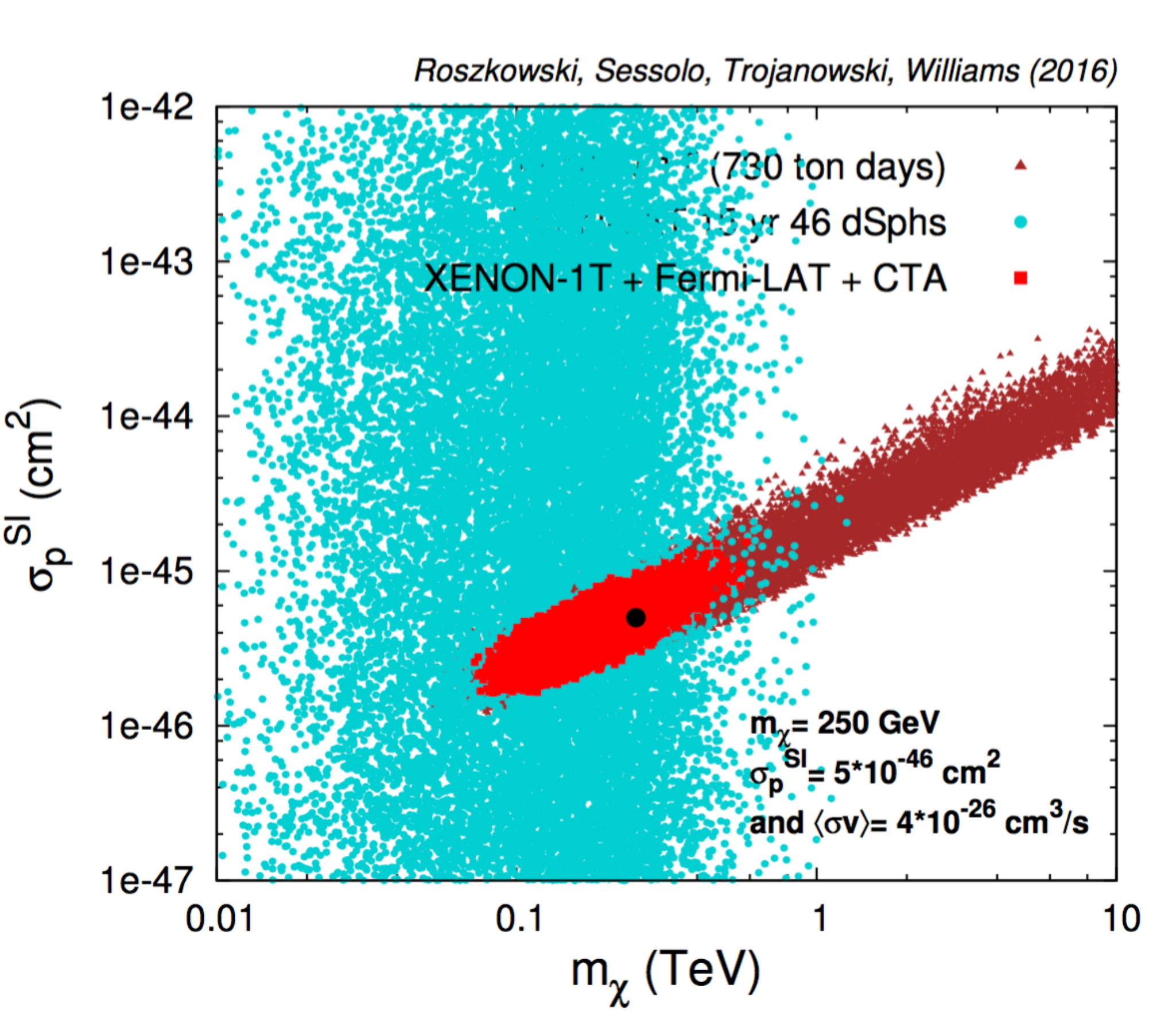}}%
\hfill
\subfloat[]{%
\label{fig:b}%
\includegraphics[width=0.44\textwidth]{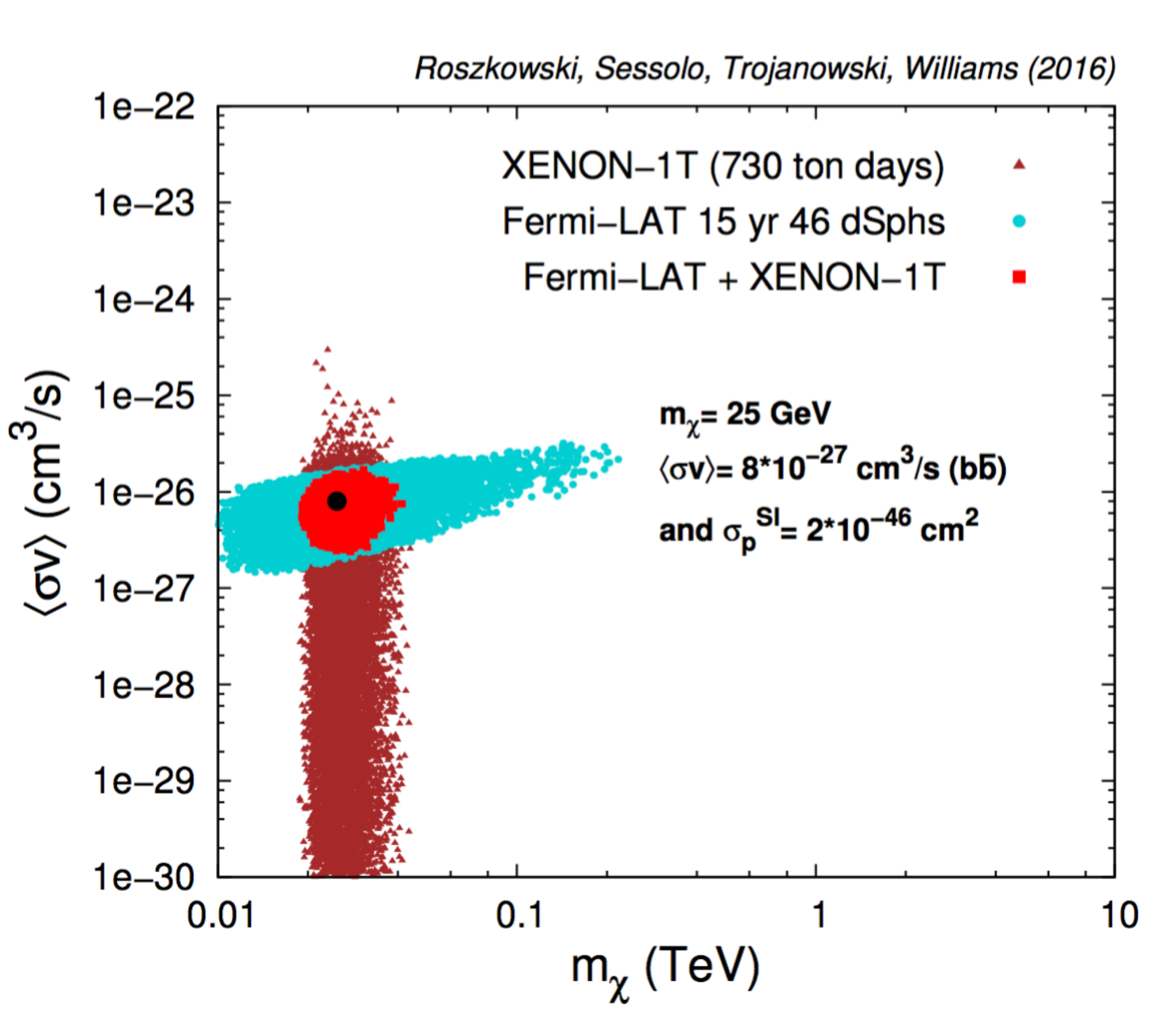}
}%
\caption{\footnotesize \protect\subref{fig:a} Comparison of DD
  (XENON1T) and ID (CTA and Fermi-LAT) experiments in reconstructing
  the DM properties for the benchmark point with $\mchi = 250\gev$  and
  $\sigsip = 5\times 10^{-46}\,\textrm{cm}^2$, while the annihilation
  cross section is equal to $\sigv = 4\times
  10^{-26}\,\textrm{cm}^3/\textrm{s}$ for a pure $b\bar{b}$ final
  state. The brown region corresponds to $2\sigma$ reconstructed
  region for only XENON1T simulated data, the light blue one for Fermi-LAT data (assuming 15 years of exposure and 46 dSphs in the stacked analysis), while the red region
  was obtained for XENON1T+CTA+Fermi-LAT joint analysis. 
\protect\subref{fig:b} Similar to (a), but for the benchmark point
defined by $\mchi = 25\gev$, $\sigsip = 2\times
10^{-46}\,\textrm{cm}^2$, $\sigv = 8\times
10^{-27}\,\textrm{cm}^3/\textrm{s}$ and pure $b\bar{b}$ annihilation
final state. Light blue region corresponds to the the Fermi-LAT
reconstruction, while brown one to XENON1T. Combined analysis leads to  
improved reconstruction of \sigv\ as indicated by the red
region. Both figures are taken from\cite{Roszkowski:2016bhs}. 
\label{fig:DDrecon}}
\end{figure}

However, this can be partially overcome by a possible complementarity
between DD and ID
searches\cite{Bernal:2008zk,Arina:2013jya,Kavanagh:2014rya,Roszkowski:2016bhs},
provided of course that a DM-induced signal is found in both types of
experiments. We illustrate this in \reffig{fig:DDrecon}\subref{fig:a}
where we show an interplay between putative signals from XENON1T and
from two ID experiments: CTA\cite{Consortium:2010bc} and
Fermi-LAT\cite{Atwood:2009ez} (for which we assume 15 years of
exposure and 46 dSphs). We consider a benchmark point with $\mchi =
250\gev$ and $\sigsip = 5\times 10^{-46}\,\textrm{cm}^2$, as well as
the annihilation cross section lying just below the current exclusion
bound\cite{Ahnen:2016qkx}, $\sigv = 4\times
10^{-26}\,\textrm{cm}^3/\textrm{s}$ assuming a pure $b\bar{b}$ final
state. As can be seen, an improved mass reconstruction in the ID
experiments allows one to strongly constrain \sigsip\ which remains
unconstrained from above using the XENON1T (and in fact, any DD) data alone.

For a low DM mass a good reconstruction of \mchi\ in DD
can help interpret the results of ID searches. This is because it is difficult to
distinguish among different DM scenarios based  on results from
ID only, due to an a priori unknown nature of the annihilation final
state and the lack of
characteristic spectral features for typical final states channels,
\eg, $b\bar{b}$ or $\tau^+\tau^-$. However, different
annihilation final states that provide a good fit to the same
signal observed in ID are often associated with different \mchi\ and
\sigv. This is, \eg, the case of a well-known Galactic
Center excess discussed in Section~\ref{sec:gammarays}. Hence,
improved DM mass reconstruction in DD experiments could help in
better discriminating among various annihilation final states and,
eventually, constrain the annihilation cross section. We illustrate
this in \reffig{fig:DDrecon}\subref{fig:b} for the benchmark point
characterized by $\mchi = 25$ GeV, $\sigsip = 2\times
10^{-46}\,\textrm{cm}^2$, $\sigv = 8\times
10^{-27}\,\textrm{cm}^3/\textrm{s}$ and pure $b\bar{b}$ annihilation
final state. As can be seen the DM mass reconstruction in Fermi-LAT is
limited and it is a consequence of the aforementioned degeneracy in
annihilation spectra.\footnote{Specifically, in reconstruction we take
  into account $b\bar{b}$, $\tau^+\tau^-$, $hh$ and $W^+W^-$ final
  states.} On the other hand, a DD measurement of a WIMP signal, which is obviously not sensitive
to \sigv, helps to reconstruct \mchi. As a result also the
annihilation cross section that fits to the assumed ID signal from the
benchmark point is constrained better (for a more detailed discussion
see\cite{Roszkowski:2016bhs}). The reconstructed value of the annihilation cross section could then be mapped into the values of the DM relic density upon additional general assumptions about the WIMP interactions or within the framework of specific models\cite{Roszkowski:2017dou}.


\subsection{Cosmic rays: limits and AMS02/Pamela\label{sec:cosmicrays}}
Charged cosmic rays (CRs) as a tool for DM searches\cite{Silk:1984zy}
play an important and complementary role to $\gamma$-rays as both are
typically produced jointly when the DM particles annihilate or decay
(for recent review see, \eg,\cite{Gaskins:2016cha}). The most
common types of charged cosmic rays are evidently electrons, $e^-$,
and protons, $p$, that can originate from many astrophysical
sources\cite{Blasi:2013rva}. On the other hand, in the case of the
annihilations or decays of neutral DM particles, one expects to
produce an equal number of both matter and antimatter particles. The
latter, including energetic positrons, $e^+$, antiprotons, $\bar{p}$,
antideuterons, $\bar{d}$ (see, \eg,\cite{Fornengo:2013osa}),
as well as heavier nuclei, \eg,
anti-helium\cite{Carlson:2014ssa,Cirelli:2014qia} are particularly
promising tools for DM ID due to relatively low astrophysical
background.

\paragraph{Production and propagation} Charged cosmic rays, similarly
to photons, can be produced both directly in the annihilation and
decay processes as well as in the DM-induced cascades of
particles. As a result one obtains a diffuse spectrum of cosmic rays
with a cut-off at energies close to \mchi\ or $\mchi/2$ for DM
annihilations or decays, respectively. A sharp cut-off can be a
``smoking-gun'' for DM detection since astrophysical sources are
expected to result in a more gradual fall. However, both scenarios can
be distinguished only if sufficient amount of data is collected.

Prompt spectra of DM-induced cosmic rays are subsequently modified due
to diffusion in the Galactic magnetic field during their propagation
to the Earth (for a detailed discussion see,
\eg,\cite{Cirelli:2010xx} and references therein). In the
case of electrons and positrons one also needs to take into account
various mechanisms of energy-loss, including synchrotron radiation and
the inverse Compton scattering on CMB photons or galactic starlight
(for a more detailed discussion see\cite{Strong:2007nh}). These
mechanisms typically play a less
important role for antiprotons and antideuterons, since the
corresponding terms in the diffusion loss equation are
suppressed by the proton or deuteron mass, respectively. However, the
spectrum of $\bar{p}$ and $\bar{d}$ is affected by their possible
interactions with protons in the interstellar medium, as well as
by convective Galactic winds that push antiparticles away from
the Galactic plane, by diffusive reacceleration and by the solar
modulation (see,
\eg,\cite{Gleeson:1968zza,Maurin:2001sj,Donato:2001ms}).

\paragraph{Experiments and anomalies} Searches for charged cosmic rays
employ several detection techniques including balloon-type
(\eg, HEAT\cite{Barwick:1995gv}, ATIC\cite{Chang:2008aa})
and ground-based telescopes (\eg, Pierre Auger
Observatory\cite{Abraham:2004dt}, the Telescope
Array\cite{Kawai:2008zza})
, as well as satellite-based experiments including, \eg,
PAMELA\cite{Picozza:2006nm}, AMS-02\cite{Aguilar:2013qda},
Fermi-LAT\cite{Atwood:2009ez}.\footnote{Note that $\gamma$-rays and
  electron/positron signals cannot be distinguished based on the
  recorded air shower, but ground-based telescopes are still capable
  of studying heavier charged CRs, \eg, protons.}

In particular, in 2009 the PAMELA Collaboration reported an excess in
the positron spectrum\cite{Adriani:2008zr} which was subsequently
confirmed by the Fermi-LAT\cite{FermiLAT:2011ab} and the
AMS-02\cite{Aguilar:2013qda} experiments. The excess was observed for
the energies between $\sim 20$ GeV and $\sim 200$
GeV\cite{Accardo:2014lma}. The distribution of the high-energy
positrons detected by the PAMELA experiment was found to be
isotropic\cite{Adriani:2015kfa}. This could be consistent with their DM origin, but can also be explained by astrophysical sources, 
especially given the uncertain impact of magnetic field configurations on the positron trajectories. Indeed, a DM-related interpretation of the signal was
extensively studied both at the level of general
WIMPs\cite{Cirelli:2008pk} and within the framework of particular
models (see,
\eg,\cite{Bergstrom:2008gr,Hooper:2009fj,Nezri:2009jd,Zurek:2008qg,Chen:2008qs}). As
the DM interpretation of the excess typically requires large
annihilation cross section and/or boost factors, it was constrained to
leptophilic DM models\cite{Chen:2008dh,Donato:2008jk} by null results
of searches for DM-induced antiprotons of similar
strength\cite{Adriani:2008zq}. However, the leptophilic models
themselves seem to be in tension with limits from gamma-ray and X-ray
backgrounds \cite{Profumo:2009uf,Cirelli:2009dv}, the optical depth of
the Universe\cite{Huetsi:2009ex,Cirelli:2009bb}, as well as from radio
data\cite{Bertone:2008xr} and the observations of the CMB
radiation\cite{Ade:2015xua}. The tension is even more
  pronounced in light of recent limits on $\gamma$-rays discussed in
  Section~\ref{sec:gammarays}. On the other hand, one needs to note
that viable astrophysical scenarios were proposed to accommodate for
the observed excess (see, \eg,\cite{Profumo:2008ms}).

Recently an excess in antiproton flux has been confirmed in the AMS-02
data\cite{Aguilar:2016kjl} (see, however,\cite{Giesen:2015ufa}). It can be explained by the annihilation of
the DM particles with the cross section into hadronic final states of
order $3\times 10^{-26}\,\textrm{cm}^3/\textrm{s}$ and the mass that
can be compatible with the $\gamma$-ray
GCE\cite{Cuoco:2016eej,Cui:2016ppb}. On the other hand, it was
shown\cite{Cuoco:2016eej} that these AMS-02 results interpreted in terms of upper limits
lead to an improvement with respect to the limits from $\gamma$-rays
coming from dSphs discussed in
Section~\ref{sec:gammarays}.

All the observed anomalies, as well as other searches for DM-induced
charged cosmic rays will be subject to further studies in currently
operating or future experiments, \eg,
CALET\cite{Torii:2015lck}, DAMPE\cite{Gargano:2017avj},
GAPS\cite{Fuke:2008zz}, in addition to AMS-02.

\subsection{Neutrinos: limits and anomalies\label{sec:neutrinos}}
Attempting to discover one elusive particle by capturing another very
weakly interacting particle is definitely a very challenging
task. However, neutrino detectors proved their unquestionable
usefulness as a tool for DM searches thanks to an enormous
experimental progress. In particular, current best limits on DM-nuclei
spin-dependent cross section \sigsdp\ are based on the results
obtained by several neutrino detectors, including
ANTARES\cite{Adrian-Martinez:2016gti}, IceCube\cite{Aartsen:2016zhm}
and
Super-Kamiokande\cite{Choi:2015ara}. 


\paragraph{Neutrinos from DM annihilations} Depending on a particular
DM model, neutrinos can be produced mainly in cascades of particles
originating from DM annihilations or decays, or even directly in these
processes. However, since annihilation or decay rates are typically
very small, in order to be able to detect DM-induced neutrinos, one
needs to focus on regions in the sky where large concentration of DM
particles can be observed, \eg, the Sun, the GC, Galactic
halo, nearby galaxies and galaxy clusters or even the Earth.

In particular, DM particles are expected to accumulate inside
celestial bodies as their velocity can decrease below the escape
velocity due to scatterings off nuclei. Neutrinos are then basically
the only products of DM annihilation that can escape and reach
detectors. Therefore they can provide a unique DM
signature\cite{Krauss:1985aaa}. The expected flux of neutrinos passing
through a detector depends on the DM annihilation rate
$\Gamma_{\textrm{ann}}$. For heavy and dense celestial bodies
$\Gamma_{\textrm{ann}}$ is determined by the capture rate of DM
$\Gamma_{\textrm{cap}}$ due to the equilibrium condition
$\Gamma_{\textrm{ann}}\simeq
\Gamma_{\textrm{cap}}/2$\cite{Press:1985ug,Gould:1987ir}.\footnote{In
  principle one should also take into account the evaporation process
  of the DM particles from the Sun, but it is negligible for
  $\mchi\gtrsim 4$ GeV\cite{Busoni:2013kaa}.}

When analyzing potential signal from such neutrinos, one needs to take
into account both the neutrino spectrum at production and the
propagation of neutrinos from the center of the celestial body to the
Earth. Both these processes in principle depend on the details of how
DM-induced cascades of particles and neutrinos themselves propagate in
the dense matter. Needless to mention that proper description of
neutrino propagation should also include neutrino oscillations. As a
result neutrino fluxes from DM annihilations in the Sun (see,
\eg,\cite{Baratella:2013fya} and references therein) differ
from the ones obtained based on spectrum at production outside a dense
matter object (see, \eg,\cite{Cirelli:2010xx}).

\paragraph{Experiments and anomalies} Neutrino telescopes (for review
see, \eg,\cite{Danninger:2014xza}) can be divided into two
main categories: muon counters (BAKSAN\cite{Boliev:2013ai}) and water
Cherenkov detectors (ANTARES, IceCube, SuperK).\footnote{Neutrino
  reactor experiments can also be used to constrain DM properties
  (see, \eg, recent studies about this in the case of
  JUNO\cite{Guo:2015hsy} and KamLAND\cite{Kumar:2015nja} detectors).}
The latter technology will also be used in planned neutrino
telescopes, \eg, BAIKAL-GVD\cite{Avrorin:2013sla},
IceCube-PINGU\cite{TheIceCube-Gen2:2016cap}, HyperK\cite{Abe:2014oxa}
and KM3Net\cite{Adrian-Martinez:2016fdl}. The most important
background in searches for DM-induced neutrinos originates from
neutrinos produced in scatterings off cosmic rays in the Earth's
atmosphere\cite{Gaisser:2002jj}. On the other hand muons produced in
the Earth's atmosphere can be vetoed more easily in analysis focusing
on upward going events. An additional source of background is
associated with neutrinos produced in the Sun's
atmosphere\cite{Ingelman:1996mj}, though it is expected to be a
subdominant contribution\cite{Danninger:2014xza}.

Recently, some interest was raised by an observation of neutrinos with
very high energies from tens of TeV up to several PeV reported by the
IceCube Collaboration\cite{Aartsen:2013bka,Aartsen:2013jdh,Aartsen:2014gkd}. Various
possible explanations were proposed including neutrino production
through annihilations\cite{Zavala:2014dla} or
decays\cite{Feldstein:2013kka} of the DM particles. However, the
observed signal seems to be isotropic and is consistent with the
Waxman-Bahcall bound\cite{Waxman:1998yy}, which can be derived from
the spectrum of high-energy cosmic rays emitted by astrophysical
sources when one takes into account a fraction of energy carried away
by neutrinos. This points towards a non-DM origin of the reported
anomaly (for recent review see, \eg,\cite{Murase:2016gly}
and references therein). Future generation of neutrino telescopes
should allow to collect more data and therefore clarify this issue.

\paragraph{Limits} As mentioned above, searches for DM-induced
neutrinos can provide the strongest up-to-date limits on the
spin-dependent cross section \sigsdp\ while current limits on the
spin-independent component \sigsip, as well as on the annihilation
cross section \sigv, remain weaker than the ones derived from DD and
ID experiments, respectively.

\subsection{X-rays: limits and the 3.5 keV line\label{sec:xray}}

In 2014 a possible excess in the X-ray emission near $3.5$ keV was
reported after analyzing the XMM-Newton data from observations of the
Andromeda galaxy and various galaxy clusters with possible connection
to decays of sterile neutrino DM\cite{Bulbul:2014sua,Boyarsky:2014jta}
(see also
\eg,\cite{Ishida:2014dlp,Abazajian:2014gza}). Subsequently,
the signal was confirmed in the data obtained by the Suzaku telescope
for core of the Perseus cluster\cite{Urban:2014yda}, in the XMM-Newton
data for the GC\cite{Boyarsky:2014ska} and in the deep field
observations by NuSTAR\cite{Neronov:2016wdd} and
Chandra\cite{Cappelluti:2017ywp}.  Interestingly, such a signal was
predicted as a smoking gun for WDM in an early
study\cite{Abazajian:2001vt}. In addition to sterile neutrinos,
further DM interpretations of the $3.5$ keV line were proposed
employing, \eg, axion-like
particles\cite{Higaki:2014zua,Jaeckel:2014qea},
axinos\cite{Kong:2014gea,Choi:2014tva} or
gravitinos\cite{Bomark:2014yja}.

The DM interpretation of the line observed in XMM-Newton data was,
however, undermined by some of later studies
(see,\eg,\cite{Jeltema:2014qfa, Jeltema:2015mee}), as well
as by results obtained for several galaxy clusters by the Suzaku
telescope\cite{Urban:2014yda}, XMM-Newton observations of the Draco
dwarf galaxy\cite{Ruchayskiy:2015onc} and the HITOMI data for the
Perseus cluster\cite{Aharonian:2016gzq}. Other explanations were then
discussed in the literature including known emission lines from the
transitions in potassium and chlorine atoms\cite{Jeltema:2014qfa}
(see, however,\cite{Bulbul:2014ala}) and charge exchange between bare
sulfur ions and neutral hydrogen
atoms\cite{Gu:2015gqm,Shah:2016efh}. In addition,
in\cite{Jeltema:2014qfa} it was argued that a similar spectral feature
is present in the data from the Tycho supernova remnant, where one does not
expect to see significant amounts of DM. On the other hand, it was
pointed out that the aforementioned null results of experimental
searches for the $3.5$ keV line are still consistent with the decaying
DM interpretation discussed above while other astrophysical explanations might be insufficient to explain the observed excess\cite{Abazajian:2017tcc}.

At this stage both the DM interpretation, as well as astrophysical
explanation of the observed $3.5$ keV line cannot be fully excluded.

\subsection{LHC mono-X searches\label{sec:lhcdm}}

The third classical strategy for WIMP dark matter searches, after direct and indirect detection, 
is to directly produce a neutral stable particle in high-energy colliders. 
Since the typical coupling and mass range expected for the WIMP in most scenarios is around or just above EWSB, 
the LHC can provide in principle an optimal instrument for pursuing this experimental venue. 

In fact, the vast majority of the searches for new physics at the LHC are designed to look for events that, 
besides the rich hadronic/leptonic activity emerging from the decay chain of the produced visible particle, 
are also characterized by a large amount 
of missing energy, as this simplifies the task of separating them from the SM backgrounds and optimize the chances for detection.  
In this sense, then, the discovery of one or more visible particles in a channel characterized by highly energetic jets or leptons, 
and large missing momentum, would also imply the discovery of a neutral and stable (at least within the detector bounds) particle, 
which could be part of the dark matter or even all of it.

In many scenarios, however, one contemplates the possibility that the dark matter WIMP is the only new field around the electroweak scale, 
while additional visible particles, if existing, are sitting beyond the realistic reach of the detector. 
In this case the detection strategy must involve the identification in the scattering event of one (or a few) 
isolated, highly energetic, object(s) from initial state radiation (ISR), accompanied by large missing momentum. 
The object recoiling against the produced invisible particles can be a jet, a gauge boson, or a lepton, so that 
searches of this typology are commonly referred to in the literature as Mono-X and have 
generated a great amount of activity and excitement in recent years.

While the LHC mono-X search results have been recast in and applied to numerous models with EW dark matter interactions, 
and they proved particularly useful in probing compressed spectra in supersymmetry, 
most official comparisons with the bounds from direct and indirect detection have been presented by ATLAS and CMS 
in two preferential frameworks: EFT and simplified model spectra (SMS). 

In the EFT framework\cite{Cao:2009uw,Beltran:2010ww,Goodman:2010yf,Bai:2010hh,Goodman:2010qn,
Rajaraman:2011wf,Fox:2011pm,Cheung:2012gi,Matsumoto:2014rxa}, which was predominantly used by the LHC collaborations for their interpretations in Run~1\cite{Aad:2013oja,Aad:2014vka,Aad:2014tda,Aad:2015zva,Khachatryan:2014rra,Khachatryan:2014rwa}, 
one derives bounds on the strength of several contact operators, which can be then employed 
for a comparison with the limits on \sigsip\ and \sigsdp\ from direct detection searches and neutrino detectors.

The EFT can in principle provide a good approximation as long as the interaction is mediated by particles with mass well above the collision energy. It was however pointed out in several papers\cite{Hill:2011be,Hill:2013hoa,Vecchi:2013iza,Crivellin:2014qxa,Hill:2014yxa,DEramo:2014nmf} that one should use special care when comparing the Wilson coefficient bounds arising from mono-X searches with those from underground direct detection searches, as the processes involved happen at widely different scales (the EWSB scale in the former case, and the nuclear scale in the latter). 
The effects of renormalization group running should be properly taken into account, particularly in cases when operator mixing introduces 
non-negligible corrections to the expected event rates in underground detectors.     

Moreover, at the center-of-mass energies typically probed in a
collider environment it is often necessary to consider models defined
in terms of renormalizable interactions.  By making use of
SMS\cite{Fox:2011fx,Shoemaker:2011vi,Busoni:2013lha,Busoni:2014sya,Busoni:2014haa,Racco:2015dxa,Xiang:2015lfa},
one introduces simple renormalizable Lagrangians, characterized by a
limited number of free parameters, like the couplings of the dark
matter to the visible sector, or the mass of the particles assumed to
mediate the interaction between the dark matter and the partons in the
nucleons.

\begin{figure}[t]
\centering
\subfloat[]{%
\includegraphics[width=0.47\textwidth]{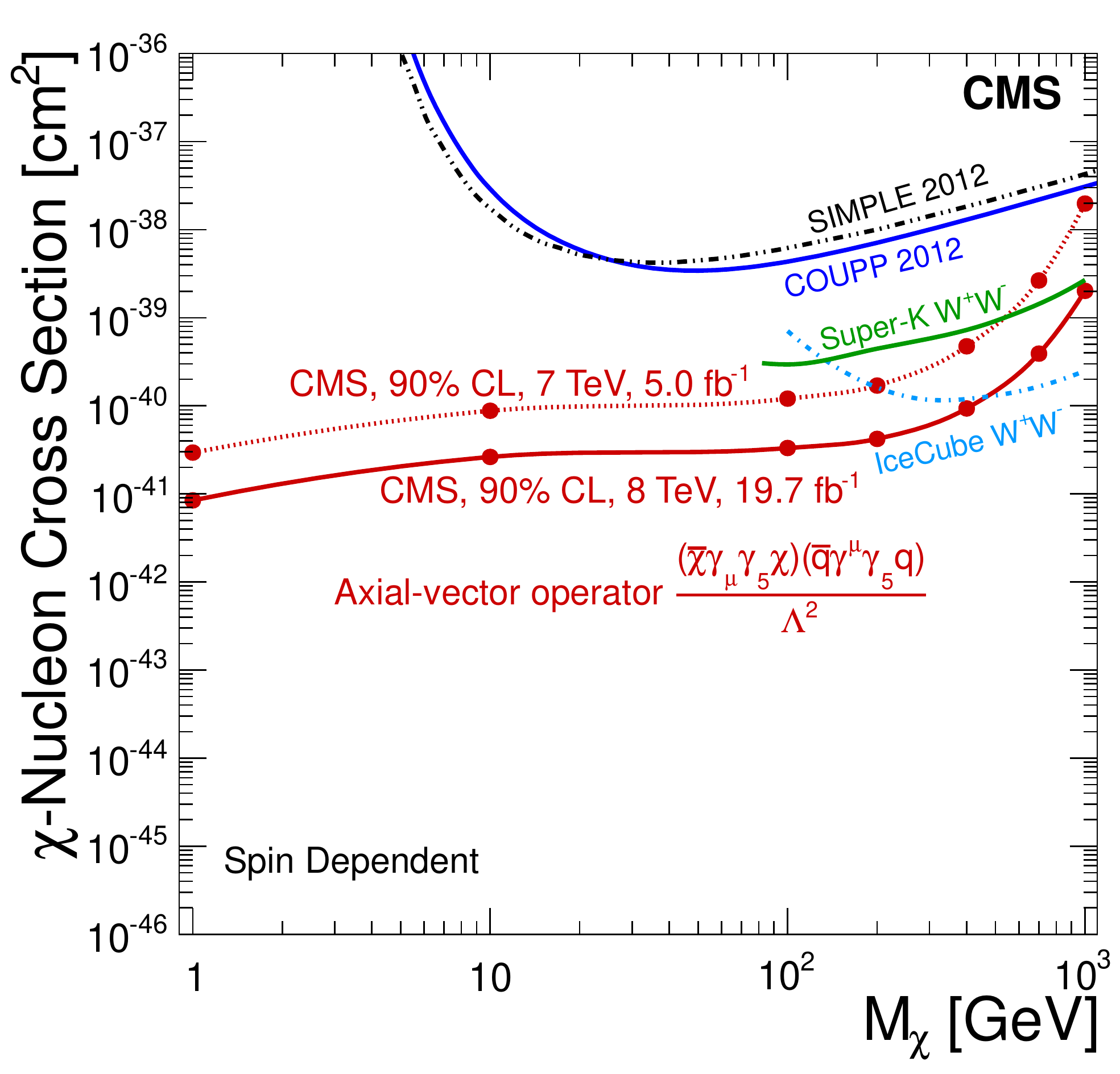}
}%
\hspace{0.02\textwidth}
\subfloat[]{%
\includegraphics[width=0.47\textwidth]{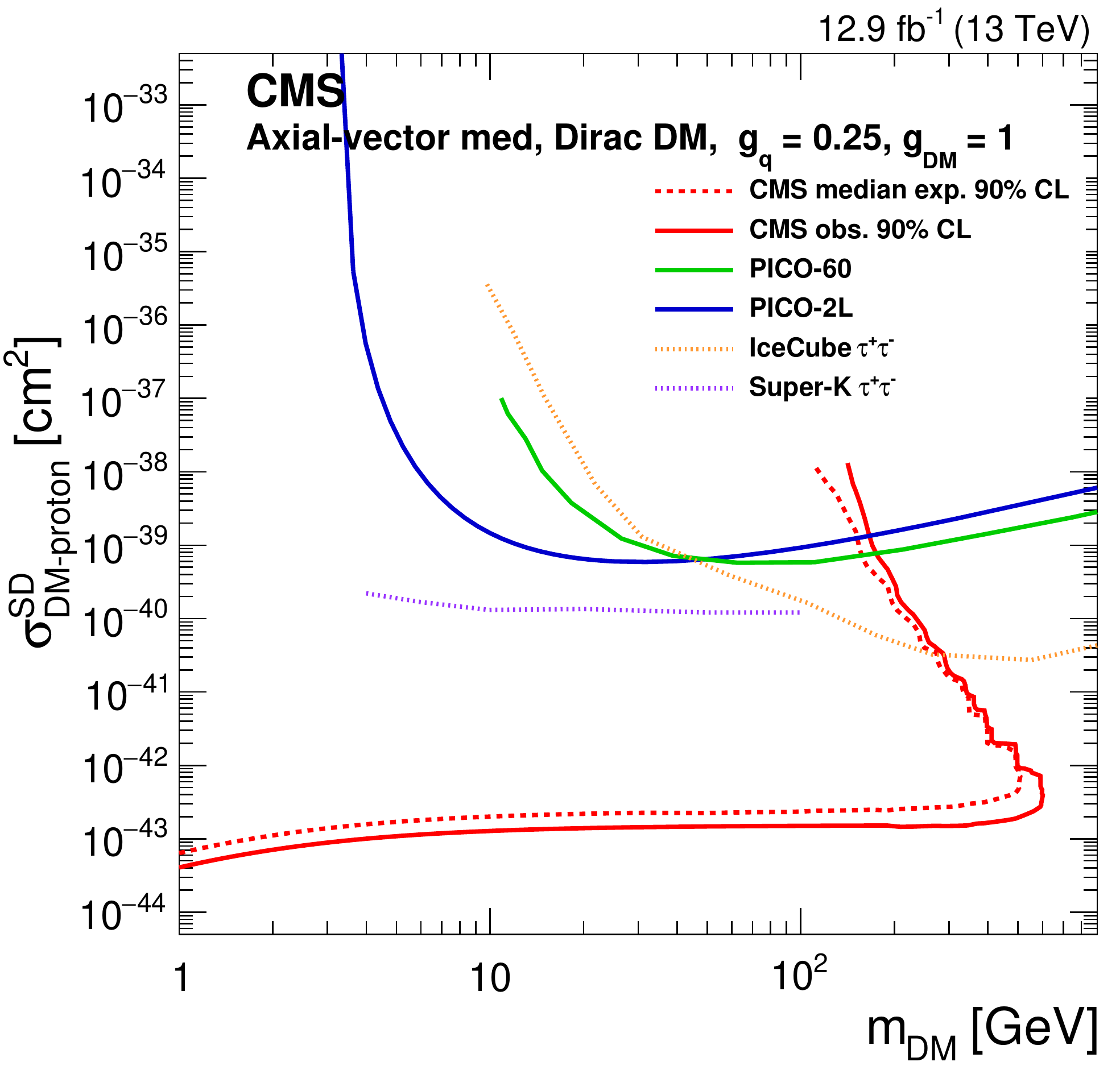}
}%
\caption{\footnotesize (a) The CMS mono-jet bound in the (\mchi, \sigsdp) plane at the end of Run~1, interpreted in the EFT framework. Plot taken from Ref.\cite{Khachatryan:2014rra}. (b) The most recent CMS Run~2 mono-jet bound in the (\mchi, \sigsdp) plane, interpreted in the SMS framework. Plot taken from Ref.\cite{Sirunyan:2017hci}.}  
\label{fig:MonoX}
\end{figure}

In \reffig{fig:MonoX}(a) we show the bounds from the CMS mono-jet search\cite{Khachatryan:2014rra} 
in the (\mchi, \sigsdp) plane at the end of Run~1.
Note that the limit is actually placed on the effective coupling, $1/\Lambda^2$, of a dimension-6 axial-vector operator 
(where running and operator-mixing effects are neglected).
The equivalent CMS bound at the end of Run~2\cite{Sirunyan:2017hci} is shown in \reffig{fig:MonoX}(b). 
It is now expressed in terms of a simplified model with Dirac dark matter, axial-vector mediator, and specific coupling strengths (see\cite{Aaboud:2016tnv} for the equivalent ATLAS bound). 
The typical ``hook'' shape of the exclusion contour is due to the fact that 
for any specific dark matter mass the data excludes a range of mediator masses. Additional bounds on the spin-independent 
cross section, \sigsip, and the annihilation cross section, \sigv, can be found in the experimental papers.

It is worth pointing out that the upper bounds of \reffig{fig:MonoX}
are especially competitive in the low range of the dark matter mass
spectrum, as the probability of emission of a high $p_T$ jet drops
drastically when \mchi\ approaches the $p_T$ cut.  Since in the remainder of this
review we focus predominantly on WIMPs of mass in the hundreds of GeV
to a few TeV range, we will avoid discussing mono-jet bounds in
greater detail.  Excellent reviews exists in the literature exploring
the LHC bounds on a large range of light and not-so-light dark matter
scenarios.  For a very recent one see, e.g.,\cite{Arcadi:2017kky} and
references therein. 

\section{The neutralino WIMP as DM\label{sec:neutralino}}

Despite the disappointing failure to discover any superpartners at the
LHC (and, in fact, any trace of ``new physics''), low scale
superymmetry (SUSY) still remains arguably by far the best motivated
scenario for ``new physics'' beyond the Standard Model.  A detailed
review of SUSY exceeds the purpose of this work, but numerous extensive
reviews exist in the literature (see, e.g.,\cite{Martin:1997ns,Drees:2004jm,Baer:2006rs}).  We
limit ourselves to recalling a few basic notions that will be relevant
for the connection to DM.

\subsection{Brief review of supersymmetry and the neutralino as dark matter}

SUSY is a space-time symmetry relating each particle of the SM to a
partner whose spin differs by 1/2.  We only consider here
$\mathcal{N}=1$ SUSY, where the space-time algebra is extended by
exactly one spinorial SUSY generator, as this is the only case that
admits a phenomenology with chiral fermions, and constitutes a
straightforward extension of the particle content of the SM.

Initially developed on the basis of aesthetic and
``proof-of-existence'' considerations, in the eighties it became arguably
the most popular solution to the gauge-hierarchy problem. Roughly
speaking, if a low-energy effective theory -- as the SM is thought to
be -- includes light fundamental scalar fields, like the Higgs boson,
the mass of the scalar particles is subject to strong renormalization
by the fields of the UV completion. If the UV completion typical scale
is close to the scale of quantum gravity, one needs a fine tuning of
approximately 28 orders of magnitude to justify a scalar mass of the
order of the EW scale.  SUSY provides an attractive solution to this
problem thanks to the non-renormalization
theorem\cite{Grisaru:1979wc,Seiberg:1993vc} which precludes
one-particle irreducible loop corrections to the superpotential so
that, as a consequence, mass terms do not get renormalized. In other
words, SUSY ``protects'' the Higgs mass of the SM and makes it
technically natural.

Apart from solving the gauge hierarchy problem, SUSY provides a
framework that naturally accommodates at the same time several
theoretical expectations and a number of experimental data.
Low-energy SUSY, in particular, the Minimal Supersymmetric Standard
Model (MSSM), provides
the right particle content for high-scale gauge coupling unification;
it furnishes a rationale for the measured values of the mass of the
Higgs boson and of the top quark; it provides a natural framework for
models of inflation and baryo/lepto-genesis; radiative electroweak
symmetry breaking (EWSB) can easily be achieved in the MSSM; and,
finally, but perhaps most importantly for the scope of this review,
some superpartners are weakly interacting and, if stable (or very long
lived) are a natural candidate for a WIMP and DM. Among them the most
popular one is the lightest neutralino, which we will refer to simply
as \textit{the neutralino} and denote with a symbol $\chi$ hereafter. Countless studies, 
starting from\cite{Goldberg:1983nd,Ellis:1983ew}, showed it to be an excellent
thermal DM candidate.

We remind the reader that the low-energy Lagrangian of the $R$
parity-conserving MSSM consists of two parts.  One comes from the
superpotential, expressed in terms of superfields (marked by carets),
which essentially provides a direct supersymmetrization of the Yukawa
part of the SM Lagrangian:
\begin{equation}
W_{\textrm{MSSM}}=\mathbf{y_u} \hat{U}^c \hat{H}_u\hat{Q}-\mathbf{y_d} \hat{D}^c\hat{H}_d\hat{Q}-\mathbf{y_e} \hat{E}^c\hat{H}_d\hat{L}+\mu \hat{H}_u\hat{H}_d\,,\label{superpot}
\end{equation}
where the $\mathbf{y_{u,d,e}}$ are $3\times 3$ Yukawa matrices and $\mu$ is the Higgs/higgsino mass 
parameter. 

The other part is the so-called ``soft'' SUSY-breaking Lagrangian,
which includes mass terms for the gauginos (Majorana fermion
superpartners of the gauge bosons) and for the scalar superpartners of
the SM fermions:
\begin{eqnarray}
\mathcal{L}_{\textrm{soft}}&=&-\frac{1}{2}\left(M_3 \gluino\gluino+M_2 \wino\wino+M_1 \bino\bino+\textrm{c.c.}\right)\nonumber\\
 & &-\left(\mathbf{a_u}\tilde{u}^{\dag}H_u\tilde{Q}+\mathbf{a_d}\tilde{d}^{\dag}H_d\tilde{Q}+\mathbf{a_e}\tilde{e}^{\dag}H_d\tilde{L}+\textrm{c.c.}\right)\nonumber\\
 & &-\tilde{Q}^{\dag}\mathbf{m_Q^2}\tilde{Q}-\tilde{L}^{\dag}\mathbf{m_L^2}\tilde{L}-\tilde{u}^{\dag}\mathbf{m_u^2}\tilde{u}
-\tilde{d}^{\dag}\mathbf{m_d^2}\tilde{d}-\tilde{e}^{\dag}\mathbf{m_e^2}\tilde{e}\nonumber\\
 & &-\mhusq H_u^{\ast}H_u-\mhdsq H_d^{\ast}H_d-\left(bH_uH_d+\textrm{c.c.}\right)\,,\label{softlagr}
\end{eqnarray}
where $M_1$ is the mass of the bino, $M_2$ of the wino, and $M_3$ of
the gluino, which are the fermionic partners of the $B$, the $W$
triplet and the gluon octet.  The matrices $\mathbf{m_Q^2}$, \etc,
$\mathbf{a_u}$, \etc, and $b$, stand for mass squared, trilinear, and
bilinear coefficients for the scalar fields, respectively.

%
%

The gauginos transform under the adjoint representation of the
respective gauge groups so that, in particular, the bino transforms as
a $U(1)$ phase and three wino states form a triplet of $SU(2)$.  In
addition, there exist two more Majorana fermions, the higgsinos of
mass $\mu$, which belong to $SU(2)$ doublets.  The bino and the
neutral degrees of freedom of the the winos and the higgsinos have the
same quantum numbers and the neutralinos are their mass eigenstates.
The masses are obtained by diagonalizing the mass matrix $\boldmath M_\chi$  given by 
\bea\label{neutmatr} \mathbf{M_\chi}=
  \begin{bmatrix}
    M_1 & 0 & -\frac{g'}{\sqrt{2}}v_d & \frac{g'}{\sqrt{2}}v_u \\
    0 & M_2 & \frac{g}{\sqrt{2}}v_d & -\frac{g}{\sqrt{2}}v_u \\
   -\frac{g'}{\sqrt{2}}v_d & \frac{g}{\sqrt{2}}v_d & 0 & -\mu \\
   \frac{g'}{\sqrt{2}}v_u & -\frac{g}{\sqrt{2}}v_u & -\mu & 0 
 \end{bmatrix}, \eea  
where $g$ and $g'$ are $SU(2)$ and $U(1)$ gauge couplings,
 respectively, and $v_u$ and $v_d$ are the vevs of the neutral
 components of the Higgs doublets $H_{u}$ and $H_d$.

 While it is the mass eigenstates that are the physical states, they
 can be dominated by some gauge eigenstates which allows one to make
 convenient approximations.  In the limit where one among $M_1$,
 $M_2$, and $\mu$ is much smaller than the other parameters, the lowest
 eigenvalue approximately coincides with the lightest of these masses.
 In other words, $\mchi\approx M_1$ when $M_1\ll M_2, \mu$, and so on
 for interchanging orders. When two or more masses are instead
 comparable, mixing effects come into play and can change the
 phenomenology.
In this context, it became clear after the LHC Run~1 and beginning of Run~2 that gaugino and higgsino 
masses are likely to be well above the Higgs vevs, i.e., $M_1, M_2, \mu\gg v_u, v_d$.  

The most popular extension of the MSSM is arguably the 
  Next-to-Minimal Supersymmetric Standard Model (NMSSM), which can provide an elegant
solution to the $\mu$ problem (see, e.g.,\cite{Ellwanger:2009dp} for a comprehensive review).
In the NMSSM there is one additional kind of neutralino, the singlino, 
which is the fermionic partner of the gauge singlet Higgs field.\footnote{In the NMSSM the superpotential 
includes additional terms involving a gauge singlet chiral superfield $\hat{S}$, i.e., $W\supset\lambda\hat{S}\hat{H}_u\hat{H}_d+\kappa/3\,\hat{S}^3$.\label{foot:nmssm}}
Because of the presence of the singlino, the NMSSM can sometimes present complementary DM signatures with respect to the MSSM.

 The neutralinos, being electric and color charge-neutral, interact
 with the SM with the strength of the weak interaction. The lightest
 among them, if it is the lightest supersymmetric particle (LSP), is
 stable provided an additional discrete symmetry ($R$-parity) is
 assumed.  Note, incidentally, that $R$-parity violation is in general
 strongly constrained by bounds on proton decay and precision tests of
 the SM\cite{Olive:2016xmw}.  Below we will review the properties
 of the neutralino and the present status of this important candidate
 that has become over time the paradigm of WIMP DM.

\subsection{Neutralino relic abundance\label{sec:reldens}}

We will now discuss in more detail the mechanisms that can lead to the
correct value of the neutralino DM relic density.  As we will see,
this often requires going beyond the simplest WIMP picture that we
discussed in Section~\ref{sec:thermalWIMPs}.  In particular, for the
bino-like neutralino (i.e., for $\mchi\approx M_1$), which
early on\cite{Roszkowski:1991ng} was
the most favored scenario in SUSY models with gaugino
mass unification at the GUT scale, one typically obtains too small an
annihilation cross section and, therefore, exceedingly large values of
the DM relic density.  However, this can be improved by assuming
specific mass patterns for the neutralinos and some other SUSY
particles. 

\subsubsection{Coannihilations}\label{sec:cannihilations}

One of the most important mechanisms where specific mass relations
between the LSP and some other states determine the relic abundance is
coannihilations (see Sec.~\ref{sec:wimpproduction}).  In
phenomenologically interesting scenarios the lightest neutralino can
be mass degenerate with some heavier supersymmetric species (which
usually is the next-to-lightest supersymmetric particle, or NLSP) thus
fulfilling some of the conditions where coannihilations can play a major
role in determining the DM relic density. The mass degeneracy can
lead to either a decrease or an increase of the final relic abundance,
depending on whether the NLSP freeze-out occurs later or earlier than
for the LSP. 

The neutralino-chargino mass degeneracy plays a major role in
determining \abundchi\ for a wino- or higgsino-like neutralino.  This
is due to characteristic mass degeneracies between the higgsino-like
(in which case $\mchi\approx \mu$) or the wino-like (with $\mchi\approx
M_2$) neutralino and the lightest chargino or the second lightest
neutralino that appear naturally in the MSSM -- recall that higgsinos
are $SU(2)$ doublets and winos are $SU(2)$ triplets.

The correct value of \abundchi\ can be obtained
without assuming any special mass relations.  This is mainly due to the fact that the most
efficient annihilation and coannihilation channels are in this case
determined by the processes whose strength is set by the respective gauge
couplings. As a result, the relic abundance $\abundchi\sim
1/\sigma_{\rm ann}\sim \mchi^2/g^4$ shows a simple parabolic
dependence on the neutralino mass. 

The wino and, to a much lesser extent, higgsino relic density are also
influenced by the Sommerfeld enhancement which is the enhancement of
the annihilation cross section due to a modification of the Yukawa
potential induced by the electroweak gauge bosons\cite{Hisano:2006nn}
(see also\cite{Hryczuk:2010zi,Beneke:2014hja} for a recent
discussion).  This effect is particularly important in the wino mass
range for which one obtains $\Omega_{\chi}\,h^2\approx 0.12$, which is
then broadened to $M_2\approx 2-3\tev$.  Although the Sommerfeld
enhancement is most important for a nearly pure wino, it can also play
a role for a mixed wino/higgsino state, thus modifying the relevant
area of the so-called relic neutralino surface in the parameter space
at which the correct value of the relic density is
achieved\cite{Bramante:2015una}.  In the case of the higgsino-like
neutralino the relic density is reduced mainly thanks to a triple mass
degeneracy between the two lightest neutralinos and the lightest
chargino\cite{Mizuta:1992qp}.  As a result, the correct \abundchi\ is
obtained for $m_\chi\approx 1\tev$ -- we will refer to it as the $\sim1\tev$
higgsino region.

In the case of bino-like lightest neutralino an important role in
determining the relic density is played by stau
coannihilation\cite{Ellis:1998kh, Nihei:2002sc,Roszkowski:2001sb} when
the LSP is mass-degenerate with the lightest stau, $m_{\chi}\approx
m_{\tilde{\tau}_1}\lesssim 400-500\gev$ for $\abundchi\approx
0.12$. The same effect can be obtained for other sleptons. On the
other hand, coannihilations of higgsino-like or wino-like DM with
sleptons lead to an increase of the relic
density\cite{Profumo:2006bx}. 
Interestingly, thanks to this
effect one can obtain $\abundchi\approx 0.12$ for higgsino mass as
small as $m_{\chi}\approx 600\gev$\cite{Roszkowski:2014iqa}.

Coannihilations with squarks can lead to the correct value of the
lightest neutralino relic density for a significantly heavier
neutralino.  In particular, such coannihilations can occur with the
lightest stop\cite{Boehm:1999bj}, $\tilde{t}_1$, or with the lightest
sbottom, $\tilde{b}_1$, which are often the lightest squark
states\cite{Arnowitt:2001yh}.  A similar mechanism leads to a
reduction of \abundchi\ for a heavy neutralino mass-degenerate with
the gluino, i.e., when $\mchi\approx
m_{\tilde{g}}$\cite{Profumo:2004wk}.  For both the stop and gluino
coannihilation it is possible to obtain $\abundchi\approx 0.12$ for
$\mchi$ as large as 6--9\tev\ when the Sommerfeld enhancement and
gluino-gluino bound-state effects are
incorporated\cite{Ellis:2014ipa,Ellis:2015vaa}, although this should
not be treated as a strict upper limit on phenomenologically
acceptable values of \mchi. 

In the framework of the NMSSM a nearly pure singlino, which can be the
lightest neutralino, typically interacts very weakly.  It annihilates
mainly into scalar-pseudoscalar pairs, with the associated couplings
proportional to $\kappa$ or $\lambda$ (cf. Footnote~\ref{foot:nmssm}), which are typically small.  
As a result, the singlino relic density is often too large. However, this
can be improved thanks to coannihilations with an higgsino, a wino, a
stau/sneutrino, a stop, or a gluino (for a detailed discussion
see\cite{Belanger:2005kh}).

\subsubsection{Funnels\label{sec:neutralinoreldens}}

Another important mechanism that can enhance neutralino pair
annihilation and lead to the correct value of the relic density of
neutralino DM in the MSSM is due to annihilations via the resonant
$s$-channel exchange of the $Z$-boson, the light Higgs
$h$\cite{Ellis:1989pg}, and/or heavy (pseudo)scalar Higgs bosons $H$
and $A$, in the respective resonance (or funnel) regions of the
parameter space\cite{Drees:1992am} if the exchanged particle mass $m$
is roughly twice $\mchi$. In a more precise treatment this condition
is slightly modified by taking into account the thermal average of the
relative velocity of annihilating neutralinos in the early Universe.
Hence both the $Z$-resonance and the $h$-resonance regions require a
light neutralino ($\mchi<100\gev$), as $m_Z=91\gev$ and $m_h=125\gev$.
On the other hand, in the $A$-funnel region the lightest neutralino
can be much heavier. In the NMSSM, where the Higgs spectrum is richer,
accordingly more resonance channels are in principle possible.

\subsubsection{Other mass patterns}

In the absence of any accidental mass patterns, the bino annihilation
rate is typically dominated by a $t$-channel slepton exchange,
$\chi\chi\rightarrow ll$, and consequently $\abundchi\sim
m_{\tilde{l}}^4/\mchi^2$ is also sensitive to the mass of the lightest
slepton, $m_{\tilde{l}}$\cite{Roszkowski:1991ng,Drees:1992am}.  This
can lead to the bino relic density spanning a few orders of magnitude
depending on both relevant masses.  In particular, one can obtain
$\abundchi\approx 0.12$, if $m_{\widetilde{B}} < m_{\tilde{l}}
\lesssim 150\gev$, in the so-called bulk  region\cite{Griest:1990kh,Roszkowski:1991ng,Drees:1992am}.

Another important option is associated with a general mixed
bino-higgsino LSP.  In the mass range $100\gev\lesssim \mchi\lesssim
1\tev$ a nearly pure higgsino $\chi$ typically yields too small a
value of the relic density (while for a pure bino it is typically too
large).  This can be circumvented for choosing an appropriate
(``well-tempered'') admixture of $\widetilde{B}$ and
$\widetilde{H}_{u,d}$.  In the context of GUT-constrained SUSY models
such a scenario can be realized in the so-called hyperbolic
  branch/focus point region\cite{Chan:1997bi,Feng:1999zg}.  The
annihilation rate in this case is dominated by neutralino
annihilations into gauge bosons, as well as through a $t$-channel
exchange of a higgsino-like chargino and/or the second lightest
neutralino.

Among other scenarios with mixed neutralino LSP that have been
discussed in the literature one can distinguish the mixed bino-wino
(see, e.g.,\cite{BirkedalHansen:2002am,Baer:2005jq,ArkaniHamed:2006mb},
singlino-higgsino (in the context of the NMSSM)\cite{Belanger:2005kh}
or even bino-higgsino-wino (see,
e.g.,\cite{Bae:2007pa,Feldman:2009wv}) states.

One loop corrections to annihilation cross sections can provide some
improvement in the computation of the neutralino relic density that
can introduce corrections of the order of the observational error and
therefore should be taken into account when estimating the uncertainty
of the determination of \abundchi\ (see, \eg,\cite{Baro:2007em,Akcay:2012db} or
recent\cite{Beneke:2014gla,Harz:2014tma} and references therein).

\subsection{Simplest models defined at the GUT scale: the CMSSM and the NUHM\label{sec:cmssm}}

As was mentioned above,  the MSSM can feature many non-trivial phenomenological signatures and it is important to understand
that predictions in different sectors of the theory can be intertwined. 

In this subsection, we show how these relations affect
the DM predictions in two popular SUSY models with simple unified, or
constrained, boundary conditions set at the GUT scale, the constrained
MSSM (CMSSM)\cite{Kane:1993td} and the Non-Universal Higgs Model
(NUHM)\cite{Matalliotakis:1994ft}.  The CMSSM, inspired by
supergravity constructions, is characterized by a set of four
parameters defined at the GUT scale: the unified scalar and gaugino
masses, \mzero\ and \mhalf, respectively; the unified trilinear
coupling \azero; ratio of the Higgs vevs \tanb, and the sign of the
$\mu$ parameter.  In the NUHM, one instead does not assume the soft
Higgs masses $m_{H_u}^2$ and $m_{H_d}^2$ to be unified with the other
scalar masses at the GUT scale, thus introducing two additional free
parameters. For many years these models remained an important
playground for SUSY phenomenology in the context of unification. In
particular, we focus on the implications of the recently discovered
the Higgs boson with mass close to 125\gev\ and the nature and
associated discovery prospects of the DM.

In \refsec{sec:pmssm}, we extend the analysis to a more general case
of the phenomenological MSSM (pMSSM), which roughly encompasses the
remainder of signatures and possibilities for the discovery of
neutralino DM in the general MSSM.

First, however, we discuss important implications of the properties of
the Higgs boson discovered at the LHC on the expected mass range of
superpartners. As we will see, in the framework of unified SUSY models
this will have important ensuing implications for the nature and
discovery prospects of WIMP DM in this class of models.

\subsubsection{Implications of the Higgs boson for SUSY breaking scale}

In the MSSM the mass of the Higgs boson is not a free parameter of the
theory, in contrast to the SM. Its value is calculated in terms of the
parameters of the model. Since the quartic couplings of the Higgs
fields are roughly given by the EW gauge couplings, the tree-level
value of the Higgs mass presents an upper bound determined by the mass
of the $Z$ boson, $M_Z$.  As a consequence, the observed value of the
Higgs mass, $\mhl= 125\gev$, implies the presence of significant
radiative corrections, which increase logarithmically as the SUSY
scale increases.

It is convenient to express the dominant 1-loop contribution to the
radiative corrections of the Higgs mass in terms on the stop mass and
stop mixing (see, e.g.,\cite{Haber:1996fp}), 
\be
\delta\mhl^2\approx\frac{3y_t^4}{16\pi^2}v^2\left[\ln\left(\frac{\msusy^2}{m_t^2}\right)+\frac{X_t^2}{\msusy^2}
  \left(1-\frac{X_t^2}{12\msusy^2}\right)\right],\label{higgscorr} 
\ee
where $y_t$ and $m_t$ are the top Yukawa coupling and the top quark
mass computed in the $\overline{MS}$ scheme, respectively,
$v=\sqrt{v_u^2 + v_d^2}\simeq 246\gev$ is the EW vev, the SUSY scale
is set at the geometrical average of the stop masses,
$\msusy=\sqrt{\mstopone\mstoptwo}$, and $X_t=A_t-\mu \cot\beta$ gives
the main stop mass matrix off-diagonal term.

\begin{figure}[t]
\centering
\subfloat[]{%
\label{fig:a}%
\includegraphics[width=0.54\textwidth]{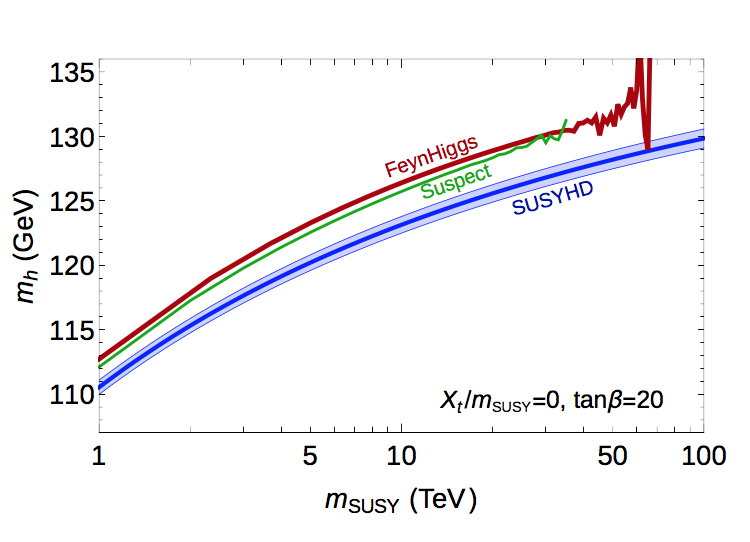}
}%
\hspace{0.02\textwidth}
\subfloat[]{%
\label{fig:b}%
\includegraphics[width=0.4\textwidth]{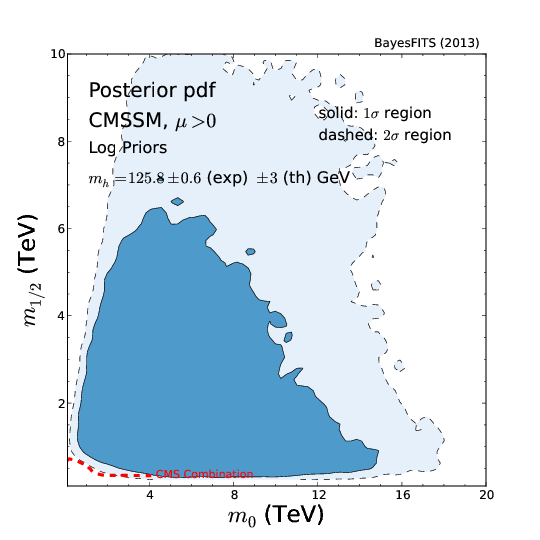}
}%
\caption{\footnotesize (a) A calculation of the Higgs mass obtained with
  different numerical programs. Taken from Ref.\cite{Vega:2015fna}.
  (b) The $1\sigma$ and
  $2\sigma$ regions of marginalized 2-dimensional posterior
  probability density function in the CMSSM that are consistent with
  the Higgs mass. Taken from Ref.\cite{Kowalska:2013hha}.
\label{fig:Higgsmass}}
\end{figure}

Equation~(\ref{higgscorr}) is sufficient to describe the qualitative
behavior of the Higgs boson mass but higher-order corrections are
necessary to obtain a more precise quantitative estimate of the
favored value of \msusy\ after the Higgs discovery.  Several numerical
codes are available to calculate the Higgs mass in terms of the MSSM
parameters.  The results are subject to significant uncertainty, which
is due to missing higher-order loop corrections, the choice of
renormalization scheme, etc.  Figure~\ref{fig:Higgsmass}(a), taken from
Ref.\cite{Vega:2015fna}, presents a comparison of the Higgs mass
obtained by different codes. The
dependence on the geometrical average of the stop masses, \msusy, is
presented in the MSSM.  \reffig{fig:Higgsmass}(b), taken from
Ref.\cite{Kowalska:2013hha}, shows in the plane $(\mzero,\mhalf)$ of
the CMSSM the $1\sigma$ and $2\sigma$ regions of marginalized
2-dimensional posterior probability density function (pdf) consistent
with the Higgs mass, taking into account errors, both experimental (which have
since decreased considerably) and theoretical (which are estimated at
the level of $2-3\gev$). 
One can see that the measured Higgs mass value alone suggests typical
superpartner masses to be in the multi-\tev\ range. Strictly speaking,
in phenomenological models like the phenomenological MSSM (pMSSM) this conclusion applies 
to the masses of the stops, but in unified models this sets the overall
scale of SUSY breaking, since in those models superpartner
masses are related by boundary conditions in the UV. 
Note, that no additional conditions, in particular of satisfying the relic density
constraint, have been imposed here. 

We stress that the above conclusion applies not only to the CMSSM, but
to a much wider class of models, some equally well inspired by supergravity,
like the NUHM and non-universal gaugino mass models, and other ones for
which SUSY breaking is transmitted to the visible sector via other
high-scale messengers, like in the case of gauge mediation.  On the
other hand, multi-\tev\ expectations for the scale of superpartner
masses are independently supported by increasingly stringent lower
limits from the LHC and by the lack of any convincing departure from
Standard Model values of rare processes involving flavor, \eg,
\bsgamma, \bsmumu, \etc.


\subsubsection{Neutralino DM in unified models  in light of LHC and other recent data}

The requirement that the neutralino relic density is close to the
observed value places an additional strong constraint on unified SUSY
models. (In contrast, the impact of limits from direct searches for DM
has not been as strong as that from collider
searches\cite{Kowalska:2015kaa}.)  In these models this additionally
implies specific properties for the neutralino LSP and, therefore, for
DM searches.

A large number of global studies 
(see, e.g.,\cite{Baer:2011ab,Kadastik:2011aa,Cao:2011sn,Ellis:2012aa,Baer:2012uya,Bechtle:2012zk,Balazs:2013qva,
  Fowlie:2012im,Akula:2012kk,Buchmueller:2012hv,Strege:2012bt,Cabrera:2012vu,Kowalska:2013hha,Dighe:2013wfa,Cohen:2013kna,Buchmueller:2013rsa,Roszkowski:2014wqa,Bechtle:2015nua,Han:2016gvr})
has been performed over the recent years in which the parameter space
of the CMSSM was confronted with a broad set of experimental
constraints on several observables: the Higgs mass and the Higgs decay
rates in different channels at the LHC; the value of the relic density
as measured by PLANCK (and previously WMAP); the lower bounds on SUSY
masses as directly measured by CMS and ATLAS; the measured values of
several $b$-physics rare decays like, e.g., \brbxsgamma,
\brbsmumu, or \brbutaunu; the measurement of the anomalous magnetic
moment of the muon, \deltagmtwomu, which shows a $\sim3\sigma$
discrepancy with the SM value. In the modern, state-of-the-art
approach, these constraints are generally implemented via a global
likelihood function, constructed to compare the measured value of the
observables with their calculated values in the SUSY parameter space.
Observables are usually calculated with sophisticated numerical codes,
and the likelihood function is used to determine statistically
preferred likelihood (if one performs a frequentist analysis based on
the profile likelihood) or, alternatively, credibility (if one
performs instead a Bayesian analysis based on the posterior
probability) regions of the parameter space.

\begin{figure}[t]
\centering
\subfloat[]{%
\includegraphics[width=0.47\textwidth]{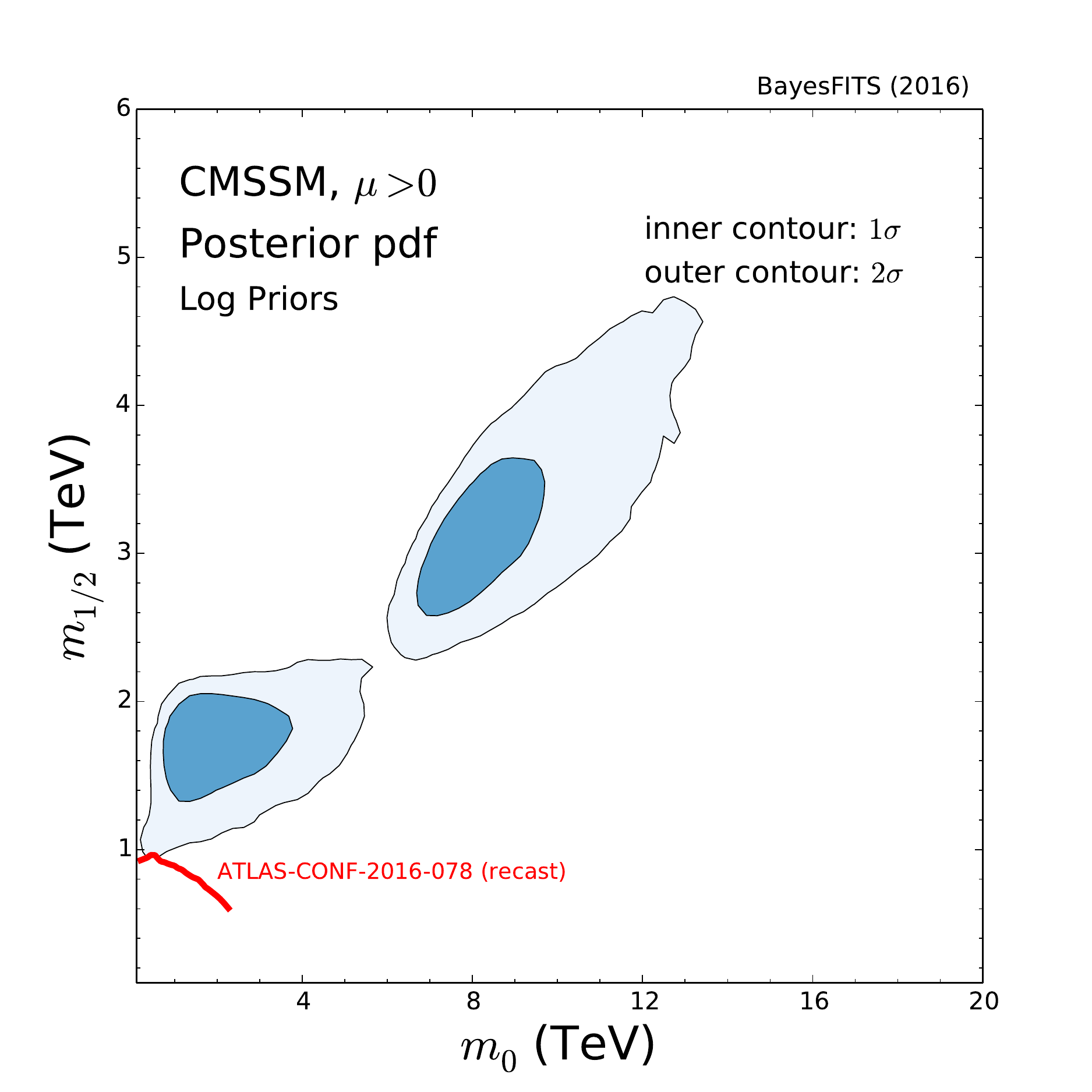}
}%
\hspace{0.02\textwidth}
\subfloat[]{%
\includegraphics[width=0.47\textwidth]{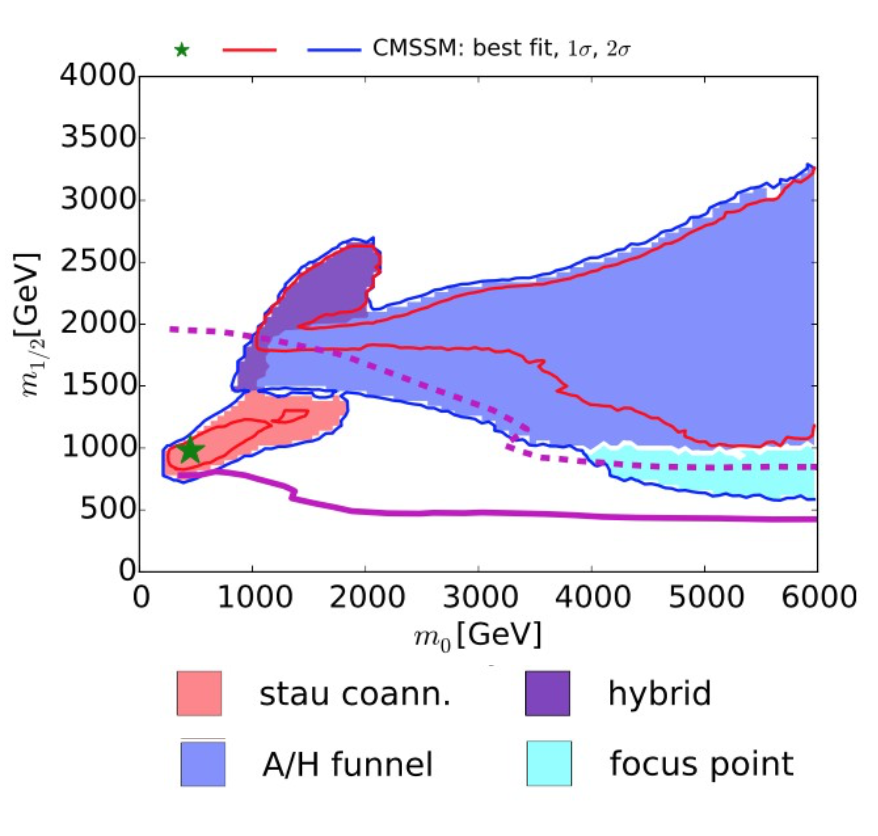}
}%
\caption{\footnotesize (a) The updated $1\sigma$ and $2\sigma$ Bayesian marginalized
  credibility  regions of the (\mzero, \mhalf) plane according to the BayesFITS
  group.\cite{Kowalska:2013hha,Roszkowski:2014wqa}. The most recent
  constraint from SUSY searches at the LHC\cite{ATLAS-CONF-2016-078},
  recast using the code of\cite{Kowalska:2016ent}, is shown as a solid
  red line for comparison. (b) The $1\sigma$ and $2\sigma$
  profile-likelihood regions in the (\mzero, \mhalf) plane of the
  CMSSM according to the MasterCode group\cite{Bagnaschi:2015eha}.
  The color code describes the mechanism for the neutralino relic
  abundance in each region. }
\label{fig:CMSSM}
\end{figure}

In \reffig{fig:CMSSM}(a) we present the $1\sigma$ and $2\sigma$
Bayesian credibility regions of the marginalized posterior probability
in the (\mzero, \mhalf) plane of the CMSSM. The figure presents an
updated version of plots previously shown in
Refs.\cite{Kowalska:2013hha} and\cite{Roszkowski:2014wqa}, obtained
now by incorporating in the likelihood function the most recent
constraints from direct squark and gluino searches at the
LHC\cite{ATLAS-CONF-2016-078} (we use the code of
Ref.\cite{Kowalska:2016ent} to recast the experimental data) and the
recent constraints from direct searches of DM in
LUX\cite{Akerib:2016vxi}.  In \reffig{fig:CMSSM}(b) we show the
$1\sigma$ and $2\sigma$ likelihood regions in the (\mzero, \mhalf)
plane of the CMSSM, following from a frequentist analysis of
Ref.\cite{Bagnaschi:2015eha}. The color code is used to indicate the
different mechanisms by which the correct relic density of the
neutralino is obtained in the early Universe, see
\refsec{sec:reldens}.
 
Note that credibility and likelihood regions are not extremely
dissimilar from one another (within the overlapping ranges of \mzero\
and \mhalf) despite the very different concepts of statistics applied in
both panels. In the
bottom left corner of \reffig{fig:CMSSM}(a) one can see a Bayesian credibility ``island'',
representing the bulk of the $A$-funnel region and a faint appearance
of the stau-coannihilation region surviving
the most recent LHC bound, in
agreement with \reffig{fig:CMSSM}(b).  This is the region of the
parameter space where the neutralino is predominantly bino-like. 

In \reffig{fig:CMSSM}(a) the parameter space is scanned to larger
values of \mzero\ and \mhalf, well into the \tev-scale region that
most easily allows one to accommodate the correct value of the Higgs
mass.  This region features the existence of a second, and actually
larger, ``island'' in the parameter space, characterized by an almost
pure higgsino-like neutralino that, as was explained in
\refsec{sec:reldens}, is characterized by the LSP higgsino-like mass
around $1\tev$ in order to give the correct relic density. Frequentist
analysis also shows the emergence of this region, despite the smaller
region of \mzero\ and \mhalf\ covered in \reffig{fig:CMSSM}(b).

Taking the view that the Higgs mass implies a multi-\tev\
scale of superpartners, having the LSP at 1\tev\ without having to
adhere to any special mechanism for obtaining the right \abundchi\
appears to be a rather intriguing and well motivated
solution\cite{Kowalska:2015kaa}, with promising prospects for DM
searches, as discussed below.

%
%
%

\subsubsection{Prospects for WIMP searches in GUT-constrained models}

\begin{figure}[t]
\centering
\subfloat[]{%
\includegraphics[width=0.47\textwidth]{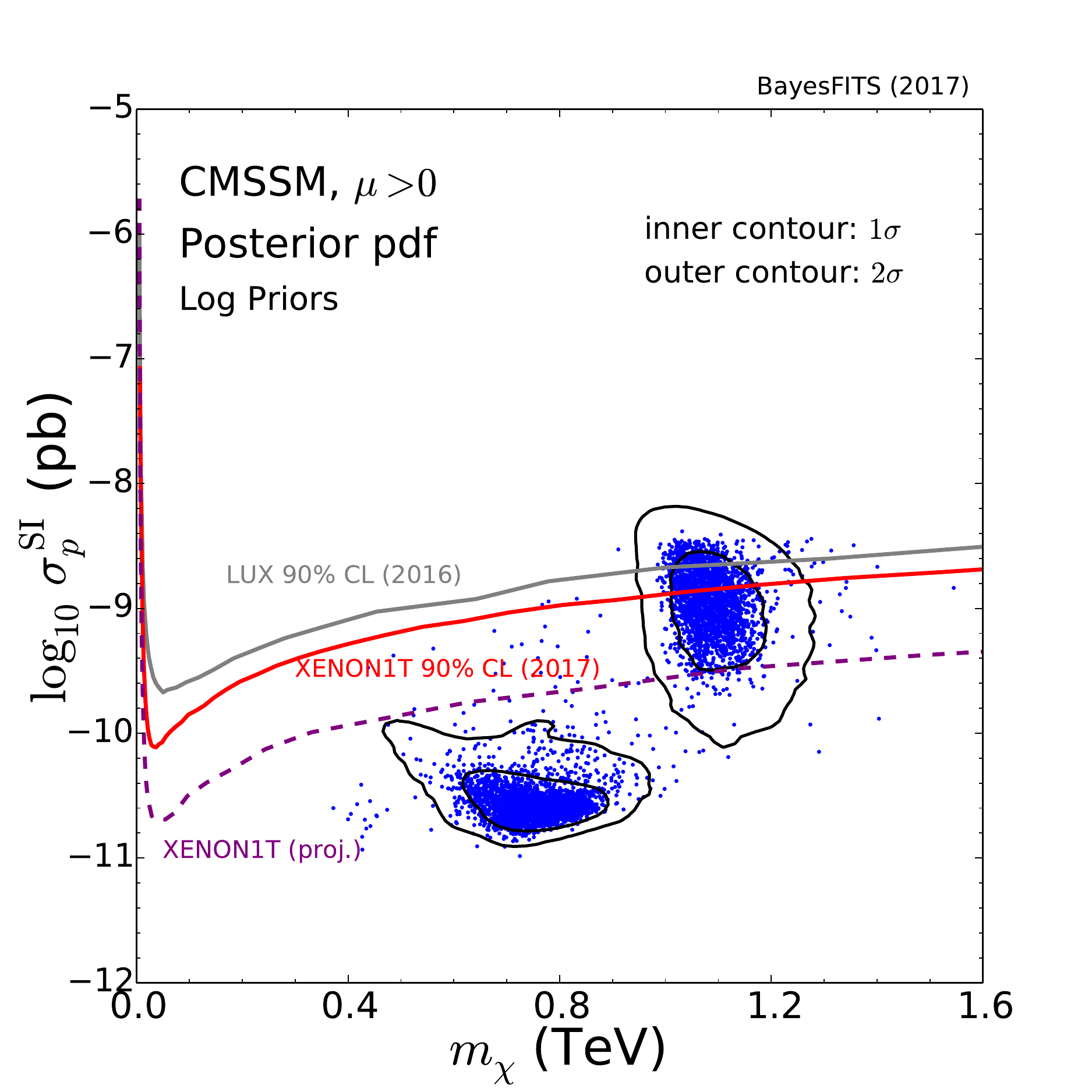}
}%
\hspace{0.02\textwidth}
\subfloat[]{%
\includegraphics[width=0.47\textwidth]{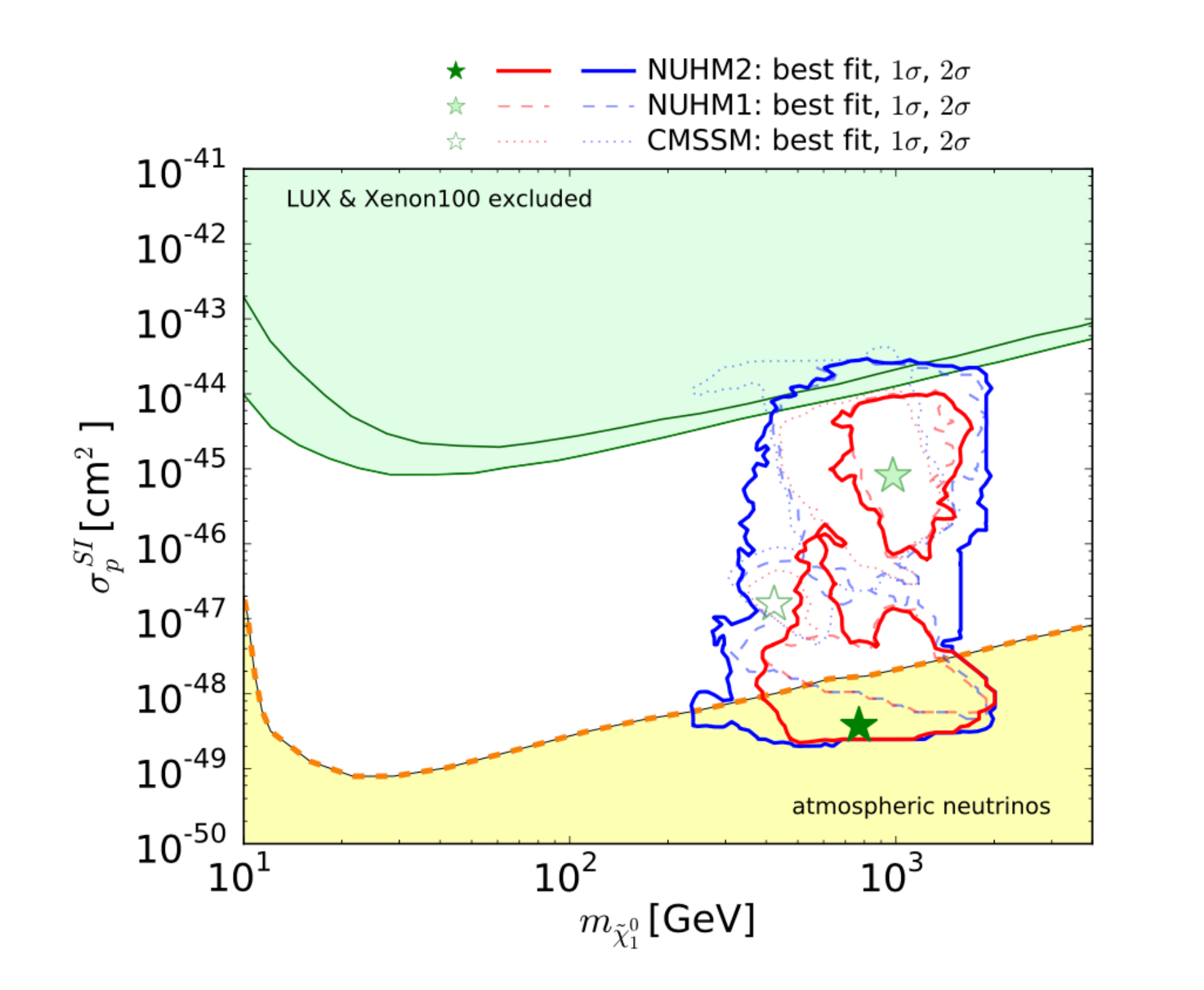}
}%
\caption{\footnotesize (a) Solid contours show
  the Bayesian $1\sigma$ and $2\sigma$ credible regions of the CMSSM
  in the (\mchi, \sigsip) plane, with LHC and DD constraints updated
  with respect to\cite{Kowalska:2013hha,Roszkowski:2014wqa}. The
  scattered blue points, sampled from the posterior probability
  distribution, belong to the $2\sigma$ region of the global profile
  likelihood. For comparison, the solid gray line marks the final
  published 90\%~C.L. LUX bound\cite{Akerib:2016vxi}, which is included in the likelihood function. 
The solid red
  line shows the recent first limit from XENON1T\cite{Aprile:2017iyp},
  whereas the dashed purple line gives the projected reach of
  XENON1T.  (b) The $1\sigma$ (red solid) and $2\sigma$ (blue solid) profile likelihood region in the (\mchi, \sigsip) plane
of the NUHM according to the MasterCode group\cite{Buchmueller:2014yva}.}
\label{fig:CMSSM_DD}
\end{figure}

Contours of the 68\% and 95\% Bayesian credible region of the CMSSM in
the (\mchi, \sigsip) plane are shown in \reffig{fig:CMSSM_DD}(a), which updates the equivalent plots presented
in Refs.\cite{Kowalska:2013hha} and\cite{Roszkowski:2014wqa}. As
stated above, the likelihood function includes the recently
published LUX data\cite{Akerib:2016vxi}, 
which we have incorporated here following a procedure similar to
Ref.\cite{Roszkowski:2016bhs}. 
Note that in recent years several
numerical codes have been devised to appropriately account for DD data
in the form of a likelihood function, see
Refs.\cite{Gresham:2013mua,Buchmueller:2014yoa,Savage:2015xta,Huang:2016pxg}.
To facilitate comparison, we mark as a solid gray line in
\reffig{fig:CMSSM_DD}(a) the 90\%~C.L. upper bound as given by the
LUX collaboration. The newest first
results from XENON1T\cite{Aprile:2017iyp} are shown instead as a solid red line.  

Note how parts of the 95\% credible posterior regions extend somewhat
above the 90\%~C.L. limit given by the experimental collaboration.
This is due, on the one hand, to the non-negligible difference that
exists between the 90\% and 95\% confidence bound (which
Ref.\cite{Baer:2016ucr} did not take into account) when the
likelihood function is not very steep over the parameter space.  
On the other hand, as the likelihood function's slope is quite gentle,
the probability density shows some sensitivity to the choice of
Bayesian priors, which in this case pull towards larger values of
\sigsip\ by favoring lighter gauginos. If, for instance, a linear (flat) prior was chosen instead, then the
$\sim1\tev$ higgsino would become even more pronounced. However, the
corresponding ranges of \sigsip\ are not as much prior-dependent. 
For completeness, we superimpose to the plot a set of viable scan points (in blue) that,
while drawn from the posterior probability density, delimit the
extension of the 95\%~C.L.  profile-likelihood region. It is important to note that in the models with unified gaugino masses at the GUT scale, e.g., the CMSSM, one typically does not obtain nearly pure higgsino DM, for which \sigsip\ could be arbitralily low, as it is seen in the low-energy MSSM. Once gaugino masses are allowed to grow large to minimize their mixing with the higgsino component, also the SM-like Higgs boson mass increases due to the impact of gaugino mass parameters on the RGE running of the stop and soft Higgs masses. 
Precise determination of the lower limit on \sigsip\ is sensitive to the accuracy of the calculation of $m_h$. The results presented in \reffig{fig:CMSSM_DD}(a) 
correspond to $m_h$ obtained with \texttt{FeynHiggs~2.10.0}\cite{Heinemeyer:1998yj,Hahn:2013ria}.

Figure~\ref{fig:CMSSM_DD}(a) shows the same two regions featured in
\reffig{fig:CMSSM}(a): the $A$-funnel on the left, and the $\sim1\tev$
higgsino on the right.  The latter clearly presents the better
prospects for DM searches, as the bulk of the parameter space falls
within the reach of tonne-scale underground detectors, which we summarize
here schematically with the dashed purple line giving, technically,
the projected 2-year sensitivity of XENON1T from Ref.\cite{Aprile:2015uzo}.

It is worth emphasizing that the $\sim1\tev$ higgsino region emerges
as a robust solution in a much wider class of constrained SUSY models -- in
fact it appears without having to employ any special mass relations to
obtain the correct relic density -- as soon as the gauginos become heavier than
1\tev, consistent with the SUSY breaking scale in the multi-\tev\
regime, as suggested by the Higgs boson mass value and LHC direct
limits on superpartners. As an example, in \reffig{fig:CMSSM_DD}(b) we
show the $1\sigma$ (red solid) and $2\sigma$ (blue solid) profile
likelihood region in the (\mchi, \sigsip) plane of the NUHM according
to the MasterCode group\cite{Buchmueller:2014yva}. The figure shows
the presence of the $\sim 1\tev$ higgsino on the right.

 In contrast, the existence or not and
the relative sizes of the stau-coannihilation and $A$-funnel regions
depend to some extent on the initial boundary conditions assumed for
the parameters of the model at hand -- as they both require in some
form the overlapping of certain mass values that could originate from
different sectors of the theory.

\subsection{The pMSSM \label{sec:pmssm} }

As we have seen in \refsec{sec:cmssm}, the $\sim1\tev$ higgsino seems
to be an attractive candidate for WIMP DM in SUSY models with boundary
conditions defined at the GUT scale, and it features very good
potential for a timely detection in one-tonne detectors.

Low energy SUSY is a very broad framework,
able to accommodate several possibilities for the spectrum of
superpartners. As SUSY must be broken in a hidden sector, little is known about the most likely mass pattern
for the supersymmetric particles, and one must rely on reasonable assumptions driven by theory considerations. 
Thus, in order to analyze DM signatures in a general and model-independent SUSY scenario we
analyze here the DM issue in the phenomenological MSSM (pMSSM).

The pMSSM\cite{Djouadi:1998di} is the most general parametrization of
the MSSM, based only on assumptions of Minimal Flavor Violation,
$R$-parity conservation, and a level of CP violation not exceeding that of the SM. 
These assumptions reduce the over hundred free parameters
potentially present in \refeq{softlagr} of the MSSM down to 19, all
defined at the SUSY scale. It is easy to see that all the
scenarios discussed in \refsec{sec:cmssm} can be described in this
framework by choosing appropriate boundary conditions. The same is true for
other popular scenarios for SUSY breaking like, e.g., anomaly
mediation\cite{Giudice:1998xp,Randall:1998uk}.

Since the number of free parameters in the pMSSM remains quite large, there is no real issue
in fitting all the constraints belonging to the standard set described above.
In particular, $\abundchi\approx 0.12$ -- in addition to all relevant collider
constraints -- can be fairly easily satisfied in different parts of
the parameter space for different neutralino WIMP
compositions. 

\begin{figure}[t]
\centering
\subfloat[]{%
\includegraphics[width=0.50\textwidth]{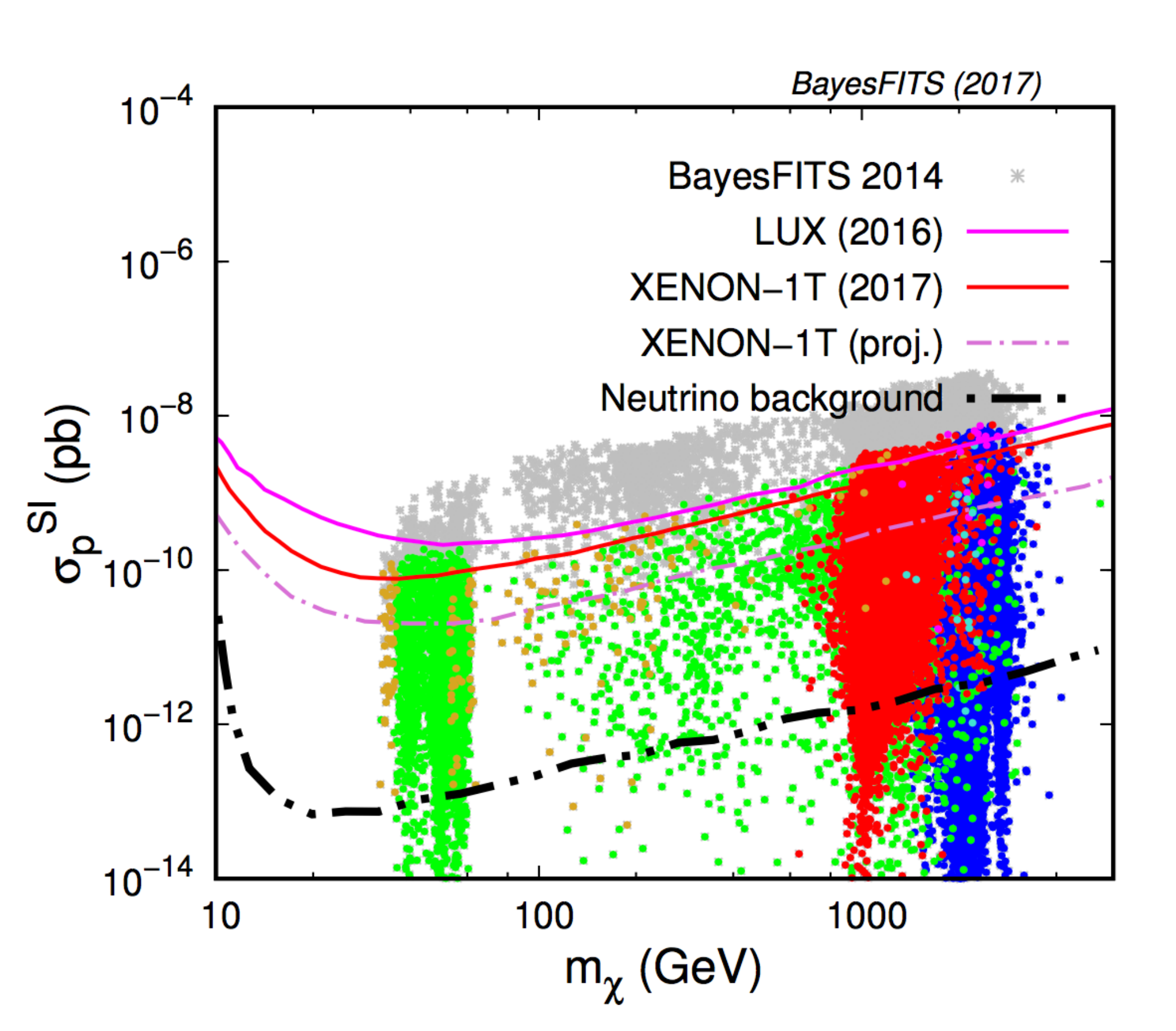}
}%
\\
\subfloat[]{%
\includegraphics[width=0.40\textwidth]{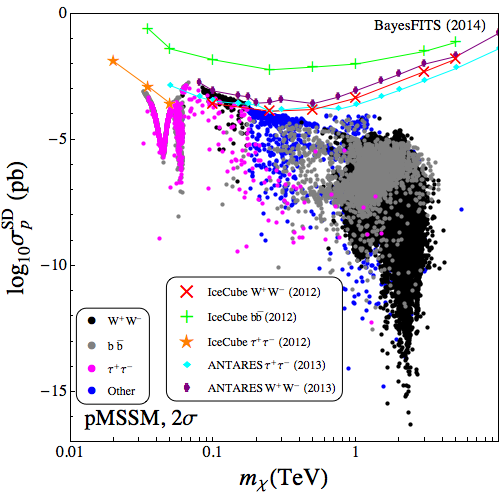}
}%
\hspace{0.02\textwidth}
\subfloat[]{%
\includegraphics[width=0.50\textwidth]{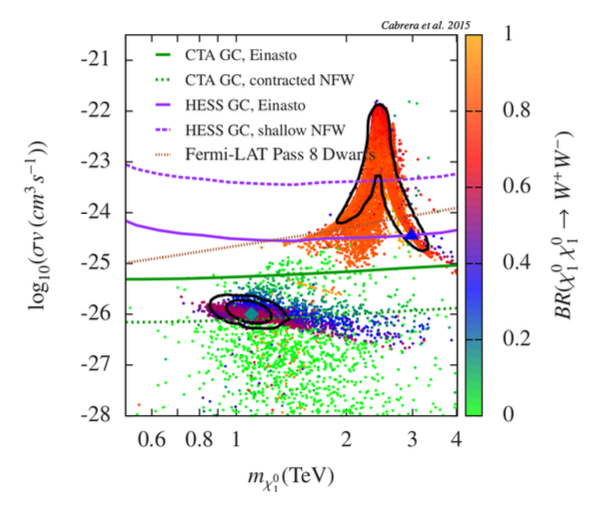}
}%
\caption{\footnotesize (a) The parameter space of the pMSSM with $\abundchi\approx0.12$ in the (\mchi, \sigsip) plane. 
Points in green are characterized by a 90\% or more bino composition of the neutralino; 
points in red are $>90\%$ higgsino; and points in blue are $>90\%$ wino. 
Bino/higgsino admixtures are shown in gold, wino/higgsino in magenta, and wino/bino in cyan. 
The plot updates the equivalent figure in Ref.\cite{Roszkowski:2014iqa} by including in the likelihood function the 
DD constraint from\cite{Akerib:2016vxi}, which we also show explicitly as a magenta solid line. 
We have also added the most recent XENON1T bound\cite{Aprile:2017iyp}, as a red solid line.
(b) A plot of the bounds on \sigsdp\ from neutrinos from the Sun at IceCube\cite{IceCube:2011aj,Aartsen:2012kia,Aartsen:2016exj} 
and ANTARES\cite{Adrian-Martinez:2013ayv,Adrian-Martinez:2016gti},
for different final states of annihilation, taken from Ref.\cite{Roszkowski:2014iqa}.
The limits are presented for the $W^+W^-$, $b\bar{b}$, and $\tau^+\tau^-$ final states.
(c) The sensitivity of several ID searches to the large mass region of the MSSM in the (\mchi, \sigv) plane, 
as a function of the branching ratio $\textrm{BR}(\chi\chi\rightarrow W^+W^-)$. The figure is taken from\cite{Catalan:2015cna}.}  
\label{fig:pmssmOh2}
\end{figure}

We show in \reffig{fig:pmssmOh2}(a) the $2\sigma$ region in the (\mchi, \sigsip) plane of the pMSSM, 
emerging from the profile likelihood of the global constraints. The neutralino composition of the points in green is 90\% or more
bino-like; points in red are for more than 90\%
higgsino-like; and points in blue are at least 90\% wino-like. 
Bino/higgsino admixtures are shown in gold, wino/higgsino in magenta, and wino/bino
in cyan. Figure~\ref{fig:pmssmOh2}(a) updates the equivalent plot of Ref.\cite{Roszkowski:2014iqa}, but 
a similar picture emerges in pMSSM global analyses by other groups, 
which can feature slightly different choices for the set of constraints or in the number of input parameters 
(see, e.g.,\cite{Cahill-Rowley:2014boa,Bagnaschi:2015eha,Aad:2015baa}).
Note that the constraints from the 2016 LUX results\cite{Akerib:2016vxi} are implemented in the likelihood function, so
that the region marked by gray points, which was belonging to the $2\sigma$ region in\cite{Roszkowski:2014iqa}, 
is now shown as excluded at the 95\%~C.L. We also show with a solid red line the recent 90\%~C.L. bound from XENON1T\cite{Aprile:2017iyp}, not included in the likelihood function.
One can see that there are countless possibilities for a neutralino DM in agreement with all the relevant constraints. 

In addition, it is possible to have accidental cancellations in the neutralino couplings to the $Z$ and $h$ bosons, as well as cancellations between the heavy and light Higgs diagrams, which result in the suppression of direct detection cross section in so-called blind spot regions of the parameter space\cite{Ellis:2000ds,Baer:2006te,Cheung:2012qy,Huang:2014xua} (see\cite{Han:2016qtc} for a recent study). 
Several points characterized by bino/higgsino admixtures, shown in gold color in \reffig{fig:pmssmOh2}(a), must belong to blind spots to evade the most recent DD bounds. It has been shown, e.g.,
in\cite{Badziak:2017the}, that LHC searches for heavy Higgs bosons in the $\tau^+\tau^-$ channel are currently extensively 
probing much of the parameter space giving rise to these special regions.   

The strongest indirect limits on the spin-dependent scattering cross section for neutralino DM with mass exceeding the 
$\sim 100\gev$ range are given by IceCube/DeepCore\cite{IceCube:2011aj,Aartsen:2012kia,Aartsen:2016exj} and 
ANTARES\cite{Adrian-Martinez:2013ayv,Adrian-Martinez:2016gti}, from observation of neutrinos from the Sun. 
In \reffig{fig:pmssmOh2}(b) we show the bounds as presented in Ref.\cite{Roszkowski:2014iqa}. The color code highlights the 
main annihilation final state of the scan points. The plot shows that, while a measurement of \sigsdp\ remains a very important 
complementary test, the parameter space of the MSSM is likely to be probed more deeply by other means. 

In \reffig{fig:pmssmOh2}(c) we show the reach of several $\gamma$-ray indirect detection searches in the (\mchi, \sigv) plane of 
the pMSSM with $\abundchi\approx 0.12$, as a function of the branching fraction $\textrm{BR}(\chi\chi\rightarrow W^+W^-)$. 
The figure is taken from Ref.\cite{Catalan:2015cna}.   
The points with $\mchi\approx 2-3\tev$ and large branching ratio to $W^+W^-$ are those characterized by a large wino composition (cf. \reffig{fig:pmssmOh2}(a)). As these points are subject to the Sommerfeld enhancement\cite{Hisano:2002fk,Hisano:2004ds}, 
a non-perturbative effect that can give a significant boost to the annihilation cross section, 
they appear to be in tension with observations from the Galactic Center at the Cherenkov telescope H.E.S.S.\cite{Abramowski:2013ax}.
The extent of the tension depends of course on the choice of halo profile. This was observed first in\cite{Cohen:2013ama,Fan:2013faa,Hryczuk:2014hpa}. 
The $\sim 1\tev$ higgsino region can also be seen in \reffig{fig:pmssmOh2}(c), for slightly lower \sigv, and characterized by 
$\textrm{BR}(\chi\chi\rightarrow W^+W^-)\approx 0.5$, as the remaining 50\% is dominated by the $Zh$ final state.

One can see in \reffig{fig:pmssmOh2}(c) that the Cherenkov Telescope Array (CTA), with $\sim 500\,\textrm{h}$ 
of observation of the Galactic Center, will probe most of the pMSSM 
parameter space with DM mass in the TeV range. We make the point again that in the majority of 
SUSY models with parameters defined at some high scale this is the region emerging as favored after 
the discovery of the Higgs boson at 125\gev. 
Thus, CTA will prove to be an indispensable instrument to probe ranges of SUSY-model parameters that would otherwise be entirely 
out of reach by other direct means. 

\begin{figure}[t]
\centering
\subfloat[]{%
\includegraphics[width=0.33\textwidth]{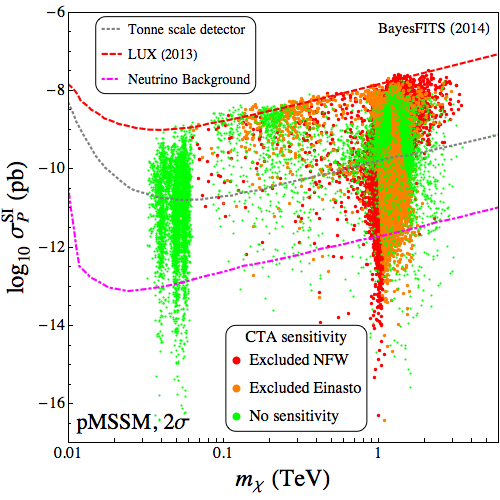}
}%
\subfloat[]{%
\includegraphics[width=0.33\textwidth]{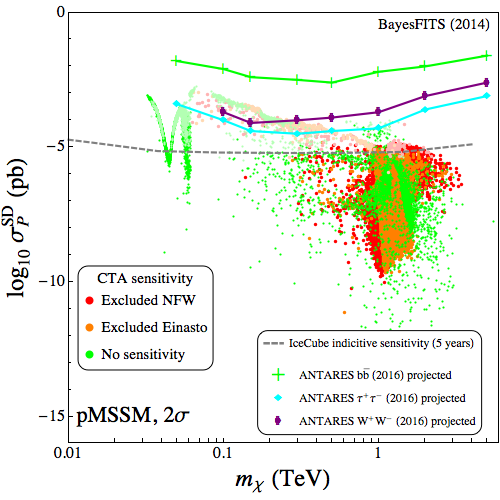}
}%
\subfloat[]{%
\includegraphics[width=0.33\textwidth]{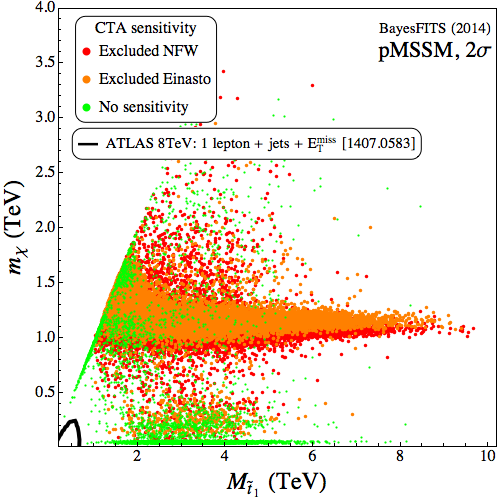}
}%
\caption{\footnotesize (a) The sensitivity of CTA $500\,\textrm{h}$ Galactic Center observation in the (\mchi, \sigsip) plane
of the pMSSM, for two choices of halo profile: 
NFW (red points), or Einasto (red+orange points).
The approximate projected sensitivity of 1-tonne detectors is shown as a dotted gray line. 
The onset of the atmospheric and diffuse supernova neutrino background is shown with a dot-dashed magenta line.
(b) The sensitivity of CTA to the pMSSM in the (\mchi, \sigsdp) plane. 
Lighter shaded points are within the projected sensitivity of IceCube/DeepCore. 
The dashed gray line is indicative of IceCube's future sensitivity.
(c) Sensitivity of CTA in the (\mstopone, \mchi) plane. The thick black line shows the approximate LHC lower bound on stop/neutralino masses. All figures are taken from\cite{Roszkowski:2014iqa}.}  
\label{fig:complem}
\end{figure}

To highlight the idea of complementarity, we show in \reffig{fig:complem}(a) 
the reach of CTA with $500\,\textrm{h}$ of observation of the Galactic Center, 
compared to the reach of 1-tonne detectors in the (\mchi, \sigsip) plane. The color code is explained in the caption.
In \reffig{fig:complem}(b) we present the equivalent picture in the (\mchi, \sigsdp) plane, compared to the estimated IceCube reach.
And finally, we show in \reffig{fig:complem}(c) the reach of CTA compared to the present limits on stop mass 
obtained in simplified models at the LHC. Figure~\ref{fig:complem} is taken from\cite{Roszkowski:2014iqa}.
Improvements in the LHC limits are not expected to have any effect on the sensitivity of CTA. 
Indeed, CTA remains sensitive to spectra where the gluinos and squarks lie well beyond the reach of present and future colliders. 

\subsection{Going beyond standard assumptions}

In this topical review we have focused on reasonable but simplest
underlying assumptions about DM that are usually made in
phenomenological studies of the subject. One is that the DM in
the Universe comprises (or is dominated by) just one species. This
translates into insisting that its relic density saturates the
measured value of about 0.12. However, as we already mentioned in
Section~\ref{sec:departuresDM}, the correct WIMP DM relic density can
be obtained even if the relevant annihilation rate varies from the
canonical thermal value.

Another usually made assumption, or actually set of assumptions, is
that, in the early Universe DM particles were generated only (or
mostly) through their freeze-out out of thermal equilibrium.  Although
these assumptions are certainly sensible, neither of them is
necessarily correct. It is therefore interesting to see how various results and conclusions derived in the literature can be affected by going beyond the standard freeze-out paradigm.  In this section we will briefly illustrate this with a few examples in the context of neutralino DM. For a more comprehensive review see, \eg,\cite{Baer:2014eja}.

\subsubsection{Multi-component DM\label{sec:multidm}}

There is really no reason, other than simplicity, to insist that the
whole of DM is made up of just one species of particles. There exist
several scenarios, in SUSY or not, where this is not necessarily the
case. In fact, the idea that, for instance, the neutralino and the
axion -- arguably the two DM candidates most strongly motivated by particle physics -- could easily co-exist in the Universe in basically any
proportion has been around for decades.  As was mentioned in
\refsec{sec:reldens}, this can be motivated by insisting on keeping
the scale of SUSY breaking as low as possible, in order to reduce the
level of fine tuning among SUSY parameters.  In such regions of the
parameter space where the neutralino is close to a pure higgsino (or a
pure wino), the relic density of DM is too low as DM mass is not large
enough.  This can also be consistently realized in specific, well
motivated models\cite{Baer:2011hx,Baer:2011uz,Bae:2015rra}. Other
possibilities include employing an additional, non-thermal
component\cite{Allahverdi:2012wb}.

\begin{figure}[t]
\centering
\includegraphics[width=0.60\textwidth]{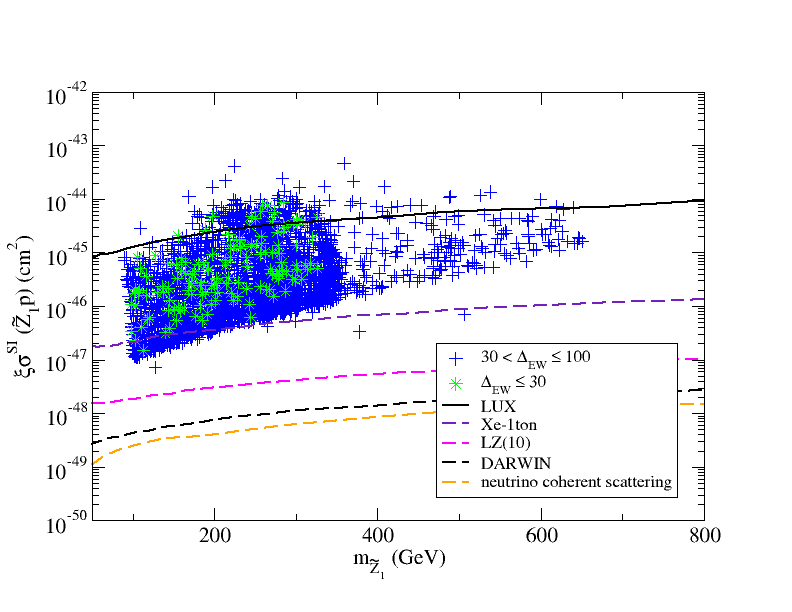}
\caption{\footnotesize 
Reach of the spin-independent direct detection searches for neutralino DM in the scenario with two-component DM (neutralino and axino). The plot is taken from Ref.\cite{Baer:2015zda}.} 
\label{fig:baerlowmchi}
\end{figure}

We illustrate this in \reffig{fig:baerlowmchi}.
We can see that prospects for WIMP detection remain good, despite
lower number density. For earlier works reaching similar conclusions,
see, \eg,\cite{Duda:2001ae}.


\subsubsection{Low reheating temperature\label{sec:lowrehchi}}

It is usually assumed that, when WIMPs freeze out the thermal plasma,
the Universe has already reached radiation dominated (RD) thermal
equilibrium. In other words, the value of the reheating temperature
$T_R$, which marks the onset of the RD epoch, is assumed to be much
larger than the freeze-out temperature. This does not  have
to be the case and can strongly alter our conclusions about WIMP DM
properties.

Low $T_R$ can result from an extended reheating period in the
evolution of the Universe after an inflationary epoch. In addition to
modifying the DM population from freeze-out, it can also be changed by
an additional entropy production from decays of some heavy species
that took place after the DM freeze-out. As a result one can either
reduce or increase the DM relic density depending on whether the
additional entropy production is accompanied by efficient direct
and/or cascade decays of the heavy field to the DM particles.

\begin{figure}[t]
\centering
\subfloat[]{%
\label{fig:a}%
\includegraphics[width=0.47\textwidth]{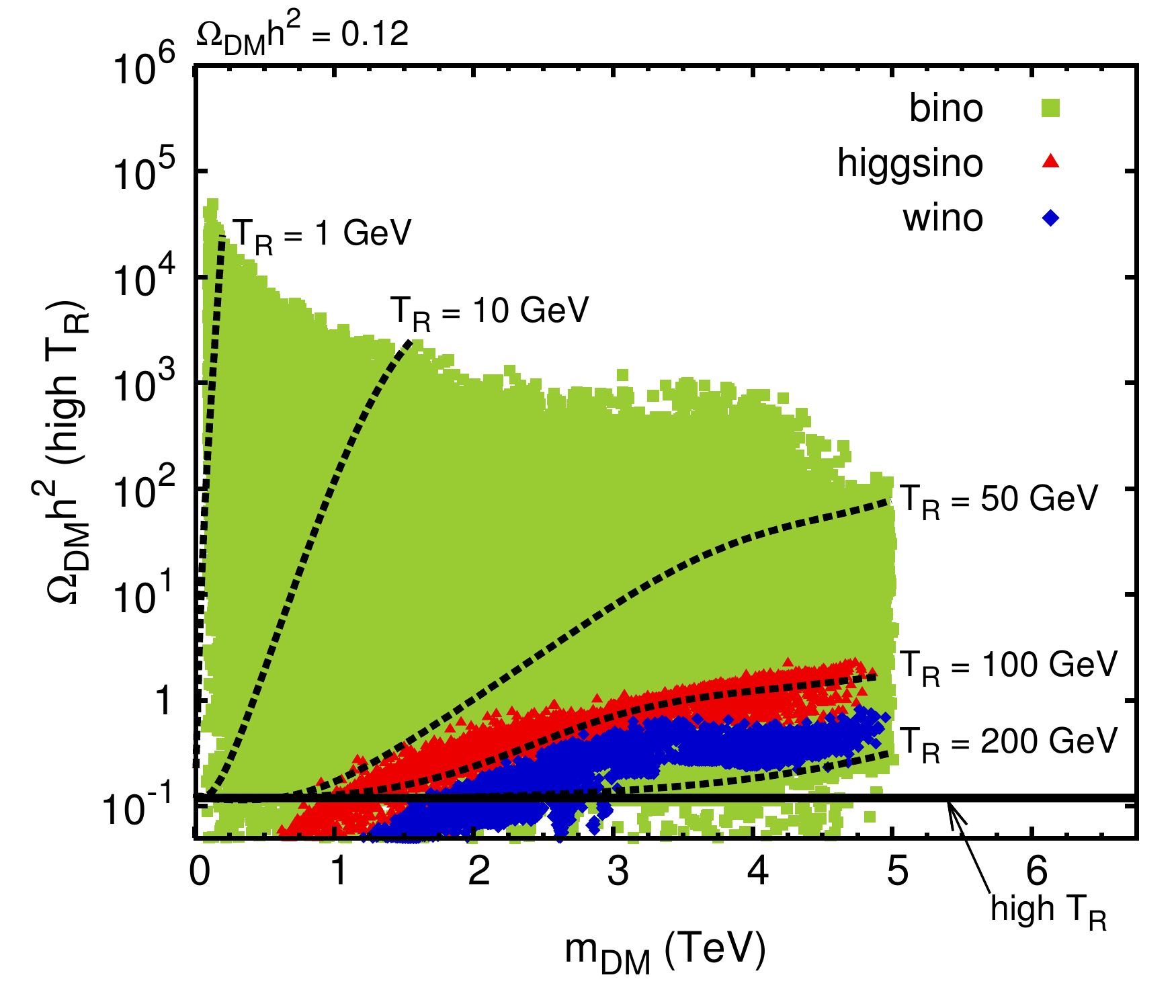}
}%
\hspace{0.02\textwidth}
\subfloat[]{%
\label{fig:b}%
\includegraphics[width=0.47\textwidth]{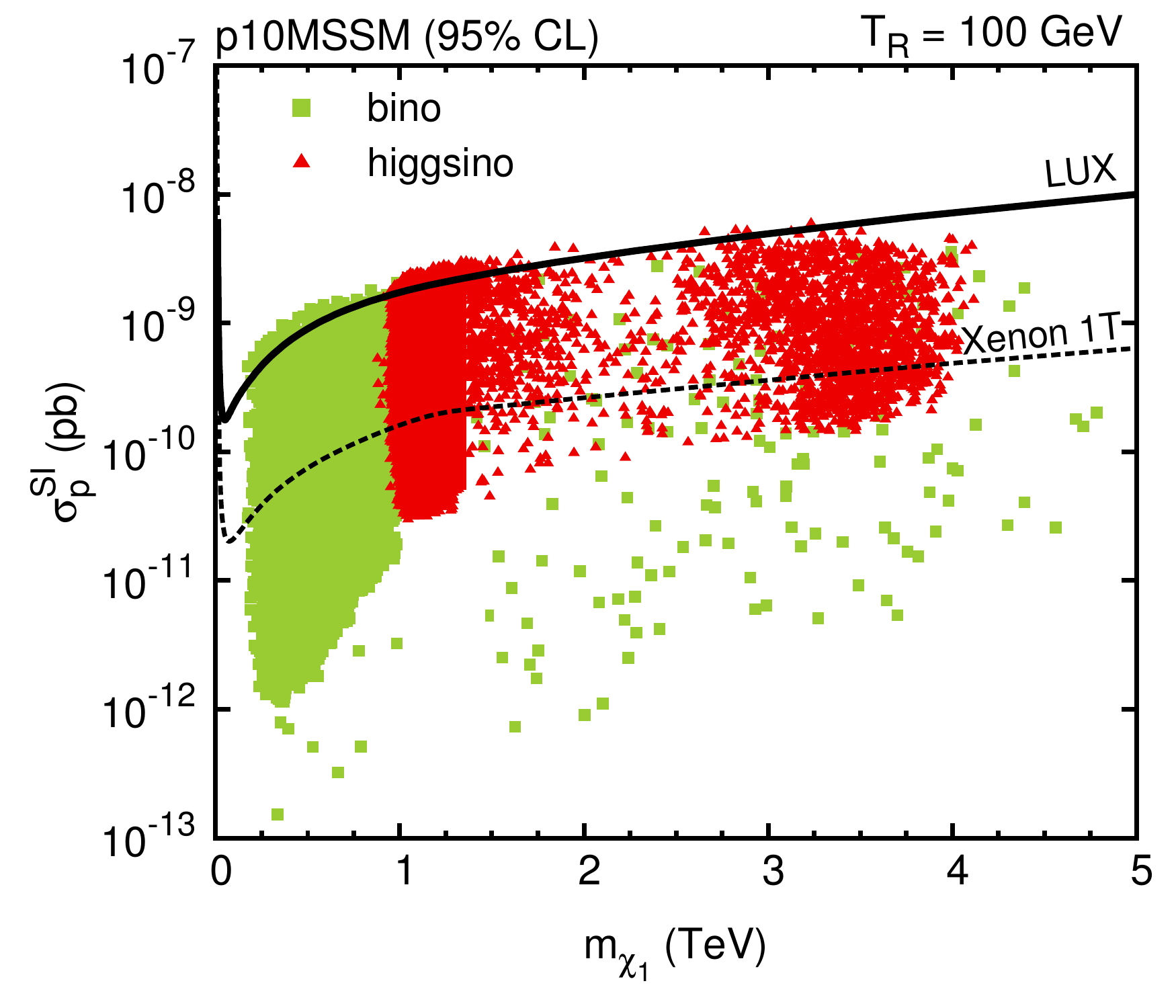}
}%
\caption{\footnotesize \protect\subref{fig:a} Contours (black dotted) of constant $\abunddm = 0.12$ for different values of the reheating temperature, $T_R$, in the MSSM, in the $(m_{\rm{DM}},
  \abunddm(\textrm{high }T_R)\,)$ plane, where $\abunddm(\textrm{high }T_R)$ corresponds to the standard cosmological scenario for which the correct value of the neutralino relic density is obtained along the the solid black
  horizontal line. Green squares
  correspond to bino DM, red triangles to higgsino DM and blue diamonds to the wino DM case. Negligible direct and/or cascade decays of the inflaton field are assumed. \protect\subref{fig:b} Direct detection cross section, \sigsip, as a function of
  \mchi\ in the ten-parameter subset of the MSSM (p10MSSM) for which 95\%~C.L. region (including the relic density constraint) for $T_R = 100$\gev\ is shown. The solid (dashed) black
  lines correspond to LUX (projected XENON1T) limit on \sigsip. 
  Color coding as in the left panel. Taken from Ref.\cite{Roszkowski:2014lga}.}
\label{fig:MSSMlowTR}
\end{figure}

From the phenomenological point of view, this mechanism allows one to
fit the relic density constraint for almost any scenario with
neutralino DM\cite{Gelmini:2006pw,Gelmini:2006pq} (for a recent
discussion, see\cite{Roszkowski:2014lga}). We illustrate this in
\reffig{fig:MSSMlowTR}\subref{fig:a} where the lines of constant
$\abundchi =0.12$ are shown for several values of the reheating
temperatures, $T_R=1,10,50,100,200$\gev\ as a function of the
neutralino DM mass and $\abundchi\ (\textrm{high }T_R)$, \ie, the value
of the DM relic density corresponding to the standard cosmological
scenario with high $T_R$. In particular one can see that for $T_R\approx
100\gev$ the correct value of \abundchi\ for higgsino DM can be
obtained for masses significantly larger than 1\tev. Such a heavy
higgsino can still be within the reach of one-tonne detectors, as can
be seen in \reffig{fig:MSSMlowTR}\subref{fig:b}, where we show the
direct detection spin-independent cross section, \sigsip, as a
function of \mchi, for phenomenologically favored points in the
p10MSSM obtained assuming $T_R=100\gev$\cite{Roszkowski:2014lga}.

\section{Summary and conclusions \label{sec:conclusions}}

It is not easy to look for the invisible but, in the case of DM, it is
certainly worth the effort. A detection of a DM signal is likely not
only to confirm the common belief that most of the dark mass in
the Universe is made up of WIMPs but to hopefully shed some light on the particle physics framework that it is part of. In this topical review we have
provided an overview of the current experimental situation, paying
particular attention to current bounds and recent claims and hints of
a possible signal in a wide range of experiments. On the particle
physics side, we reviewed several candidates for explaining the DM,
concentrating mostly on the class of WIMPs that could be produced
mostly through the freeze-out mechanism. We have paid attention to the
neutralino of SUSY since it remains the most strongly motivated
candidate that additionally shows excellent detection prospects. We
have emphasized that the currently most interesting -- in our opinion
-- case of $\sim1\tev$ higgsino-like neutralino in unified SUSY models
will nearly fully be tested in the new tonne-scale underground
detectors which are coming online. However, one should remember that,
even if eventually a genuine DM signal is detected, then it is likely
that several measurements will have to be made in both direct and
indirect detection experiments -- and this will likely be possible
only under rather favorable conditions -- in order to shed some light
on the actual nature of the WIMP.

\section*{Acknowledgements \label{sec:acks}}

We are indebted to Andrew J. Williams for profusely contributing to
the early stages of this work. We would also like to thank Kamila
Kowalska for her input on the recasting of the recent LHC bounds and
for discussions. ST would like to thank 
Kevork~Abazajian, Jonathan~L.~Feng, Dan~Hooper, Manoj~Kaplinghat,
Christopher~Karwin, Simona~Murgia and Timothy~M.P.~Tait for helpful
discussions and comments. LR is supported in part by the National Science
Centre (NCN) research grant No. 2015-18-A-ST2-00748 and by the
Lancaster-Manchester-Sheffield Consortium for Fundamental Physics
under STFC Grant No.~ST/L000520/1.  The work of EMS is supported by
the Alexander von Humboldt Foundation. ST is partly supported by the Polish
Ministry of Science and Higher Education under research grant
1309/MOB/IV/2015/0 and by NSF Grant No.~PHY-1620638.

\bibliographystyle{JHEP}

\bibliography{dmreview_ropp_2017}

\end{document}